\documentclass[preprint,12pt,number]{elsarticle}
\usepackage{amssymb}
\usepackage{amsmath}
\usepackage{booktabs}
\usepackage{multirow}
\usepackage{tabularx} 
\usepackage{graphicx}
\usepackage{rotating}  
\usepackage{placeins}
\usepackage{afterpage}
\usepackage{caption} 
\usepackage{lineno}
\usepackage{float}
\usepackage{subcaption}  
\usepackage{threeparttable}

\biboptions{sort&compress}
\begin{document}

\begin{frontmatter}

\begin{titlepage}
    \begin{center}
        {\LARGE Bridging Computational Fluid Dynamics Algorithm and Physics-Informed Learning: SIMPLE-PINN for Incompressible Navier-Stokes Equations}\\[0.5cm]

        \textbf{Authors:}\\[0.5cm]
        Chang Wei$^{1}$, Yuchen Fan$^{1}$, Chin Chun Ooi$^{2}$, Jian Cheng Wong$^{2}$, Heyang Wang$^{1,*}$, Pao-Hsiung Chiu$^{2,*}$\\[0.5cm]

        \textbf{Affiliations:}\\[0.3cm]
        $^{1}$Laboratory of Efficient Utilization of Low and Medium Grade Energy, School of Mechanical Engineering, Tianjin University, Tianjin 300350, China\\
        $^{2}$Institute of High Performance Computing, Agency for Science, Technology and Research (A*STAR), Singapore 138632\\[0.5cm]

        \textbf{Corresponding authors:}\\[0.2cm]
        Heyang Wang: heyang.wang@tju.edu.cn\\
        Pao-Hsiung Chiu: chiuph@a-star.edu.sg\\[0.5cm]

        \textbf{Acknowledgment:}\\[0.2cm]
        Chang Wei and Yuchen Fan would like to acknowledge support from the China Scholarship Council for the scholarship to conduct research at Agency for Science, Technology and Research (A*STAR). This research was in part supported by the National Research Foundation, Singapore through the AI Singapore Programme, under the project ``AI-based urban cooling technology development" (Award No. AISG3-TC-2024-014-SGKR)%, in part supported by the National Research Foundation, Singapore and the National Environment Agency under the Weather Science Research Programme (Award No.: WSRP-2025-1R-02-02).

        \vfill
        % \today
    \end{center}
\end{titlepage}


%\title{A physics-informed neural network incorporating velocity-pressure correction inspired by the SIMPLE algorithm for the Navier-Stokes Equations}
\title{Bridging Computational Fluid Dynamics Algorithm and Physics-Informed Learning: SIMPLE-PINN for Incompressible Navier-Stokes Equations}

% \author[inst1]{Chang Wei}
% \author[inst1]{Yuchen Fan}
% \author[inst2]{Chin Chun Ooi}
% \author[inst2]{Jian Cheng Wong}
% \author[inst1]{Heyang Wang\corref{bbb}}
% \author[inst2]{Pao-Hsiung Chiu\corref{aaa}}

% \cortext[bbb]{Corresponding author at: Laboratory of Efficient Utilization of Low and Medium Grade Energy, School of Mechanical Engineering, Tianjin University, Tianjin, 300350, China. \textit{E-mail address:} heyang.wang@tju.edu.cn (H. Wang).}
% \cortext[aaa]{Corresponding author at: Institute of High Performance Computing, Agency for Science, Technology and Research (A*STAR), 138632, Singapore. \textit{E-mail address:} chiuph@a-star.edu.sg (P-H. Chiu).}

% \affiliation[inst1]{organization={Laboratory of Efficient Utilization of Low and Medium Grade Energy, School of Mechanical Engineering, Tianjin University},%Department and Organization
%             % addressline={Address One}, 
%             city={Tianjin},
%             postcode={300350}, 
%             % state={State One},
%             country={China}}
            
% \affiliation[inst2]{organization={Institute of High Performance Computing, Agency for Science, Technology and Research (A*STAR)},%Department and Organization
%             postcode={138632}, 
%             country={Singapore}}

\begin{abstract}
Physics-informed neural networks (PINNs) have shown promise for solving partial differential equations (PDEs) by directly embedding them into the loss function. Despite their notable success, existing PINNs often exhibit training instability and slow convergence when applied to strongly nonlinear fluid dynamics problems. To address these challenges, this paper proposes a novel PINN framework, named as SIMPLE-PINN, which incorporates velocity and pressure correction loss terms inspired by the semi-implicit pressure link equation. These correction terms, derived from the momentum and continuity residuals, are tailored for the PINN framework, ensuring velocity-pressure coupling and reinforcing the underlying physical constraints of the Navier-Stokes equations. Through this, the framework can effectively mitigate training instability and accelerate convergence to achieve accurate solution. Furthermore, a hybrid numerical-automatic differentiation strategy is employed to improve the model's generalizability in resolving flows involving complex geometries. The performance of SIMPLE-PINN is evaluated on a range of challenging benchmark cases, including strongly nonlinear flows, long-term flow prediction, and multiphysics coupling problems. The numerical results demonstrate SIMPLE-PINN’s high accuracy and rapid convergence. Notably, SIMPLE-PINN achieves, for the first time, a fully data-free solution of lid-driven cavity flow at $Re$=20000 in just 448~s, and successfully captures the onset and long-time evolution of vortex shedding in flow past a cylinder over $t$=0-100. These findings demonstrate SIMPLE-PINN’s potential as a reliable and competitive neural solver for complex PDEs in intelligent scientific computing, with promising engineering applications in aerospace, civil engineering, and mechanical engineering.
\end{abstract}

\begin{highlights}
\item SIMPLE-PINN framework to bridge classical numerical idea and PINNs.
\item Derive velocity-pressure coupling correction loss function for PINNs.
\item Precise data-free simulation of lid-driven cavity flow at $Re=20000$ in 448~s.
\item Stable long-term simulation of vortex shedding in flow past a cylinder over $t$=0-100.
\item Time-window-free simulation of Rayleigh-Taylor instability ($Ra=10^6$) for $t$=0-6.
\end{highlights}


\end{frontmatter}

% \linenumbers

\section{Introduction}
Physics-informed neural networks (PINNs) incorporate governing equations and boundary conditions into the loss function, reducing the reliance on large labeled datasets and mitigating the risk of nonphysical predictions \citep{raissi2018deep,raissi2019physics,karniadakis2021physics}. PINNs have been widely applied across diverse domains \citep{mahmoudabadbozchelou2022nn,lee2022applications,raissi2020hidden}, establishing a new paradigm for scientific computing \citep{cai2021physics,cuomo2022scientific}.
Despite their considerable potential, PINNs still encounter significant limitations when applied to complex fluid flow scenarios. Fluid dynamics problems are governed by the incompressible Navier-Stokes (N-S) equations, which are intrinsically challenging to solve. The convection terms in the momentum equations introduce strong nonlinearity, leading to increased numerical stiffness and instability at high Reynolds ($Re$) numbers. In PINNs, this nonlinearity also results in a highly non-convex optimization landscape \citep{lu2025enhanced,basir2022investigating,chen2024df}, complicating the search for optimal network parameters and increasing the likelihood of suboptimal convergence \citep{wang2022and}. In addition, PINNs enforce physical laws by minimizing the residuals of PDEs at a discrete set of collocation points. Under the sparse sampling scenario, the satisfaction of the physical constraints may not be ensured globally. As a result, the solution may only satisfy the physical constraints at sampled points while violating them at unsampled locations \citep{wei2025fvmpinn}. The aforementioned challenges become even more pronounced in multiphysics scenarios \citep{temam2024navier}, where strong nonlinear interactions and complex couplings among physical fields often lead to training difficulties, erroneous convergence to local minima, and gradient instabilities \citep{cao2023tsonn,wang2023expert,he2023artificial}, ultimately degrading model convergence and accuracy. 

Some strategies have been proposed to improve PINNs for solving the N-S equations by combining the finite volume method (FVM) with encoder-decoder networks \citep{ranade2021discretizationnet} or by incorporating time-stepping schemes \citep{cao2023tsonn}. Others seek stabilization through artificial viscosity or entropy-viscosity models \citep{he2023artificial,wang2023solution}. Adaptive training strategies have also been investigated, including dynamic loss weighting \citep{jin2021nsfnets} and multi-level datasets training methods \citep{Tsai2025MLDPINN} to improve training efficiency, and curriculum training to mitigate error accumulation in long-term simulations \citep{wang2024piratenets,wang2023expert,wang2025gradient}.

Through the aforementioned efforts, the training difficulty issues in PINNs for low to moderate $Re$ number flows have been mitigated. Nevertheless, limitations such as excessive training time and the limited capability of PINNs to accurately resolve higher $Re$ number flows remain key challenges. We postulate that one of the major issues stems from the continuity equation ($\nabla \cdot \mathbf{u} = 0$), which not only ensures a divergence-free velocity field but also implicitly governs the pressure distribution via velocity-pressure coupling \citep{temam2024navier}. Solving the velocity-pressure coupled equations is increasingly difficult for highly nonlinear and tightly coupled scenarios. PINN models that insufficiently enforce the continuity constraint may result in non-physical velocity-pressure solutions that violate the physics of incompressible fluid flow.

Even though the coupling constraint makes training more challenging, few PINN studies explicitly address this issue by accounting for their coupled nature when solving the Navier-Stokes equations. In contrast, traditional numerical methods, underpinned by decades of pioneering research in scientific computing, have yielded robust and efficient strategies for enforcing velocity-pressure coupling in fluid simulations. Notably, the projection method \citep{Chorin1968} and the semi-implicit method for pressure-linked equations (SIMPLE) algorithm \citep{patankar2018numerical} remain among the most widely adopted approaches. This motivates us to pursue a promising direction for enhancing PINN-based neural solvers by incorporating strategies from numerical methods to better capture velocity-pressure coupling, thereby improving convergence and stability.

This paper proposes SIMPLE-PINN, a fast PINN framework inspired by the SIMPLE algorithm to incorporate velocity and pressure correction loss. Within this framework, we rigorously derive the velocity and pressure correction terms and address \textit{two major challenges} associated with directly embedding the SIMPLE algorithm into neural networks: \textit{the high computational complexity due to stencil dependencies} and \textit{the inaccessibility of next-iteration variables}. We overcome these challenges through Taylor expansion and second-order extrapolation, while designing the correction terms as loss functions tailored for neural networks. In this way, SIMPLE-PINN effectively strengthens the crucial velocity-pressure coupling, a constraint that is often lacking in conventional PINNs. This targeted enforcement of physical constraints enhances the physical consistency of the solution and improves the reliability of the framework for complex fluid dynamics problems. Furthermore, a hybrid derivative computation strategy through both numerical differentiation (ND) and automatic differentiation (AD) is employed to address the issue of stencil points falling inside solid regions, thereby extending SIMPLE-PINN to flow problems involving complex geometries. 

\begin{figure}[!ht]
    \centering
    \includegraphics[width=1\linewidth]{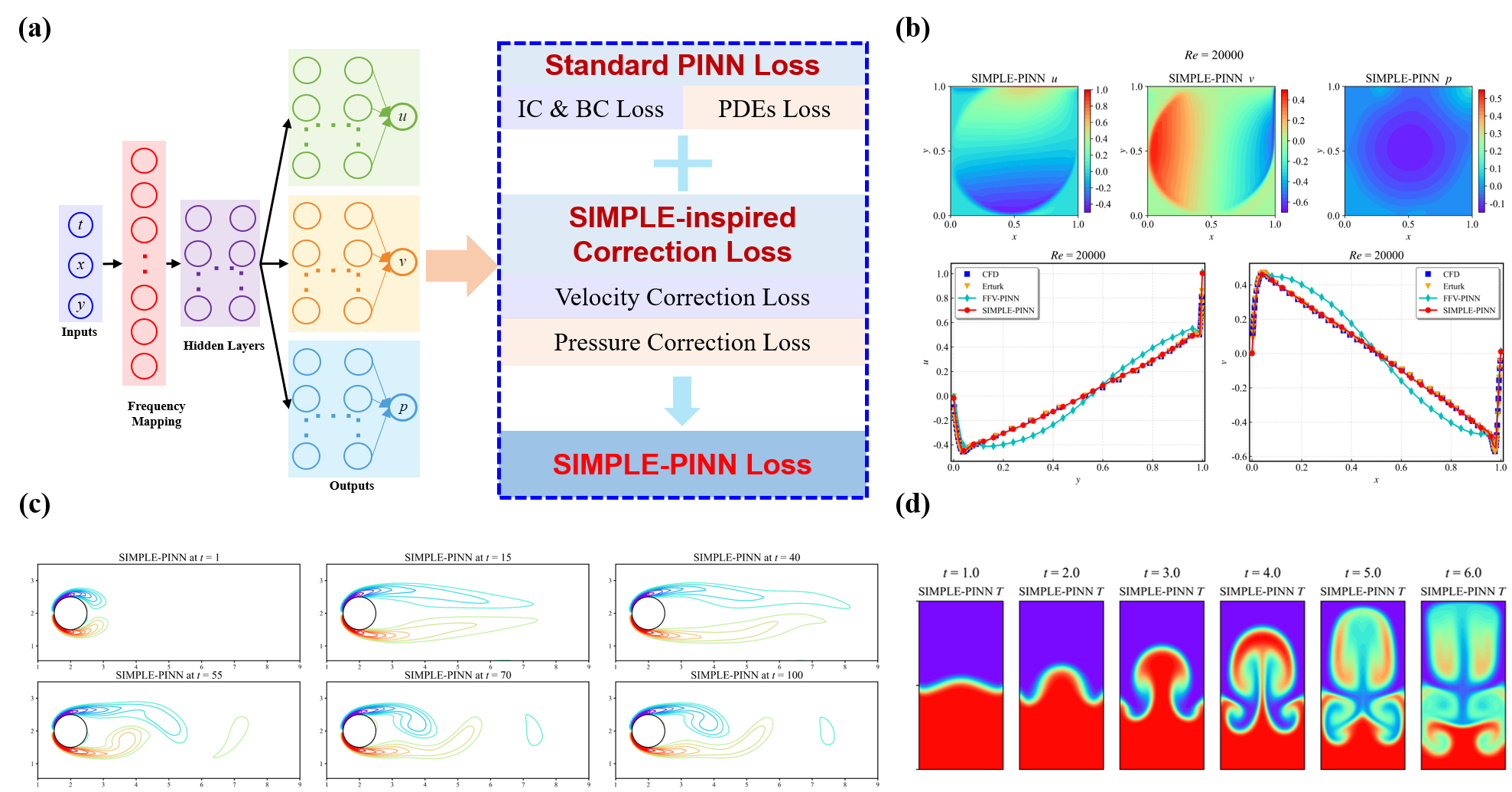}
    \caption{(a) Schematic illustration of the SIMPLE-PINN framework; (b) Application of SIMPLE-PINN to lid-driven cavity flow at $Re = 20000$; (c) Long-term prediction of vortex shedding in flow past a cylinder using SIMPLE-PINN; (d) Application of SIMPLE-PINN to multiphysics problems.}
    \label{fig:nn}
\end{figure}

The proposed framework was validated on a range of canonical fluid dynamics benchmarks, including high \textit{Re} lid-driven cavity flow, wavy channel flow, flow past a NACA0012 airfoil, flow past three square cylinders, unsteady flow past a circular cylinder, and the Rayleigh-Taylor instability at high Rayleigh (\textit{Ra}) number. Our results highlight the framework’s high accuracy, fast convergence, and robustness across diverse fluid dynamics scenarios and complex geometries. The SIMPLE-PINN framework and its performance on three challenging benchmark problems for PINNs are illustrated in Fig.~\ref{fig:nn}. 
% to address a common challenge in ND-based PINNs, where some stencil points may fall inside solid regions and lead to inaccurate derivative evaluations, the SIMPLE-PINN employs a hybrid derivative computation strategy: computing derivatives near complex boundaries while leveraging local physical information elsewhere. Specifically, AD is applied to collocation points near complex boundaries to preserve geometric flexibility, while a simplified finite volume discretization is used in the remaining regions to exploit local physical information. 

The primary contributions of this work are summarized as follows:
\begin{enumerate}
\item We developed SIMPLE-PINN, a novel physics-informed neural network framework inspired by the classical SIMPLE algorithm, by overcoming the difficulties of directly integrating the SIMPLE algorithm with PINNs. We formulated explicit correction terms, derived through Taylor expansion and second-order extrapolation, to transform the classical algorithmic insight into a tractable PINN-specific loss function. 
% By embedding these velocity and pressure correction terms, SIMPLE-PINN effectively strengthens the enforcement of continuity and momentum constraints, thereby significantly improving convergence behavior and solution accuracy for incompressible flow problems.
\item To overcome the challenges of applying finite volume discretization near irregular boundaries, SIMPLE-PINN applies automatic differentiation at collocation points in these regions. This approach preserves geometric flexibility near irregular boundaries while exploiting the conservation properties of FVM in interior domains, thereby broadening the framework’s applicability to complex flow geometries.
\item The effectiveness of SIMPLE-PINN is validated through comprehensive evaluations on several challenging benchmark problems. SIMPLE-PINN enables precise, data-free simulation of lid-driven cavity flow at $Re=20000$ in just 448~s training time (Fig.~\ref{fig:nn}b), achieves stable long-term prediction of vortex shedding in flow past a cylinder over the interval $t$=0-100 (Fig.~\ref{fig:nn}c), and performs time-window-free simulation of Rayleigh-Taylor instability at $Ra=10^6$ up to $t$ = 6 (Fig.~\ref{fig:nn}d).
\end{enumerate}

This paper is organized as follows. Section \ref{sec: method} provides a brief overview of PINNs for N-S equations, and subsequently proposes the SIMPLE-PINN framework. Section \ref{sec: numerical experiments} presents numerical experiments on benchmark problems, including high \textit{Re} flows, complex geometries, unsteady flows, and multiphysics scenarios, to demonstrate SIMPLE-PINN’s effectiveness. Section \ref{sec: conclusion} concludes and outlines directions for future work. 

\section{Methodology}\label{sec: method}
\subsection{Overview of PINNs}
PINNs have emerged as a novel paradigm in scientific computing, combining data-driven models with the governing equations of physical systems. In PINNs, the neural network acts as a surrogate model that simultaneously satisfies available observational data (if any) and the underlying physical laws, typically formulated as linear or nonlinear PDEs. Consider a general form of a PDE given by
\begin{equation}
\mathcal{N}[u(t,\mathbf{x}); \xi] = 0, \quad (t,\mathbf{x}) \in [0, T]\times\Omega  
\label{eq:pde}
\end{equation}
subject to boundary conditions (BCs)
\begin{equation}
u(t,\mathbf{x}) = g(t,\mathbf{x}), \quad (t,\mathbf{x}) \in  [0, T] \times\partial \Omega 
\end{equation}
and initial conditions (ICs)
\begin{equation}
u(0, \mathbf{x}) = u_0(\mathbf{x}), \quad \mathbf{x} \in \Omega 
\end{equation}
Here, \( \mathcal{N} \) denotes the differential operator, \( \xi \) represents the relevant physical parameters, and \( u(t,\mathbf{x}) \) corresponds to the solution of the PDE.  

In PINNs, the output \( u_{\boldsymbol{\theta}}(\mathbf{x}, t) \) is used to approximate the target solution \( u(\mathbf{x}, t) \), where \(\boldsymbol{\theta}\) denotes the trainable parameters. Those parameters are updated using gradient-based optimizers, such as Adam \citep{kingma2014adam} or L-BFGS \citep{liu1989limited}. The derivatives of \( u_{\boldsymbol{\theta}} \) with respect to the input coordinates can be computed via AD, which allows accurate evaluation of the differential operators in \(\mathcal{N}\) \citep{baydin2018automatic}. Moreover, unlike traditional numerical methods, PINNs operate in a mesh-free manner, providing high flexibility for problems involving complex geometries \citep{karniadakis2021physics}.

The training of PINNs is formulated as the minimization of a composite loss function, which generally consists of three primary components:
\begin{equation}
    \mathcal{L} = 
    W_{\text{PDE}} \, \mathcal{L}_{\text{PDE}} 
    + W_{\text{IC}} \, \mathcal{L}_{\text{IC}} 
    + W_{\text{BC}} \, \mathcal{L}_{\text{BC}}
    \label{eq:loss}
\end{equation}
where \(\mathcal{L}_{\text{PDE}}\), \(\mathcal{L}_{\text{IC}}\), and \(\mathcal{L}_{\text{BC}}\) denote the residual losses corresponding to the PDE, initial conditions, and boundary conditions, respectively. The coefficients $W_{\text{PDE}}$, $W_{\text{IC}}$ and $W_{\text{BC}}$ act as weighting factors to balance the contributions of each term in the loss function. 

\subsection{PINNs for Navier-Stokes equations} 
The governing equations for two-dimensional unsteady incompressible flow are formulated by the N-S equations, as follows:
\begin{subequations} \label{eq:2d_NS_unsteady}
    \begin{align}
        \frac{\partial u}{\partial x} + \frac{\partial v}{\partial y} &= 0 \label{eq:2d_NS_unsteady-div} \\ 
         \frac{\partial u}{\partial t} +\frac{\partial \left(uu\right)}{\partial x} +  \frac{\partial \left(vu\right)}{\partial y} &= \frac{1}{Re}\left( \frac{\partial ^2u}{\partial x^2} + \frac{\partial ^2u}{\partial y^2} \right) - \frac{\partial p}{\partial x} \label{eq:2d_NS_unsteady-m1} \\ 
        \label{eq:2d_NS_unsteady-m2} 
    \end{align}
\end{subequations}
where $u$ and $v$ denote the velocity components in the $x$- and $y$-directions, respectively, $p$ is the pressure, and $Re$ denotes the Reynolds number, a dimensionless parameter that characterizes the ratio of inertial to viscous forces. 

The PINN is trained to produce predictions $[u_\theta(x,y,t), v_\theta(x,y,t), p_\theta(x,y,t)]$ that adhere to the N-S equations and the prescribed ICs and BCs. The PDE residual loss $\mathcal{L}_{\text{PDE}}$ is defined at a set of collocation points $(t_i, x_i, y_i)$, and is formulated as the sum of residuals associated with the conservation of mass and momentum:
\begin{subequations} \label{eq:pinn_loss_total}
% \small
\begin{align}
& \mathcal{L}_{\text{PDE}} = \frac{1}{N_{\text{PDE}}} \sum_{i=1}^{N_{\text{PDE}}} 
(|Res_{c}(t_i,x_i, y_i)|^2 + |Res_{u}(t_i,x_i, y_i)|^2 + |Res_{v}(t_i,x_i, y_i)|^2)\label{eq:pinn_loss_total-a} \\[1ex]
&Res_{c} = \frac{\partial u_\theta}{\partial x} + \frac{\partial v_\theta}{\partial y} \label{eq:res_continuity-b} \\[0.5ex]
&Res_{u} = \frac{\partial u_\theta}{\partial t} 
       + \frac{\partial (u_\theta u_\theta)}{\partial x} 
       + \frac{\partial (v_\theta u_\theta)}{\partial y} 
       - \frac{1}{Re} \Big( \frac{\partial^2 u_\theta}{\partial x^2} + \frac{\partial^2 u_\theta}{\partial y^2} \Big)
       + \frac{\partial p_\theta}{\partial x} \label{eq:res_momentum_x-c} \\[0.5ex]
&Res_{v} = \frac{\partial v_\theta}{\partial t} 
       + \frac{\partial (u_\theta v_\theta)}{\partial x} 
       + \frac{\partial (v_\theta v_\theta)}{\partial y} 
       - \frac{1}{Re} \Big( \frac{\partial^2 v_\theta}{\partial x^2} + \frac{\partial^2 v_\theta}{\partial y^2} \Big)
       + \frac{\partial p_\theta}{\partial y} \label{eq:res_momentum_y-d}
\end{align}
\end{subequations}
where $Res_c$, $Res_u$, and $Res_v$ denote the residuals of the continuity equation and the momentum equations in the $x$- and $y$-directions, respectively. $N_{\text{PDE}}$ denotes the number of collocation points used to evaluate the PDE residuals. 

The initial condition loss $\mathcal{L}_{\text{IC}}$ penalizes deviations between the predictions and the prescribed initial conditions at  $t=0$. Likewise, the boundary condition loss $\mathcal{L}_{\text{BC}}$ enforces consistency between the predictions and the specified boundary values across all sampled time instances:
\begin{subequations} \label{eq:ic_bc_loss}
\begin{align}
\mathcal{L}_{\text{IC}} &= \frac{1}{N_{\text{IC}}} \sum_{i \in \Omega_0} 
\Big( |u_\theta^i - u_0^i|^2 + |v_\theta^i - v_0^i|^2 \Big) \\
\mathcal{L}_{\text{BC}} &= \frac{1}{N_{\text{BC}}} \sum_{i \in \partial \Omega} 
\Big( |u_\theta^i - g_u^i|^2 + |v_\theta^i - g_v^i|^2 \Big)
\end{align}
\end{subequations}
where $N_{\text{IC}}$ and $N_{\text{BC}}$ denote the number of collocation points for the initial and boundary conditions, $u_0^i$ and $v_0^i$ are the given initial velocities, $g_u^i$ and $g_v^i$ are the boundary velocity, $\Omega_0$ is the set of initial points, and $\partial \Omega$ represents the set of boundary points.

\subsection{The SIMPLE-PINN framework} \label{sec_SIMPLE-PINN}
To overcome the limitations of existing PINNs in solving incompressible N-S equations, especially at high \textit{Re} number,  we propose the SIMPLE-PINN framework. The framework integrates insights from classical FVM into PINN by introducing a velocity and pressure correction strategy as inspired by the SIMPLE algorithm:
\begin{equation}
    \mathcal{L}_{SIMPLE} = 
    W_{\text{PDE}} \, \mathcal{L}_{\text{PDE}} 
    + W_{\text{IC}} \, \mathcal{L}_{\text{IC}} 
    + W_{\text{BC}} \, \mathcal{L}_{\text{BC}}
    + W_{\text{RC}} \, \mathcal{L}_{\text{RC}}
    \label{eq:loss-SIMPLE}
\end{equation}
In the above, the newly introduced loss term $\mathcal{L}_{\text{RC}}$, representing as correction terms, will be derived directly from the governing equations and reformulated as loss functions suitable for neural networks, enabling explicit enforcement of velocity-pressure coupling.

\subsubsection{Derivation of discretized divergence-free correction equations}
We start by employing a simplified FVM \citep{wei2025ffv}, chosen for its enhanced numerical stability, to discretize the N-S equations (Eq.~\eqref{eq:2d_NS_unsteady}) within the control volumes shown in Fig.\ref{fig:fvm_cv}. The corresponding formulations are presented below.
\begin{subequations}\label{eq:n-s_fvm_implicit}
% \small
\begin{align}
&u_{e}^{t} - u_{w}^{t} + v_{n}^{t}- v_{s}^{t} = 0
\label{eq:continuity_fvm_implicit} \\[6pt]
&\frac{u_{P}^{t} - u_{P}^{t - \delta t}}{\delta t}\, \Delta x \Delta y  
+ a_{P} u_{P}^{t} + \sum a_{NB} u_{NB}^{t} + \sum a_{nb} u_{nb}^{t}  
+ (p_{e}^{t} - p_{w}^{t}) \Delta y = 0 
\label{eq:momentum-u_fvm_implicit} \\[6pt]
&\frac{v_{P}^{t} - v_{P}^{t - \delta t}}{\delta t}\, \Delta x \Delta y  
+ a_{P} v_{P}^{t} + \sum a_{NB} v_{NB}^{t} + \sum a_{nb} v_{nb}^{t}  
+ (p_{n}^{t} - p_{s}^{t}) \Delta x = 0
\label{eq:momentum-v_fvm_implicit}
\end{align}
\end{subequations}
where the coefficients $a_P$, $a_{NB}$, and $a_{nb}$ result from the discretization of the convection and diffusion terms. Among them, $a_P$ and $a_{NB}$ (\textit{E}, \textit{W}, \textit{N}, and \textit{S}) are constant coefficients and thus remain unchanged during training, whereas $a_{nb}$ (\textit{e}, \textit{w}, \textit{n}, and \textit{s}) depends on the velocity on control faces and varies as the neural network predictions are updated. The specific values of these coefficients are listed in \ref{sec: appendix coefficient fvm}, while the detailed derivation can be found in \citep{wei2025ffv}. The unsteady term is discretized using an implicit Euler scheme; for simplicity, unless otherwise specified, the time superscript $t$ will be omitted in the following discussion. It should be noted that in Eq.~\eqref{eq:continuity_fvm_implicit}, the velocity values on the control surfaces are directly obtained from the neural network predictions.  Similarly, in Eqs.~\eqref{eq:momentum-u_fvm_implicit} and \eqref{eq:momentum-v_fvm_implicit}, the discretized pressure gradient terms are evaluated directly from the outputs of the neural network.
\begin{figure}[ht]
    \centering
    \includegraphics[width=0.65\linewidth]{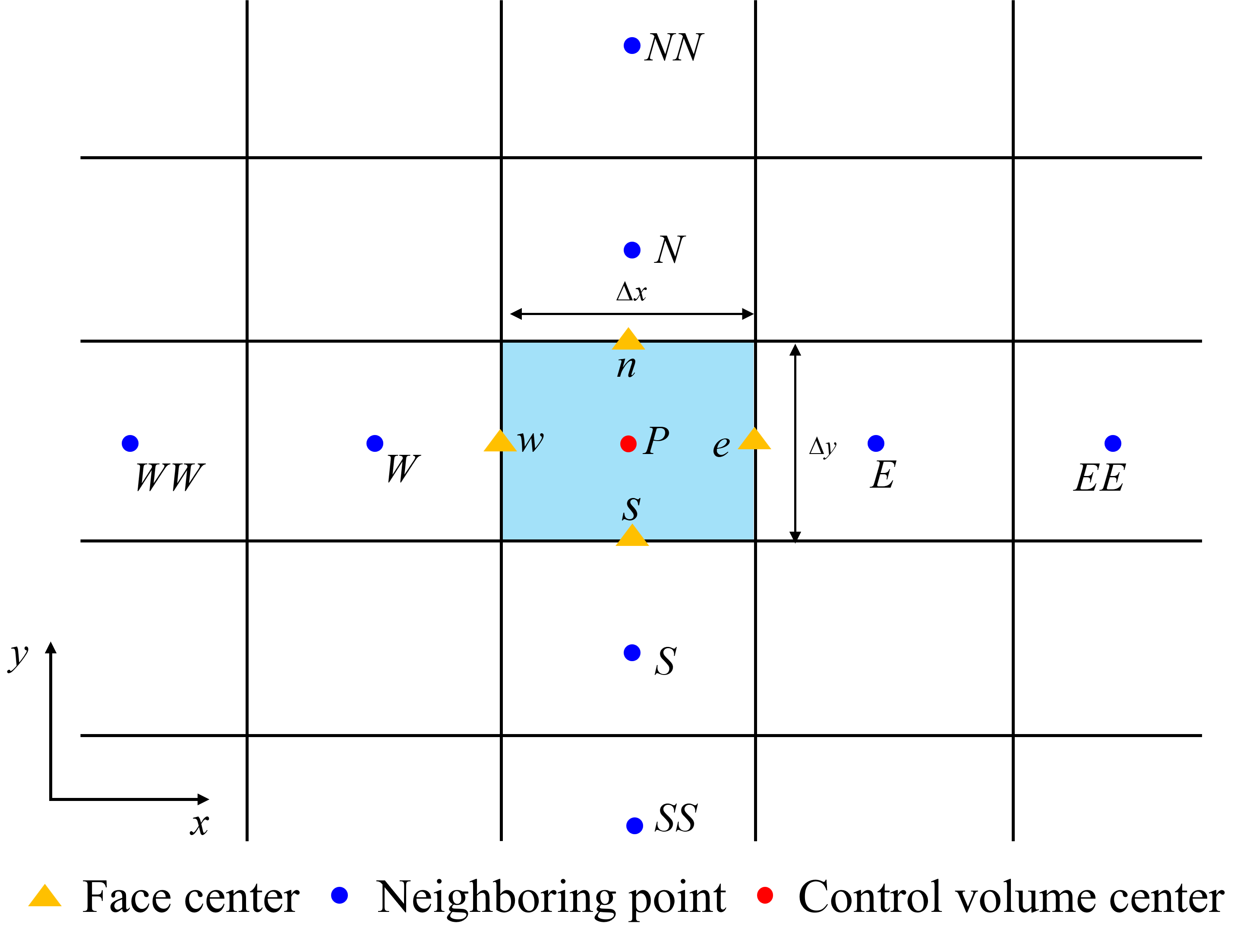}
\caption{Schematic of a control volume in a two-dimensional domain.  
The uppercase letters (\textit{E}, \textit{W}, \textit{N}, \textit{S}, \textit{EE}, \textit{WW}, \textit{NN}, and \textit{SS}) indicate the centers of the neighboring control volumes,  
whereas the lowercase letters (\textit{e}, \textit{w}, \textit{n}, and \textit{s}) correspond to the control surfaces. The letter \textit{P} marks the center of the current control volume.  
Grid spacing is assumed uniform ($\Delta x = \Delta y$) for illustration, though not required.}
    \label{fig:fvm_cv}
\end{figure}

Although the PDE loss (Eq.~\eqref{eq:pinn_loss_total-a}) ideally vanishes upon full convergence of the PINN, finite residuals $r^{n}$ inevitably arise at the $n$-th training iteration due to approximation, optimization, and generalization errors \citep{lu2021deepxde}.  
Consequently, the residuals corresponding to the continuity and momentum equations at $n$-th iteration can be expressed as follows:
\begin{subequations}\label{eq:momentum-n}
% \small 
\begin{align}
& u_{e}^{n} - u_{w}^{n} + v_{n}^{n} - v_{s}^{n} = r_c^{n}  \label{eq:c_n} \\[2mm]
& \frac{u_{P}^{n} - u_{P}^{n,t - \delta t}}{\delta t} \Delta x \Delta y  
+ a_{P} u_{P}^{n} + \sum a_{NB} u_{NB}^{n} + \sum a_{nb}^{n} u_{nb}^{n}  
+ (p_e^{n} - p_w^{n}) \Delta y = r_u^{n} \\[2mm]
& \frac{v_{P}^{n} - v_{P}^{n,t - \delta t}}{\delta t} \Delta x \Delta y  
+ a_{P} v_{P}^{n} + \sum a_{NB} v_{NB}^{n} + \sum a_{nb}^{n} v_{nb}^{n}  
+ (p_n^{n} - p_s^{n}) \Delta x = r_v^{n}
\end{align}
\end{subequations}
After rearrangement, we obtain:
\begin{subequations}\label{eq:residuals-fvm-n}
% \footnotesize
% \small
\begin{align}
& u_{e}^{n} - u_{w}^{n} + v_{n}^{n} - v_{s}^{n} = r_c^{n} \label{eq:residuals:continuity-n} \\[1mm]
& \left( \frac{\Delta x \Delta y}{\delta t} + a_{P}\right) u_{P}^{n} 
+ \sum a_{NB} u_{NB}^{n} + \sum a_{nb}^{n} u_{nb}^{n} 
+ b_{P,u}^n = r_u^{n} \label{eq:residuals:momentum-u-n} \\[1mm]
& \left( \frac{\Delta x \Delta y}{\delta t} + a_{P}\right) v_{P}^{n} 
+ \sum a_{NB} v_{NB}^{n} + \sum a_{nb}^{n} v_{nb}^{n} + b_{P,v}^n 
 = r_v^{n} \label{eq:residuals:momentum-v-n}
\end{align}
\normalsize
\end{subequations}
Here, the terms $b_{P,u}^n$ and $b_{P,v}^n$ represent the contributions from the pressure gradient and the implicit time-stepping, and are defined as follows:
\begin{subequations}
% \small
\label{b_source}
\begin{align}
&b_{P,u}^n = (p_e^n - p_w^n) \, \Delta y - \frac{u_P^{n,t-\delta t}}{\delta t} \, \Delta x \, \Delta y \\
&b_{P,v}^n = (p_n^n - p_s^n) \, \Delta x - \frac{v_P^{n,t-\delta t}}{\delta t} \, \Delta x \, \Delta y
\end{align}
\end{subequations}

At the $(n+1)$-th training iteration, assuming the physical constraints are satisfied, Eq.~\eqref{eq:residuals-fvm-n} can be expressed as:
\begin{subequations}\label{eq:momentum-nplus1}
% \small 
\begin{align}
& u_{e}^{n+1} - u_{w}^{n+1} + v_{n}^{n+1} - v_{s}^{n+1} = 0 \label{eq:momentum-nplus1:continuity} \\[1.5mm]
& \left( \frac{\Delta x \Delta y}{\delta t} + a_{P}\right) u_{P}^{n+1} + \sum a_{NB} u_{NB}^{n+1} + \sum a_{nb}^{n+1} u_{nb}^{n+1} + b_{P,u}^{n+1} = 0 \label{eq:momentum-nplus1:u} \\[1.5mm]
& \left( \frac{\Delta x \Delta y}{\delta t} + a_{P}\right) v_{P}^{n+1} + \sum a_{NB} v_{NB}^{n+1} + \sum a_{nb}^{n+1} v_{nb}^{n+1} +b_{P,v}^{n+1}  = 0 \label{eq:momentum-nplus1:v} 
\end{align}
\end{subequations}
By subtracting the momentum equation in Eq.~\eqref{eq:residuals-fvm-n} from that in Eq.~\eqref{eq:momentum-nplus1}, the following relations are obtained,
\begin{subequations}\label{eq:momentum-correction}
\footnotesize
\begin{align}
% & (u_e^{n+1} - u_w^{n+1}) - (u_e^n - u_w^n) 
%   + (v_n^{n+1} - v_s^{n+1}) - (v_n^n - v_s^n) = -r_c^n \label{eq:continuity-correction} \\[1mm]
a(u_P^{n+1} - u_P^n) + \sum a_{NB}(u_{NB}^{n+1} - u_{NB}^{n}) + \sum \bigl(a_{nb}^{n+1} u_{nb}^{n+1} - a_{nb}^{n} u_{nb}^{n}\bigr)
  + b_{P,u}^{n+1} - b_{P,u}^{n} = -r_u^n \label{eq:mom_cor_u} \\[1mm]
a(v_P^{n+1} - v_P^n) + \sum a_{NB}(v_{NB}^{n+1} - v_{NB}^{n}) + \sum \bigl(a_{nb}^{n+1} v_{nb}^{n+1} - a_{nb}^{n} v_{nb}^{n}\bigr)
  + b_{P,v}^{n+1} - b_{P,v}^{n} = -r_v^n \label{eq:mom_cor_v} 
\end{align}
\end{subequations}
\normalsize
where $a = \frac{\Delta x \Delta y}{\delta t} + a_P$ is a constant coefficient. By rearranging Eqs.~\eqref{eq:momentum-correction}, the expressions can be rewritten in the following form:
% \begin{subequations}\label{eq:momentum-correction-delta}
% \small
% \begin{align}
% & u_P^{n+1} - u_P^n 
% = -\frac{ \sum a_{NB}(u_{NB}^{n+1} - u_{NB}^{n}) 
% + \sum \bigl(a_{nb}^{n+1} u_{nb}^{n+1} - a_{nb}^{n} u_{nb}^{n}\bigr) 
% + b_{P,u}^{n+1} - b_{P,u}^{n} + r_u^n }{\frac{\Delta x \Delta y}{\delta t} + a_P} 
%   \label{eq:momentum-correction-delta:u} \\[1mm]
% & v_P^{n+1} - v_P^n 
% = -\frac{\sum a_{NB}(v_{NB}^{n+1} - v_{NB}^{n}) 
% + \sum \bigl(a_{nb}^{n+1} v_{nb}^{n+1} - a_{nb}^{n} v_{nb}^{n}\bigr) 
% +b_{P,v}^{n+1} - b_{P,v}^{n} + r_v^n}{\frac{\Delta x \Delta y}{\delta t} + a_P} 
%  \label{eq:momentum-correction-delta:v}
% \end{align}
% \end{subequations}

% Eqs \ref{eq:momentum-correction-delta} can be rearranged into the following form:
\begin{subequations}\label{eq:momentum-correction-delta-re}
\small
\begin{align}
& u_P^{n+1}  
=  u_P^n -\frac{ \sum a_{NB}(u_{NB}^{n+1} - u_{NB}^{n}) 
+ \sum \bigl(a_{nb}^{n+1} u_{nb}^{n+1} - a_{nb}^{n} u_{nb}^{n}\bigr) 
+ r_u^n }{a} -\frac{ b_{P,u}^{n+1} - b_{P,u}^{n}}{a}
  \label{eq:momentum-correction-delta-re:u} \\[1mm]
& v_P^{n+1} 
=   v_P^n -\frac{\sum a_{NB}(v_{NB}^{n+1} - v_{NB}^{n})
+ \sum \bigl(a_{nb}^{n+1} v_{nb}^{n+1} - a_{nb}^{n} v_{nb}^{n}\bigr) 
 + r_v^n}{a} -\frac{b_{P,v}^{n+1} - b_{P,v}^{n} }{a}
 \label{eq:momentum-correction-delta-re:v}
\end{align}
\end{subequations}
We further assume intermediate velocity $u^{*}$ and $v^{*}$, i.e.,
\begin{subequations}\label{eq:uv-intermediate}
\footnotesize
\begin{align}
u_P^* &= u_P^n - \frac{ \sum a_{NB}(u_{NB}^{n+1} - u_{NB}^{n}) 
+ \sum \bigl(a_{nb}^{n+1} u_{nb}^{n+1} - a_{nb}^{n} u_{nb}^{n}\bigr) 
 + r_u^n }{a} = u_p^n + S_{P,u}
  \label{eq:u-intermediate} \\[1mm]
v_P^* &= v_P^n -\frac{\sum a_{NB}(v_{NB}^{n+1} - v_{NB}^{n})
+ \sum \bigl(a_{nb}^{n+1} v_{nb}^{n+1} - a_{nb}^{n} v_{nb}^{n}\bigr) 
 + r_v^n}{a} = v_p^n + S_{P,v}
 \label{eq:v-intermediate}
\end{align}
\end{subequations}
where $S_{P,u}$ and $S_{P,v}$ are the collected terms obtained from neighboring contributions of velocity and residual.
Now, we substitute Eqs. \eqref{eq:uv-intermediate} into Eqs. \eqref{eq:momentum-correction-delta-re},
\begin{subequations}\label{eq:uv_n+1}
\small
\begin{align}
& u_P^{n+1}  
=  u_P^* -\frac{ b_{P,u}^{n+1} - b_{P,u}^{n}}{a} 
\label{eq:u_n+1} \\[1mm]
& v_P^{n+1} 
=  v_P^*-\frac{ b_{P,v}^{n+1} - b_{P,v}^{n}}{a} 
\label{eq:v_n+1}
\end{align}
\end{subequations}
% We neglect the contributions of velocities from neighboring nodes as well as the $n$-th residuals $r_u$ and $r_v$, which leads to
% \begin{subequations}\label{eq:simplified_uv_n+1}
% \small
% \begin{align}
% & u_P^{n+1}  
% \approx u_P^*-\frac{b_{P,u}^{n+1} - b_{P,u}^{n} }{a} 
% \label{eq:simplifiedu_n+1} \\[1mm]
% & v_P^{n+1} 
% \approx v_P^*-\frac{ b_{P,v}^{n+1} - b_{P,v}^{n} }{a} 
% \label{eq:simplifiedv_n+1}
% \end{align}
% \end{subequations}
The above relations define the velocity correction at the cell centers. This formulation is then extended to the face velocities, leading to the following expressions,
\begin{subequations}\label{eq:u-face}
% \small
\begin{align}
u_e^{\,n+1}
& = u_e^* -\frac{b_{e,u}^{\,n+1} - b_{e,u}^{\,n}}{a}
\label{eq:face-ue} \\[1mm]
u_w^{\,n+1} 
& = u_w^* -\frac{b_{w,u}^{\,n+1} - b_{w,u}^{\,n}}{a}
\label{eq:face-uw} \\[1mm]
v_n^{\,n+1}  
& = v_n^* -\frac{b_{n,v}^{\,n+1} - b_{n,v}^{\,n}}{a}
\label{eq:face-vn} \\[1mm]
v_s^{\,n+1}  
& = v_s^* -\frac{b_{s,v}^{\,n+1} - b_{s,v}^{\,n}}{a}
\label{eq:face-vs}
\end{align}
\end{subequations}
By substituting the face velocities given in Eq.~\eqref{eq:u-face} into Eq.~\eqref{eq:momentum-nplus1:continuity}, we obtain
\begin{equation}\label{eq:pressure-correction-b}
\begin{split}
(b_{e,u}^{\,n+1} - b_{w,u}^{\,n+1}) + (b_{n,v}^{\,n+1} - b_{s,v}^{\,n+1}) 
 - (b_{e,u}^{\,n} - b_{w,u}^{\,n}) - (b_{n,v}^{\,n} - b_{s,v}^{\,n}) 
=  a r_c^* \\
\end{split}
\end{equation}
where 
\begin{equation}\label{eq:r*}
r_c^* = u_e^* - u_w^* + v_n^* - v_s^*
\end{equation}
The detailed expressions for each of the $b$ terms in Eq.~\eqref{eq:pressure-correction-b} can be found in \ref{sec: b_terms}. Further simplification leads to
\begin{equation}\label{eq:pressure-SIMPLE-b}
(A_P^{\,n+1} - DIV_P^{\,n+1}) - (A_P^n - DIV_P^n) = a r_c^*
\end{equation}
where
\begin{subequations}\label{eq:AP-BP-terms}
\begin{align}
A_P^{\,n} &= (p_E^{\,n} - 2 p_P^{\,n} + p_W^{\,n}) \, \Delta y 
           + (p_N^{\,n} - 2 p_P^{\,n} + p_S^{\,n}) \, \Delta x
           \label{eq:AP-n}\\[1mm]
DIV_P^{\,n} &= \frac{\big[(u_e^{\,n,t-\delta t} - u_w^{\,n,t-\delta t}) 
              + (v_n^{\,n,t-\delta t} - v_s^{\,n,t-\delta t})\big] \Delta x \Delta y}{\delta t}
              \label{eq:BP-n}\\[1mm]
A_P^{\,n+1} &= (p_E^{\,n+1} - 2 p_P^{\,n+1} + p_W^{\,n+1}) \, \Delta y 
             + (p_N^{\,n+1} - 2 p_P^{\,n+1} + p_S^{\,n+1}) \, \Delta x
             \label{eq:AP-np1}\\[1mm]
DIV_P^{\,n+1} &= \frac{\big[(u_e^{\,n+1,t-\delta t} - u_w^{\,n+1,t-\delta t}) 
              + (v_n^{\,n+1,t-\delta t} - v_s^{\,n+1,t-\delta t}) \big] \Delta x \Delta y}{\delta t}
              \label{eq:BP-np1}
\end{align}
\end{subequations}

Based on the assumption that the physical constraints are fully satisfied in the $(n+1)$-th training iteration, the term $DIV_P^{\,n+1}$, which corresponds to the divergence of the velocity field, vanishes. Therefore, Eq.~\eqref{eq:pressure-SIMPLE-b} reduces to
\begin{equation}\label{eq:pressure-SIMPLE-b-reduced}
% \small
A_P^{\,n+1} - (A_P^n - DIV_P^n) = a r_c^*
\end{equation}
% \normalsize
Substituting the definitions of $A_P^{\,n+1}$ and $A_P^n$ from Eq.~\eqref{eq:AP-BP-terms}, we obtain the following equation:
\begin{equation}\label{eq:pressure-Poisson-discrete}
% \small
\begin{aligned}
&\big[(p_E^{\,n+1} - 2 p_P^{\,n+1} + p_W^{\,n+1}) \, \Delta y 
+ (p_N^{\,n+1} - 2 p_P^{\,n+1} + p_S^{\,n+1}) \, \Delta x \big] \\
&- \big[ (p_E^{\,n} - 2 p_P^{\,n} + p_W^{\,n}) \, \Delta y 
+ (p_N^{\,n} - 2 p_P^{\,n} + p_S^{\,n}) \, \Delta x - DIV_P^n \big] 
= a r_c^*
\end{aligned}
\end{equation}
% \normalsize
Rearranging terms gives
\begin{equation}\label{eq:pressure-correction-pp}
% \small
\begin{aligned}
& -2 (\Delta x + \Delta y) \, (p_P^{\,n+1} - p_P^n) 
+ \Delta y \, \big( p_E^{\,n+1} - p_E^n + p_W^{\,n+1} - p_W^n \big) \\
& + \Delta x \, \big( p_N^{\,n+1} - p_N^n + p_S^{\,n+1} - p_S^n \big) 
= a r_c^* - DIV_P^n
\end{aligned}
\end{equation}
% \normalsize
Dividing both sides by $-2 (\Delta x + \Delta y)$, we obtain the expression for pressure correction:
\footnotesize
\begin{equation}\label{eq:pressure-correction-pp-final}
p_P^{\,n+1} - p_P^n
= \frac{ a \, r_c^* - DIV_P^n
- \Delta y \, (p_E^{\,n+1} - p_E^n + p_W^{\,n+1} - p_W^n)
- \Delta x \, (p_N^{\,n+1} - p_N^n + p_S^{\,n+1} - p_S^n) }{ -2 (\Delta x + \Delta y) }
\end{equation}
\normalsize
Assuming $\Delta x = \Delta y = h$, Eq.~\eqref{eq:pressure-correction-pp-final} can be expressed as:
\footnotesize
\begin{equation}\label{eq:p-cor}
p_P^{\,n+1} - p_P^n =\frac{a \, r_c^* - DIV_P^n - h \, \Big[(p_E^{\,n+1} + p_W^{\,n+1} + p_N^{\,n+1} + p_S^{\,n+1}) 
           - (p_E^n + p_W^n + p_N^n + p_S^n)\Big]}{-4h}
\end{equation}
\normalsize

Eq~\eqref{eq:p-cor} determines the pressure update step from the current iteration $n$ to $n+1$ through the pressure correction. This step ensures that the corrected pressure field satisfies the continuity equation. However, merely updating the pressure field is insufficient, as pressure and velocity are strongly coupled. Based on our previous fundamental assumption (Eqs.\eqref{eq:momentum-nplus1}) that the velocity components $u^{n+1}$ and $v^{n+1}$  simultaneously satisfy both the discretized momentum and continuity equations, a corresponding velocity correction (Eq.\eqref{eq:momentum-correction-delta-re}) is employed to maintain the consistency and physical accuracy of the flow field.

% By subtracting $u^n$ from both sides of Eq.~\eqref{eq:u_n+1}, and similarly for Eq.~\eqref{eq:v_n+1}, we obtain
% \begin{subequations}\label{eq:uv_cor}
% \small
% \begin{align}
% & u_P^{n+1}  - u_P^n 
% =  u_P^* -\frac{ b_{P,u}^{n+1} - b_{P,u}^{n}}{a} - u_P^n
% \label{eq:u_cor} \\[1mm]
% & v_P^{n+1} - v_P^n
% =  v_P^*-\frac{ b_{P,v}^{n+1} - b_{P,v}^{n}}{a} - v_P^n
% \label{eq:v_cor}
% \end{align}
% \end{subequations}
% Substituting the definitions of $u^*$ and $v^*$ given in Eq.~\eqref{eq:uv-intermediate} into Eq.~\eqref{eq:uv_cor} yields the expression for velocity correction
% \begin{subequations}\label{eq:mom-cor}
% \small
% \begin{align}
% & u_P^{n+1} - u_P^n 
% = R_u =-\frac{ \sum a_{NB}(u_{NB}^{n+1} - u_{NB}^{n}) 
% + \sum \bigl(a_{nb}^{n+1} u_{nb}^{n+1} - a_{nb}^{n} u_{nb}^{n}\bigr) 
% + b_{P,u}^{n+1} - b_{P,u}^{n} + r_u^n }{a} 
%   \label{eq:mom-cor:u} \\[1mm]
% & v_P^{n+1} - v_P^n 
% =R_v = -\frac{\sum a_{NB}(v_{NB}^{n+1} - v_{NB}^{n}) 
% + \sum \bigl(a_{nb}^{n+1} v_{nb}^{n+1} - a_{nb}^{n} v_{nb}^{n}\bigr) 
% +b_{P,v}^{n+1} - b_{P,v}^{n} + r_v^n}{a} 
%  \label{eq:mom-cor:v}
% \end{align}
% \end{subequations}

\subsubsection{Derivation of SIMPLE-PINN residual correction terms}
Although the pressure and velocity correction equations derived above are obtained through rigorous mathematical formulation, their direct application within neural networks remains challenging. This is also one of the reasons why current ND-PINN models generally only replace AD with ND without incorporating the pressure-velocity coupling mechanism. Specifically, the main difficulties lie in two aspects. First, the computation of the pressure and velocity correction terms depends on the physical quantities at multiple neighboring points, which significantly increases computational complexity and hinders efficient implementation within the neural network. Second, these correction terms involve variables at the next iteration (\(n+1\)), which are not directly accessible during neural network training, making it difficult to integrate them into the PINN framework. To overcome these challenges, we propose a loss function specifically designed for PINNs that incorporates the effects of pressure-velocity corrections without requiring explicit values from neighboring points and future iteration steps, thereby enabling efficient and stable training.

We begin by focusing on \(r^*_c\) in the pressure correction term, which constitutes the primary obstacle to its direct implementation within neural networks. To facilitate a more tractable computation of \(r^*_c\), several simplifications are introduced \citep{sun2008efficient}. First, only the velocity component \(u\) is explicitly considered, noting that the treatment for \(v\) follows an analogous reasoning.
\begin{equation}\label{eq:ue-uw}
u_e^* - u_w^* = u_e^n - u_w^n + S_{e,u} - S_{w,u}
\end{equation}
A Taylor expansion of \(S_{e,u}\) around \(S_{w,u}\) yields
\begin{equation}\label{eq:taylor-reu}
S_{e,u} = S_{w,u} + \Delta x \left( \frac{\partial S_{u}}{\partial x} \right)_w 
+ \frac{\Delta x^2}{2}\left( \frac{\partial^2 S_{u}}{\partial x^2} \right)_w 
+ O(\Delta x^3)
\end{equation}
Subtracting \(S_{w,u}\) from both sides of Eq.~\eqref{eq:taylor-reu} gives
\begin{equation}\label{eq:taylor-u}
S_{e,u} - S_{w,u} = O(\Delta x)
\end{equation}
Similarly, for the \(v\)-component, we have
\begin{equation}\label{eq:taylor-v}
S_{n,v} - S_{s,v} = O(\Delta y)
\end{equation}
Substituting Eqs.~\eqref{eq:taylor-u} and \eqref{eq:taylor-v} into Eq.~\eqref{eq:r*} leads to
\begin{equation}\label{eq:uv*}
r^*_c = u_e^* - u_w^* + v_n^* - v_s^* 
= u_e^n - u_w^n + v_n^n - v_s^n + O(\Delta x + \Delta y)
= r_c^n + O(\Delta x + \Delta y)
\end{equation}
under the assumption of \(\Delta x\) and \(\Delta y\) are sufficiently small, \(r_c^*\) is approximated by \(r_c^n\) in this study. 

Next, we consider the treatment of velocity values on the control surfaces within the velocity correction term. Inspired by the SIMPLE algorithm, we directly neglect the cell-face velocity contributions, $\sum \left(a_{nb}^{\,n+1} u_{nb}^{\,n+1} - a_{nb}^{\,n} u_{nb}^{\,n} \right) $ and $ \sum \left(a_{nb}^{\,n+1} v_{nb}^{\,n+1} - a_{nb}^{\,n} v_{nb}^{\,n} \right)$. Therefore, the pressure and velocity correction terms, $R_p$, $R_u$, and $R_v$ can be formulated as
\begin{subequations}\label{eq:uvp-cor}
\small
\begin{align}
p^{n+1} - p^{n} \approx R_p & =  
\frac{a \, r_c^n - DIV_P^n - h \, \Big[(p_E^{\,n+1} + p_W^{\,n+1} + p_N^{\,n+1} + p_S^{\,n+1}) 
           - (p_E^n + p_W^n + p_N^n + p_S^n)\Big]}{-4h} \label{eq:simple_p} \\[1mm]
u^{n+1} - u^{n} \approx R_u & =
-\frac{ \sum a_{NB}(u_{NB}^{n+1} - u_{NB}^{n}) 
+ b_{P,u}^{n+1} - b_{P,u}^{n} + r_u^n }{a} \label{eq:simple_u} \\[1mm]
v^{n+1} - v^{n} \approx R_v & =
-\frac{\sum a_{NB}(v_{NB}^{n+1} - v_{NB}^{n}) 
+ b_{P,v}^{n+1} - b_{P,v}^{n} + r_v^n}{a} \label{eq:simple_v}        
\end{align}
\end{subequations}

As outlined in the SIMPLE algorithm \citep{patankar2018numerical}, neglecting the influence of neighboring points does not affect the final solution; however, it may lead to an overestimation of the correction terms and influence the convergence trajectory \citep{darwish2016finite}. To address this issue and improve training stability, we introduce relaxation factors \(\alpha_p\), \(\alpha_u\), and \(\alpha_v\) to adjust the corrections, leading to
\begin{subequations}\label{cor_uv}
% \small
\begin{align}
p_{P}^{\,n+1} & = p_{P}^{n} + \alpha_p  R_p \label{cor:p} \\
u_{P}^{\,n+1} &= u_{P}^{n} + \alpha_u R_u \label{cor:u} \\
v_{P}^{\,n+1} &= v_{P}^{n} + \alpha_v R_v \label{cor:v}
\end{align}
\end{subequations}

The residual correction loss is defined using the $L_1$ norm, applied separately to the pressure and velocity components, as
\begin{subequations}\label{cor_uvp_loss}
\begin{align}
L_{rc,p} &= \frac{1}{N_{rc}}
\left\| p^{\,n+1}_{P} - p^{\,n}_{P} - \alpha_p R_p \right\|_{L_1(\Omega \times (0,T])}
\label{cor_phi_loss:p}\\[1mm]
L_{rc,u} &= \frac{1}{N_{rc}}
\left\| u^{\,n+1}_{P} - u^{\,n}_{P} - \alpha_u R_u \right\|_{L_1(\Omega \times (0,T])}
\label{cor_phi_loss:u} \\[1mm]
L_{rc,v} &= \frac{1}{N_{rc}}
\left\| v^{\,n+1}_{P} - v^{\,n}_{P} - \alpha_v R_v \right\|_{L_1(\Omega \times (0,T])}
\label{cor_phi_loss:v} 
\end{align}
\end{subequations}
where $N_{rc}$ denotes the number of residual points used to evaluate the correction term. 

Finally, since the \((n+1)\)-th iteration values are not directly accessible during training, we employ a second-order extrapolation scheme to estimate them, thereby avoiding the need to obtain future values explicitly.
\begin{subequations}\label{assume_uvpb}
\begin{align}
p^{\,n+1}_P &= 2 \, p^n_P - p^{\,n-1}_P \label{assume_extrap} \\
u^{\,n+1} &= 2 u^{\,n} - u^{\,n-1} & 
v^{\,n+1} &= 2 v^{\,n} - v^{\,n-1} \label{assume_u_extrap} \\
b_u^{\,n+1} &= 2 b_u^{\,n} - b_u^{\,n-1} & 
b_v^{\,n+1} &= 2 b_v^{\,n} - b_v^{\,n-1} \label{assume_v_extrap}
\end{align}
\end{subequations}
By substituting Eqs.~\eqref{assume_uvpb} into the residual correction loss defined in Eq.~\eqref{cor_uvp_loss}, the three correction terms employed in this study can be expressed as
\begin{subequations}\label{cor_rc_loss_uv}
\begin{align}
RC_{p} &= \frac{1}{N_{rc}}
\left\| p^{\,n}_{P} - p^{\,n-1}_{P} - \alpha_p\, R_p \right\|_{L^{1}(\Omega \times (0,T])} \\
RC_{u} &= \frac{1}{N_{rc}} 
\left\| u^{\,n}_{P} - u^{\,n-1}_{P} - \alpha_u\, R_u \right\|_{L^{1}(\Omega \times (0,T])} \\
RC_{v} &= \frac{1}{N_{rc}} 
\left\| v^{\,n}_{P} - v^{\,n-1}_{P} - \alpha_v\, R_v \right\|_{L^{1}(\Omega \times (0,T])}
\end{align}
\end{subequations}
The residuals correction terms $R_p$, $R_u$, and $R_v$ appearing above are given by
\begin{subequations}\label{R_uvp}
\small
\begin{align}
R_p &= \frac{a\, r_c^n - DIV_P^n 
    - h\Big[(p_E^n + p_W^n + p_N^n + p_S^n) 
           - (p_E^{n-1} + p_W^{n-1} + p_N^{n-1} + p_S^{n-1})\Big]}{-4h} \\[4pt]
R_u &= -\frac{\displaystyle\sum_{NB} a_{NB}\big(u_{NB}^{n} - u_{NB}^{n-1}\big) 
         + \big(b_{P,u}^{n} - b_{P,u}^{n-1}\big) + r_u^n}{a} \\[4pt]
R_v &= -\frac{\displaystyle\sum_{NB} a_{NB}\big(v_{NB}^{n} - v_{NB}^{n-1}\big) 
         + \big(b_{P,v}^{n} - b_{P,v}^{n-1}\big) + r_v^n}{a}
\end{align}
\end{subequations}
As a result, the SIMPLE-PINN loss function proposed in this study is summarized as follows:
\begin{subequations}
\begin{align}\label{eq:simple-loss}\nonumber
    &\mathcal{L}_{SIMPLE} = 
    \lambda_{\text{c}} \, Res_{c} 
    + \lambda_{\text{u}} \, Res_{u}
    + \lambda_{\text{v}} \, Res_{v}
    + \lambda_{\text{IC}} \, \mathcal{L}_{\text{IC}} 
    + \lambda_{\text{BC}} \, \mathcal{L}_{\text{BC}} \\ 
    &~~~~~~~~~~~~~~~~+ \lambda_{\text{RC}} \, (RC_{p} + RC_{u} + RC_{v})
\end{align}
\end{subequations}
where
\begin{subequations}
\begin{align}
    &Res_{c} = \frac{1}{N_{PDE}} \left\| u_e^n - u_w^n + v_n^n - v_s^n \right\|_{L^{2}}\\
    &Res_{u} = \frac{1}{N_{PDE}} \left\| \left( \frac{\Delta x \Delta y}{\delta t} + a_{P}\right) u_{P}^{n} + \sum a_{NB} u_{NB}^{n} + \sum a_{nb}^{n} u_{nb}^{n} + b_{P,u}^n \right\|_{L^{2}} \\
    &Res_{v} = \frac{1}{N_{PDE}} \left\| \left( \frac{\Delta x \Delta y}{\delta t} + a_{P}\right) v_{P}^{n} + \sum a_{NB} v_{NB}^{n} + \sum a_{nb}^{n} v_{nb}^{n} + b_{P,v}^n \right\|_{L^{2}}
\end{align}
\end{subequations}
while $RC_{p}$, $RC_{u}$ and $RC_{v}$ are computed by Eqs \eqref{cor_rc_loss_uv}.

The proposed SIMPLE-PINN framework integrates velocity and pressure correction terms $(RC_{u}$, $RC_{v}$ and $RC_{p}$)  into the loss function, effectively enforcing their coupled relationship, which is often overlooked in existing PINN models. These correction terms, derived from the residuals of the governing equations, guide network parameter updates in directions aligned with physical laws, ensuring stronger adherence to physical constraints during training. Crucially, the inclusion of these loss terms distinguishes SIMPLE-PINN from other hybrid models: it adapts a classical numerical solution algorithm into a format tailored for PINNs. By inheriting the strengths of the classical SIMPLE algorithm, these terms continuously adjust velocity and pressure to enforce the divergence-free condition at each iteration during training, thereby enhancing training stability and accelerating convergence. Furthermore, while formulated for unsteady governing equations, the correction terms can be seamlessly adapted for steady-state problems by excluding time-dependent components, enabling a unified approach for both transient and steady flow scenarios.

\subsubsection{SIMPLE-PINN for irregular domains}    
\begin{figure}[ht]
    \centering
    \includegraphics[width=0.5\linewidth]{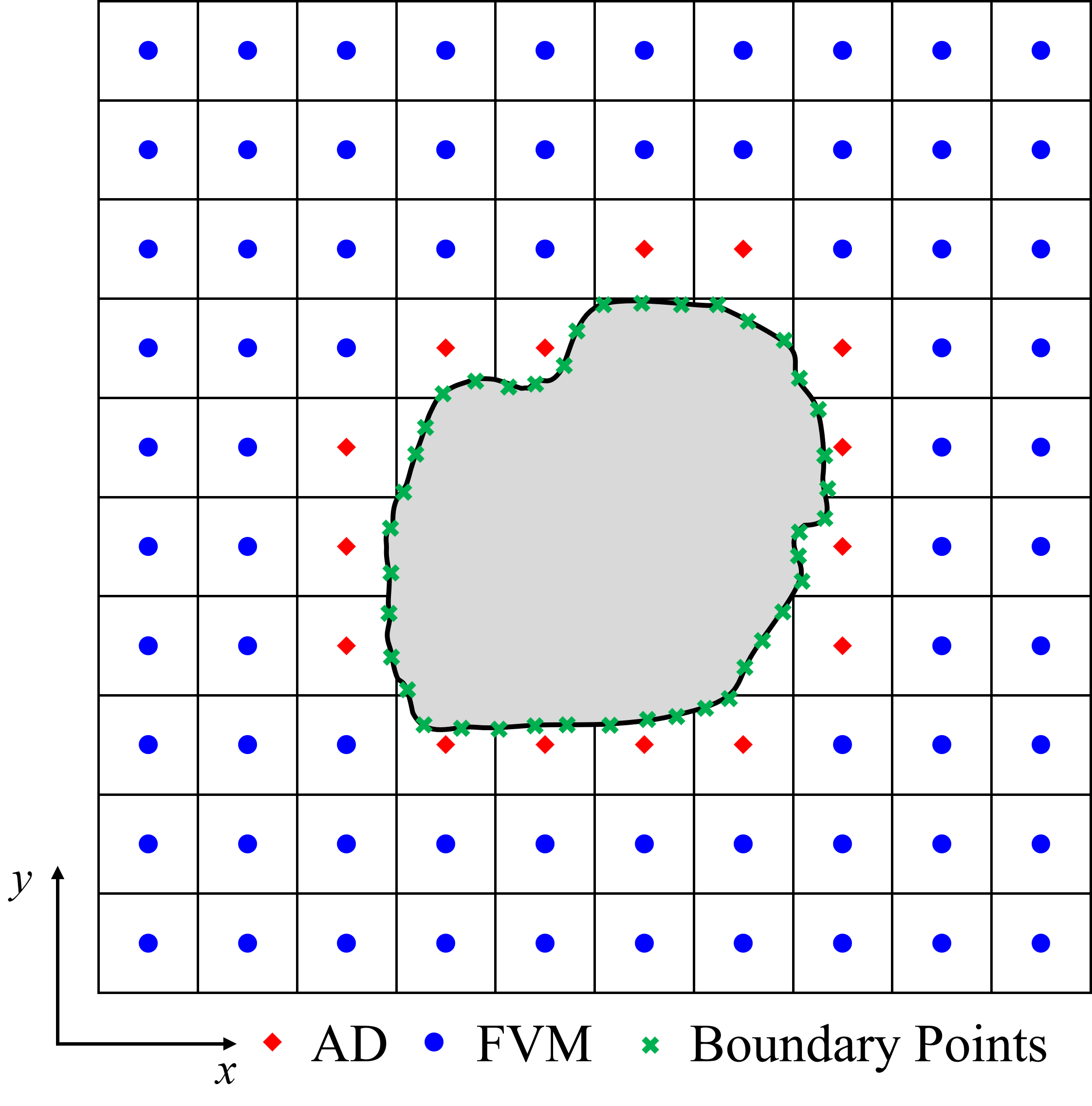}
    \caption{Two-dimensional computational domain containing an arbitrary-shaped object (grey region). Sample points are classified into three categories: boundary points (green cross), points near the complex geometric boundary where the PDE loss is evaluated using AD (red diamond), and points where the loss is computed using the simplified FVM (blue circle).}
    \label{fig: fvm_ad}
\end{figure}

Since the PDE loss in SIMPLE-PINN is discretized using a simplified FVM \citep{wei2025ffv}, difficulties arise when collocation points are located near irregular boundaries. In such cases, some of the neighboring points required for the FVM stencil may fall outside the computational domain, as illustrated by the grey region in Fig.\ref{fig: fvm_ad}. 
For complex geometries, existing PINN frameworks have employed various strategies, including coordinate or mesh transformations \citep{gao2021phygeonet, cao2024solver}, dual-network architectures for separately enforcing boundary and domain solutions \citep{sheng2021pfnn}, PINNs coupled with the immersed boundary method \citep{sundar2024physics, xiao2025immersed, huang2022direct, balam2021immersed}, linear local structure approximations \citep{wong2023lsa, wong2024soft}, and signed distance function-based FDM approaches \citep{xiang2022hybrid}.

To enhance the capability of SIMPLE-PINN in handling flows involving complex fluid domains, we design an easy-to-implement hybrid strategy that combines AD with the simplified FVM. The implementation proceeds as follows: collocation points are first generated uniformly across the entire physical domain, after which those located within the fluid region are identified as fluid points. These fluid points are then divided into two categories: if a point and its four neighbor points (\textit{E}, \textit{W}, \textit{N}, \textit{S} in Fig. \ref{fig:fvm_cv}) all lie in the fluid region, its PDE residual is computed using the simplified FVM (blue circle); otherwise, if any neighbor point falls inside the solid geometry, the residual is computed using AD (red diamond), as illustrated in Fig.~\ref{fig: fvm_ad}. Points located on complex geometric boundaries are randomly sampled and marked as boundary points (green cross), where boundary conditions are imposed as soft constraints. Finally, the AD- and FVM-based residuals, together with correction terms, are integrated to train the network.
% The detailed implementation steps are provided in \ref{sec: AD_FVM_framework}.

\subsubsection{SIMPLE-PINN architecture}    

In this study, we employ a multi-layer perceptron (MLP) as the backbone architecture for our SIMPLE-PINNs, chosen for its proven capability in approximating PINN solutions and for the simplicity and ease of implementation that this network structure offers. The SIMPLE-PINN framework, however, is not limited to MLPs; readily extended to alternative architectures offering enhanced representational capacity and training efficiency, such as residual neural networks~\citep{he2016deep,wang2021understanding,wang2024piratenets}. The first layer of SIMPLE-PINN employs a frequency annealing mapping that projects the input coordinates $(t, x, y)$ into a higher-dimensional space to better capture high-frequency features, as illustrated in Fig.~\ref{fig:nn}(a). For steady-state problems, the temporal coordinate $t$ is omitted, and only $(x, y)$ is mapped. The network then consists of two shared hidden layers followed by variable-specific layers for $u$, $v$, and $p$, each with the SiLU activation function. For the Rayleigh-Taylor instability case, an additional output branch is added for the temperature field $T$. SIMPLE-PINN is trained using the Adam optimizer with a warmup cosine decay learning rate schedule.

\section{Numerical experiments} \label{sec: numerical experiments}
In this section, we conduct a series of numerical experiments, including lid-driven cavity flow at high \textit{Re} numbers, wavy channel flow, flow past a NACA0012 airfoil,  flow past three square cylinders, the Rayleigh-Taylor instability, and flow past a cylinder. Each benchmark problem is selected to highlight a distinct capability of the proposed framework: the lid-driven cavity flow demonstrates the effect of velocity-pressure coupling in a simple geometry; the wavy channel flow evaluates the performance of SIMPLE-PINN in periodically perturbed channel flow configurations; the NACA0012 airfoil illustrates the role of AD in handling curved boundaries and its effectiveness in open-domain problems; the three square cylinders emphasize the necessity of AD for capturing sharp gradients and resolving flows with multiple obstacles; and finally, the unsteady flow past a cylinder together with  Rayleigh-Taylor instability showcase the versatility of the framework in addressing transient dynamics and multiphysics phenomena. These problems are well established in fluid dynamics and remain challenging for PINNs due to their nonlinear and multiscale characteristics. The training configurations adopted for the experiments are summarized in Table \ref{tab:training_parameters} of \ref{app_hyper}. All experiments are implemented in JAX and conducted on a single NVIDIA RTX 3090 GPU. 

The accuracy of SIMPLE-PINN is quantitatively assessed using two main metrics: the mean squared error (MSE) and the relative $L_2$ error. These measures quantify the differences between the predicted solutions and the corresponding numerical or benchmark solutions. MSE is mathematically defined as:
\begin{equation} \label{eq:mse}
\small
\text{MSE} = \frac{1}{N} \sum_{i=1}^{N} (u_i - \hat{u}_i)^2
\end{equation} 
\normalsize
where $u_i$ denotes the reference values from the numerical solution, $\hat{u}_i$ represents the predictions from the SIMPLE-PINN, and $N$ is the total number of evaluation points.

The relative $L_2$ error is mathematically formulated as:
\begin{equation} \label{eq:l2error}
\small
\text{Relative $L_2$ error} = \frac{\| u - \hat{u} \|_{L_2}}{\| u \|_{L_2}}
\end{equation} 
\normalsize
where $\| u - \hat{u} \|_{L_2}$ is the $L_2$ norm of the error vector, and $\| u \|_{L_2}$ is the $L_2$ norm of the reference values.
\subsection{Lid-driven cavity flow at high \textit{Re} numbers}
\begin{figure}[!ht]
    \centering
    \includegraphics[width=0.5\linewidth]{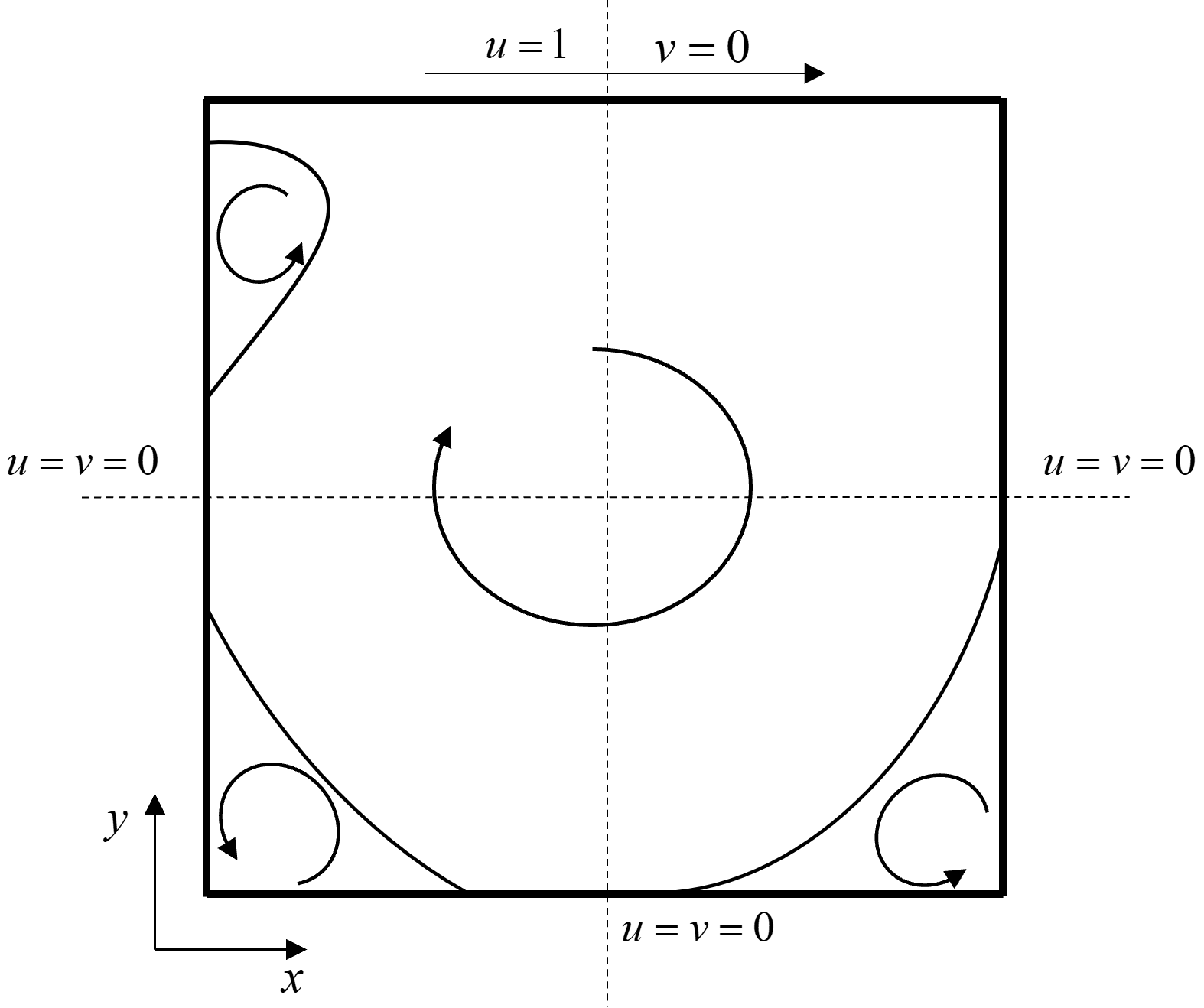}
    \caption{Geometry and boundary conditions of the lid-driven cavity flow.}
    \label{fig:ldc}
\end{figure}

The lid-driven cavity is a classical benchmark in computational fluid dynamics (CFD), characterized by its simple geometry yet rich flow dynamics. The fluid motion within the cavity is governed by the steady incompressible N-S equations, obtained from Eq.~\eqref{eq:2d_NS_unsteady} by neglecting the temporal derivatives. The domain geometry and boundary conditions are shown in Fig.\ref{fig:ldc}, where the top lid moves at a velocity of $u=1, v=0$ and the remaining walls are stationary with no-slip conditions $u=v=0$.
% \begin{subequations} \label{eq:2d_NS_steady}
% \small
%     \begin{align}
%         \frac{\partial u}{\partial x} + \frac{\partial v}{\partial y} &= 0 \label{eq:2d_NS_steady-div} \\ 
%          \frac{\partial \left(uu\right)}{\partial x} +  \frac{\partial \left(vu\right)}{\partial y} &= \frac{1}{Re}\left( \frac{\partial ^2u}{\partial x^2} + \frac{\partial ^2u}{\partial y^2} \right) - \frac{\partial p}{\partial x} \label{eq:2d_NS_steady-m1} \\ 
%          \frac{\partial \left( u v\right)}{\partial x} +  \frac{\partial \left(v v\right)}{\partial y} &= \frac{1}{Re}\left( \frac{\partial ^2v}{\partial x^2} + \frac{\partial ^2v}{\partial y^2} \right) - \frac{\partial p}{\partial y}  \label{eq:2d_NS_steady-m2} 
%     \end{align}
% \end{subequations}
% \normalsize

Although lid-driven cavity flow has been extensively studied numerically, obtaining reliable solutions at high \textit{Re} numbers (above $Re = 10000$) remains a significant challenge in CFD \citep{erturk2005numerical}. As the \textit{Re} number increases, the flow becomes increasingly complex, and conventional numerical methods generally require very fine grids to achieve convergence. Insufficient grid resolution may lead to oscillatory or unsteady solutions \citep{erturk2006fourth,erturk2009discussions}. In this context, the SIMPLE-PINN framework is applied to simulate the lid-driven cavity flow at high \textit{Re} numbers ($Re = 10000$ and $20000$), in order to demonstrate its capability to resolve intricate dynamic features while alleviating the stringent grid resolution requirements.

Table~\ref{ldc_table} provides a summary of the performance of different state-of-the-art PINN models in solving lid-driven cavity flows at high \textit{Re} numbers. Notably, most of these models are restricted to $Re$ numbers of 5000 or below and demand significant training time. In contrast, SIMPLE-PINN is capable of accurately solving flows at higher $Re$ numbers without compromising solution accuracy. In particular, the pressure field shows the most pronounced reduction in error, highlighting the efficacy of the velocity-pressure coupling correction mechanism in SIMPLE-PINN. It is worth noting that while a few previous studies have reported solutions at $Re = 10000$, they typically relied on either additional labeled data \citep{jiang2023applications} or curriculum learning strategies \citep{li2025learning}. In contrast, SIMPLE-PINN can directly solve the problem without incorporating any additional data. Moreover, to the best of the authors’ knowledge, this study represents the first successful PINN-based solution of the lid-driven cavity flow at $Re = 20000$, a problem for which obtaining a converged solution is highly challenging even for traditional numerical methods.

In terms of computational cost and training time, although SIMPLE-PINN introduces an additional loss term for pressure correction, this does not impose a substantial computational burden. As shown in Table~\ref{ldc_table}, the training times of both methods are comparable. By enforcing velocity-pressure coupling, the network’s adherence to the underlying physical constraints is enhanced, allowing SIMPLE-PINN to be trained with smaller batch sizes, which helps with rapid convergence. At $Re = 10000$, SIMPLE-PINN attained accuracy comparable to FFV-PINN~\citep{wei2025ffv}, yet it converged in roughly half the training time, demonstrating its superior computational efficiency. At $Re = 20000$, under the same training conditions, FFV-PINN failed to produce satisfactory solutions, whereas SIMPLE-PINN successfully converged within 448~s~(0.124~h), further highlighting its robustness and efficiency at higher $Re$ numbers.

\begin{table}[ht]
  \caption{Comparison of relative $L_2$ errors for velocity magnitude $V$ and pressure $p$ and training time (in hours) obtained by different PINNs variants.}
  \label{ldc_table}
  \centering
  \small
  \renewcommand{\arraystretch}{1.2}
  \begin{tabular}{lcccll}
    \toprule
    Method & $Re$ & Rel.~$L_2$ $V$ & Rel.~$L_2$ $p$ & Time (h) & Hardware \\
    \midrule
    JAXPI \citep{wang2023expert} & 3200 & $1.58\times 10^{-1}$ & $-$ & $-$ & $-$ \\
    PirateNet \citep{wang2024piratenets} & 3200 & $4.21\times 10^{-2}$ & $-$ & 11.83 & RTX 3090 \\
    % ev-NSFnet \citep{wang2023solution} & 5000 & $-$ & $-$ & $-$ & $-$ \\
    TSONN \citep{cao2023tsonn} & 5000 & $1.00\times 10^{-1}$ & $-$ & $-$ & $-$ \\
    SOAP \citep{wang2025gradient} & 5000 & $3.99\times 10^{-2}$ & $-$ & 8.25 & RTX A6000 \\
    % FFV-PINN & 3200 & $3.69\times 10^{-2}$ & $-$ & 0.0636 & RTX 3090 & $-$ \\
    % FFV-PINN & 5000 & $3.90\times 10^{-2}$ & $-$ & 0.0658 & RTX 3090 & $-$ \\
    FFV-PINN \citep{wei2025ffv} & 10000 & $3.03\times 10^{-2}$ & $5.10\times 10^{-2}$ & 0.189 & RTX 3090 \\
    SIMPLE-PINN & 10000 & $3.69\times 10^{-2}$ & $2.63\times 10^{-2}$ & 0.097 & RTX 3090 \\
    ND-PINN & 20000 & $8.92\times 10^{-1}$ & $9.86\times 10^{-1}$ & 0.113 & RTX 3090 \\
    FFV-PINN & 20000 & $2.39\times 10^{-1}$ & $4.45\times 10^{-1}$ & 0.117 & RTX 3090 \\
    SIMPLE-PINN & 20000 & $3.65\times 10^{-2}$ & $2.71\times 10^{-2}$ & 0.124 & RTX 3090 \\
    \bottomrule
  \end{tabular}
  {\footnotesize\raggedright
  \par\vspace*{1mm} 
  Note: The training time for SOAP is reported in \citep{wang2025gradient}, whereas the training time for PirateNet was obtained by executing the authors' code on an NVIDIA RTX 3090 GPU. ND-PINN refers to the standard PINN model employing FVM discretization without the inclusion of correction terms. At $Re = 20000$, both SIMPLE-PINN and FFV-PINN were trained under identical training configurations to ensure a fair comparison.\par}
\end{table}

In terms of spatial resolution, Erturk reported that a resolution as high as $1025 \times 1025$ is required to ensure stability \citep{erturk2009discussions} at $Re = 20000$, or at least $601 \times 601$ when adopting a fourth-order discretization scheme \citep{erturk2006fourth}.  In the present study, simulations performed with ANSYS Fluent utilized a $512 \times 512$ grid, achieving convergence in approximately 610 s on 24 CPUs with a convergence tolerance of $10^{-10}$. The attainment of stable convergence at this comparatively coarse resolution is enabled by the use of a coupled solver with a pseudo-transient formulation, which enhances numerical stability.
% the CFD solution obtained using ANSYS Fluent at $Re=20000$ employs a $512 \times 512$ grid and requires approximately 610 s on 24 CPUs with a convergence tolerance of $10^{-10}$. Achieving convergence also necessitates the use of a coupled solver with a pseudo-transient formulation. Moreover, Erturk reported that a resolution as high as $1025 \times 1025$ is required to ensure stability \citep{erturk2009discussions}, or at least $601 \times 601$ when adopting a fourth-order discretization scheme \citep{erturk2006fourth}. 
In contrast, SIMPLE-PINN achieves stable solutions with only $258 \times 258$ collocation points, reducing the number of required sample points by nearly 75\% compared with the CFD grid.

\begin{figure}[!ht]
    \centering
    \includegraphics[width=1.0\linewidth]{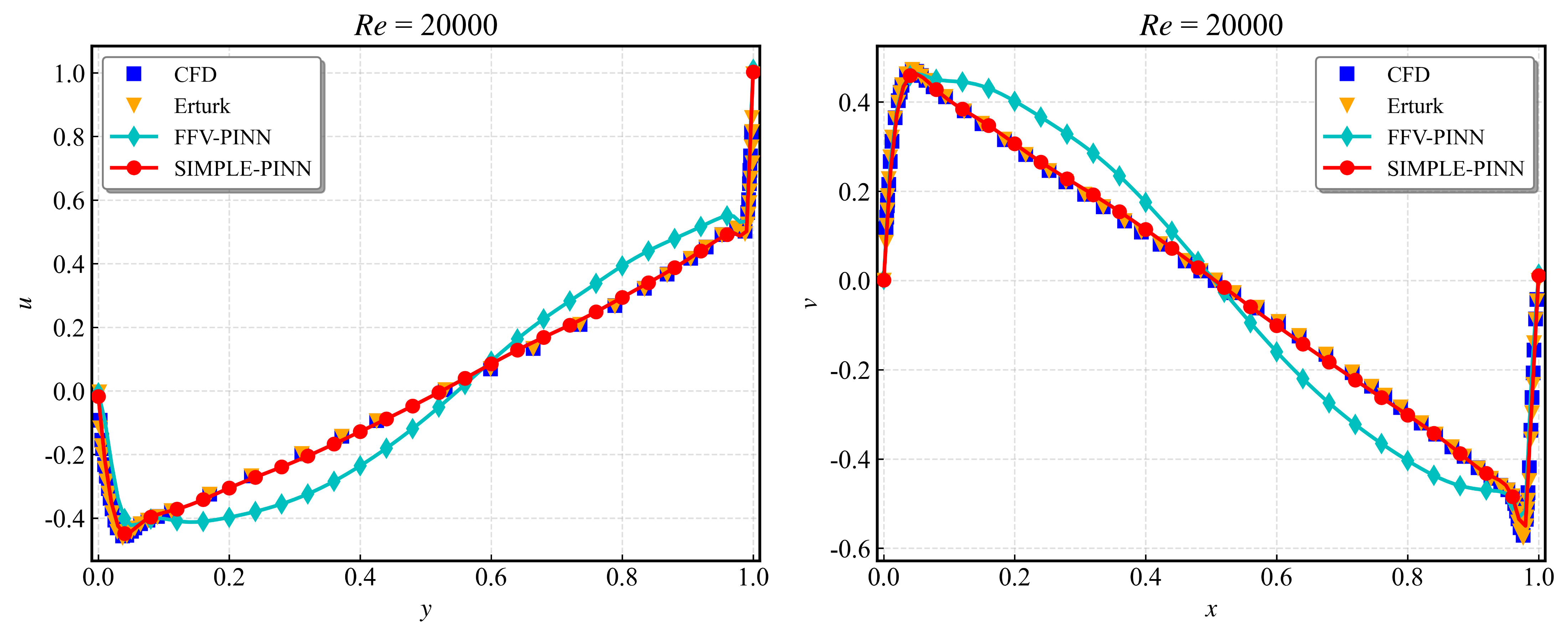}
    \caption{Velocity profiles along the cavity centerline at $Re = 20000$. The left panel shows the horizontal velocity component ($u$) along the vertical centerline ($x=0.5$), and the right panel shows the vertical velocity component ($v$) along the horizontal centerline ($y=0.5$). Results from FFV-PINN and SIMPLE-PINN are compared with CFD and Erturk's benchmark data \citep{erturk2009discussions}.}
    \label{fig:ldc_uv_lines}
\end{figure}

Fig. \ref{fig:ldc_uv_lines} shows the velocity profiles, with $u$ along the vertical line ($x=0.5$) and $v$ along the horizontal line ($y=0.5$), comparing the SIMPLE-PINN and FFV-PINN frameworks against established benchmark solutions (Erturk’s result) and CFD results at $Re=20000$. The SIMPLE-PINN results closely match both the CFD and Erturk's data for both velocity components, accurately capturing the velocity distribution patterns, including the steep velocity gradients near the walls. In contrast, the FFV-PINN results exhibit noticeable deviations from the benchmark data. Overall, these comparisons provide qualitative evidence of the superior accuracy of SIMPLE-PINN over FFV-PINN, demonstrating the effectiveness of the velocity-pressure coupling correction in solving high-\textit{Re} problems.

\begin{figure}[!ht]
    \centering
    \includegraphics[width=1.0\linewidth]{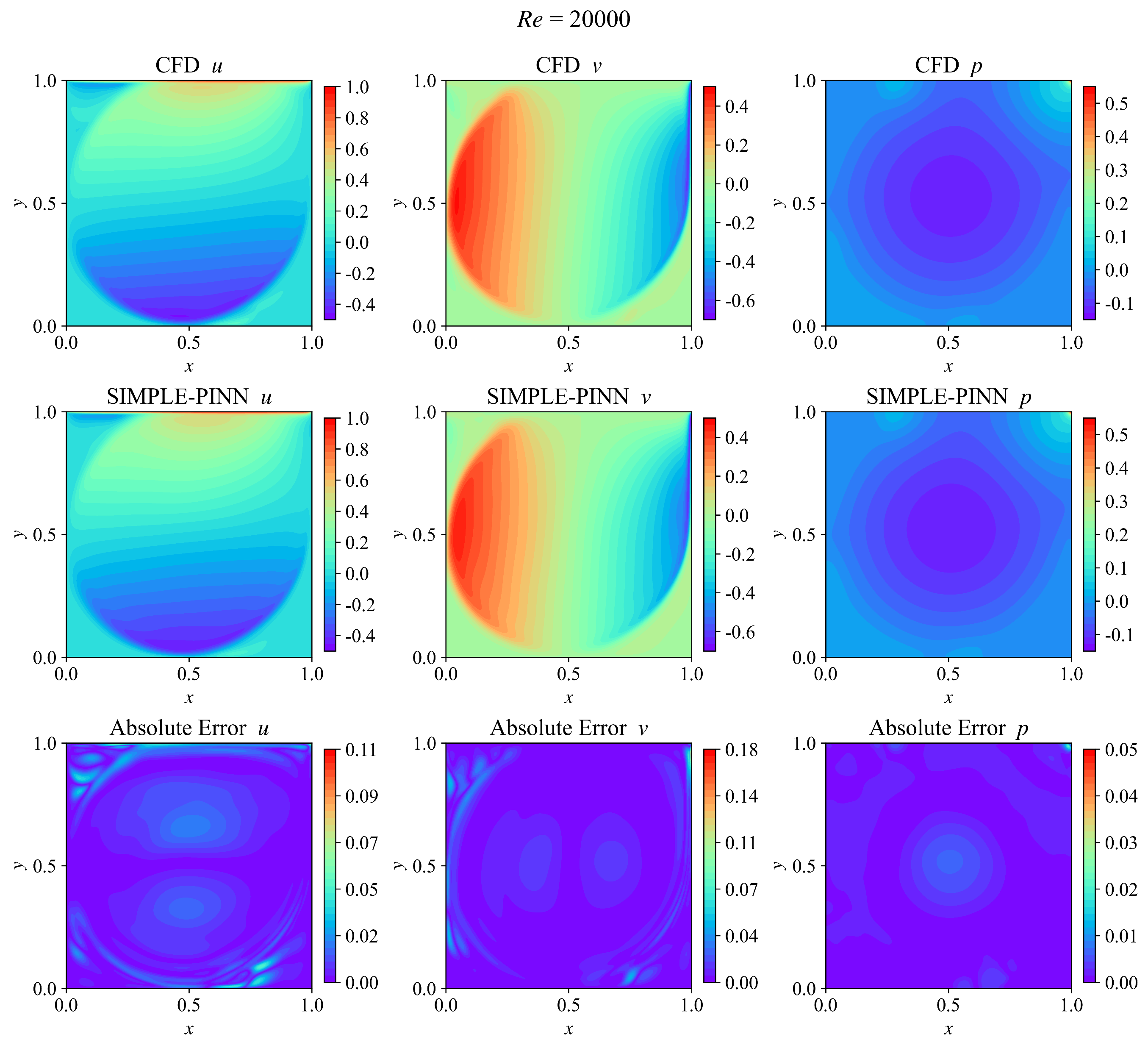}
    \caption{Comparison of velocity and pressure fields at $Re=20000$. The top row presents the benchmark CFD solutions for the horizontal and vertical velocity components ($u$ and $v$) and pressure ($p$). The middle row displays the corresponding predictions obtained using the SIMPLE-PINN framework. The bottom row shows the pointwise absolute error fields for each variable.}
    \label{fig:ldc_uv_contour}
\end{figure}

Fig. \ref{fig:ldc_uv_contour} presents the contours of the velocity components and pressure at $Re=20000$, comparing results from SIMPLE-PINN with high-fidelity CFD. The SIMPLE-PINN predictions exhibit a visual agreement that is nearly indistinguishable from the CFD results. The pointwise absolute error remains consistently low throughout most of the domain, with slightly higher values near the top lid corners due to discontinuities, which are inherently more challenging to resolve \citep{wang2024piratenets}.

\subsection{Wavy channel flow} 
\begin{figure}[!ht]
    \centering
    \includegraphics[width=1.0\linewidth]{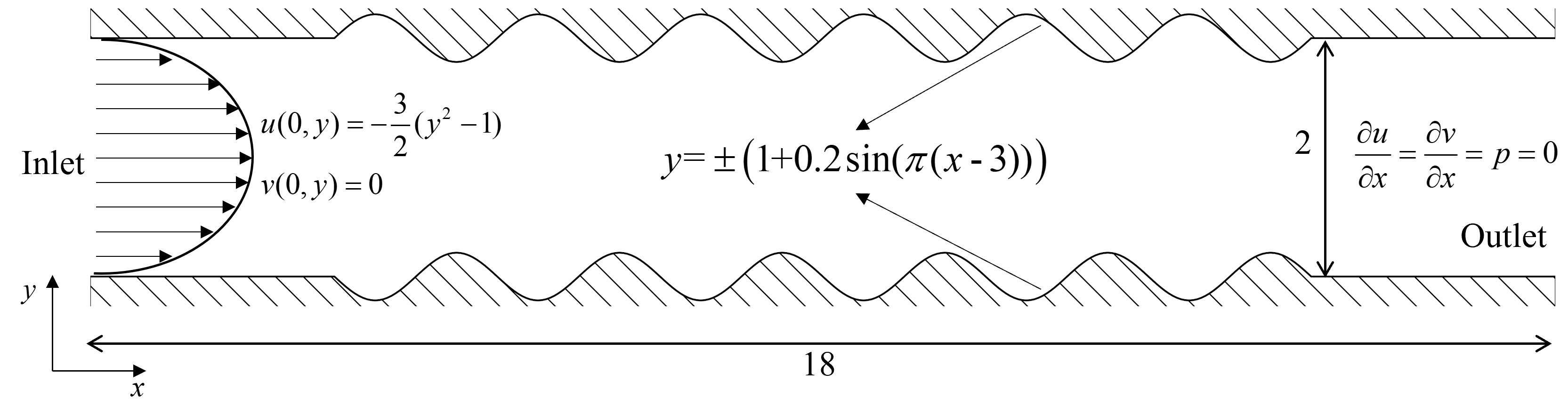}
    \caption{A schematic of the wavy channel flow problem. The top and bottom walls are defined by the sinusoidal function $y = \pm(1+0.2\sin(\pi(x-3)))$, with no-slip boundary conditions applied. The fluid enters the channel with a parabolic velocity profile, $u(0,y) = -\frac{3}{2}(y^2-1)$, and zero vertical velocity, $v(0,y)=0$. At the outlet, zero-gradient boundary conditions are applied for velocity variables, and the pressure is set to $p=0$.}
    \label{fig:wavy_channel_geo}
\end{figure}

To evaluate the capability of SIMPLE-PINN in handling flows involving complex geometries, this section considers the wavy channel flow at $Re=100$. As illustrated in Fig.~\ref{fig:wavy_channel_geo}, the channel is bounded by sinusoidally undulating walls. Such wall geometry can induce flow instabilities, often resulting in vortex formation.

\begin{figure}[!ht]
    \centering
    \includegraphics[width=1.0\linewidth]{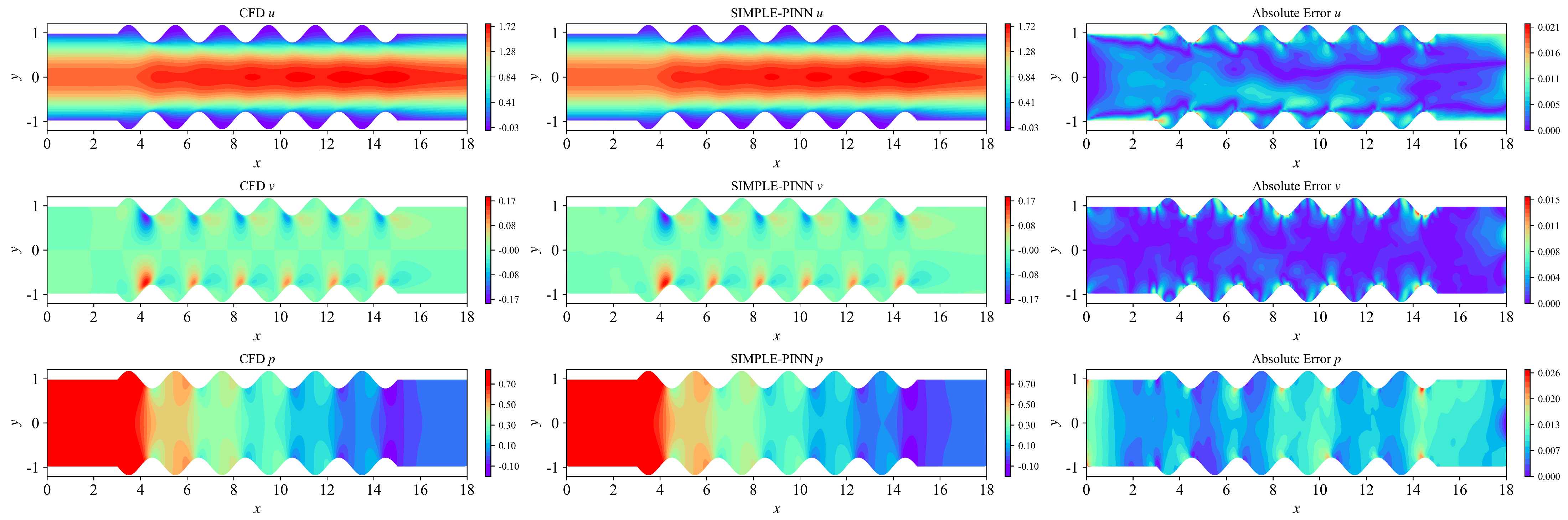}
    \caption{Comparison of velocity and pressure contours for the wavy channel flow. The first column presents the CFD solutions \citep{chiu2023cdfib}, the second column shows the corresponding results from the SIMPLE-PINN, and the third column depicts the absolute error fields. From top to bottom, the rows correspond to the horizontal velocity ($u$), vertical velocity ($v$), and pressure ($p$), respectively.}
    \label{fig:wavy_channel_contour}
\end{figure}

Fig. \ref{fig:wavy_channel_contour} compares SIMPLE-PINN predictions with high-fidelity CFD results \citep{chiu2023cdfib}. The SIMPLE-PINN results exhibit excellent agreement with the CFD solutions. Specifically, SIMPLE-PINN accurately captures the negative $u$ velocities near the top and bottom walls, the variations in $v$ induced by the sinusoidal wall geometry, and the complex pressure distribution that develops along the deepening channel. The pointwise absolute error further confirms the accuracy of SIMPLE-PINN, with errors remaining below approximately $10^{-2}$ throughout most of the domain.

\begin{figure}[!ht]
    \centering
    \includegraphics[width=0.9\linewidth]{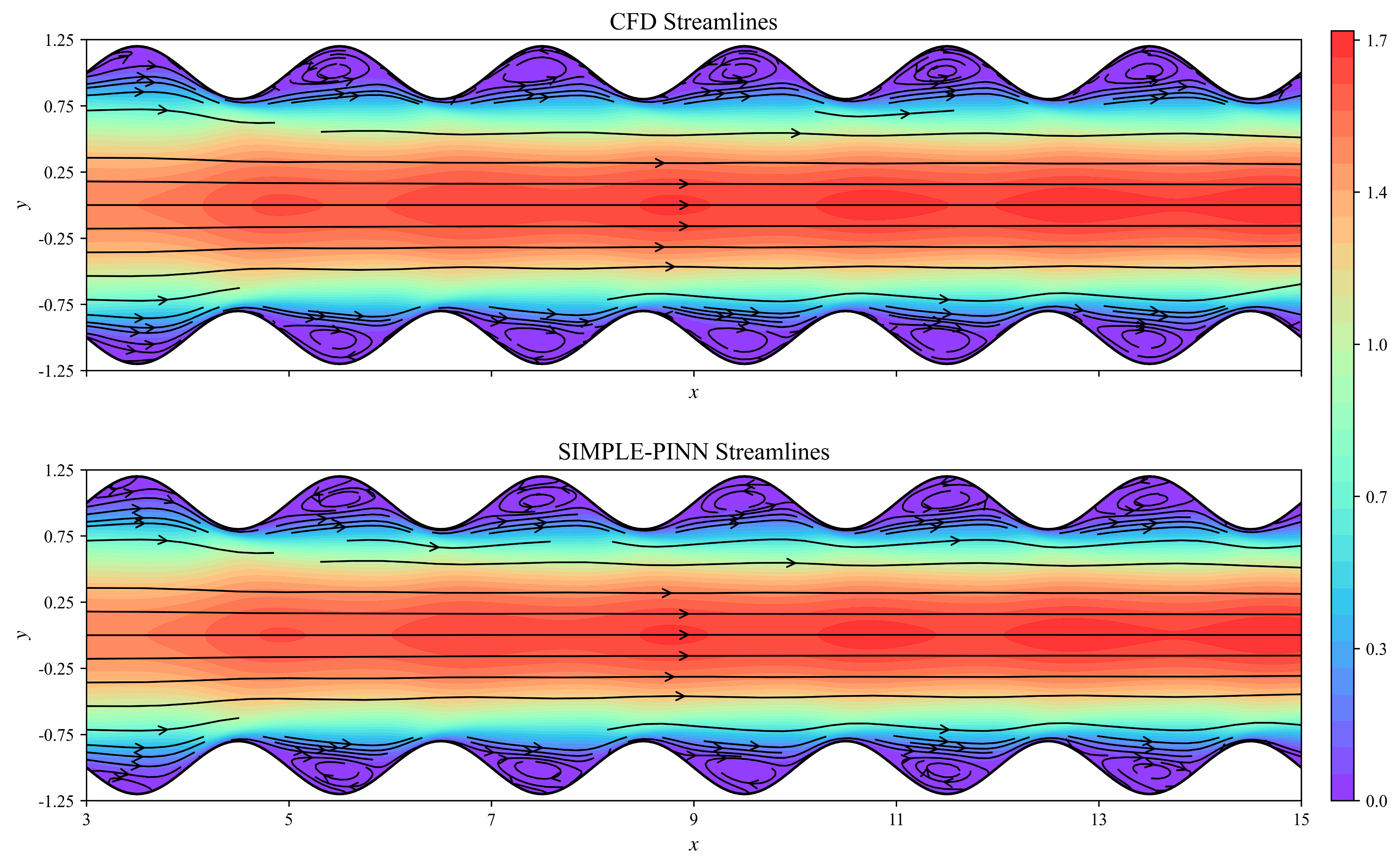}
    \caption{A comparison of streamlines and velocity magnitude for the wavy channel flow. The top panel shows the CFD results, while the bottom panel presents the SIMPLE-PINN predictions. The color map represents the velocity magnitude.}
    \label{fig:wavy_channel_stlines}
\end{figure}

Fig.~\ref{fig:wavy_channel_stlines} compares streamlines between SIMPLE-PINN and high-fidelity CFD, with velocity magnitude ($V$) represented as the background color. The undulating top and bottom walls generate vortical structures near the crests and troughs as the fluid progresses along the wavy channel. The streamlines predicted by SIMPLE-PINN exhibit excellent agreement with the CFD results, capturing the evolution of vortices along the channel from small-scale structures near the crests and troughs to larger and more stable vortices downstream. This close visual agreement highlights the high accuracy and physical consistency of SIMPLE-PINN, which reliably resolves the intricate flow features of the wavy channel, accurately reproduces both the global flow structure and the evolution of localized vortices, and demonstrates its capability to handle flows within complex, irregular geometries.

Table~\ref{tab:performance_SIMPLE-PINN} summarizes the performance metrics of SIMPLE-PINN on various benchmark problems. For the wavy channel flow, the model achieves relative $L_2$ errors of $3.91\times10^{-3}$ to $8.00\times10^{-2}$ and MSE values of $3.95\times10^{-6}$ to $2.03\times10^{-5}$ across all velocity components and pressure. The solution is obtained within a short training time of 219 s using only $160 \times 72$ collocation points, quantitatively demonstrating that SIMPLE-PINN can ensure both high accuracy and computational efficiency.
\begin{table}[!ht]
  \centering
  \small
    % \footnotesize

  \caption{Performance metrics of the SIMPLE-PINN framework on various benchmark problems}
  \label{tab:performance_SIMPLE-PINN}
  \renewcommand{\arraystretch}{1.2}
  \resizebox{\textwidth}{!}{
  \begin{tabular}{lccccc}
    \toprule
        Cases & Wavy Channel & Square  & Square  & Cylinder & RT  \\      
    \midrule
    Parameter & $Re=100$ & $Re=25$ & $Re=40$ & $Re=100$ &  $Ra=10^6$ \\
    Rel.$L_2$ $u$ & $3.94\times10^{-3}$ & $4.57\times10^{-3}$ & $5.11\times10^{-3}$ & $5.10\times10^{-2}$ & $7.75\times10^{-2}$ \\
    Rel.$L_2$ $v$ & $8.00\times10^{-2}$ & $3.00\times10^{-2}$ & $1.77\times10^{-2}$ & $2.15\times10^{-1}$ & $7.00\times10^{-2}$ \\
    Rel.$L_2$ $V$ & $3.91\times10^{-3}$ & $4.66\times10^{-3}$ & $5.15\times10^{-3}$ & $5.11\times10^{-2}$ & $6.23\times10^{-2}$ \\
    Rel.$L_2$ $p$ & $1.92\times10^{-2}$ & $4.39\times10^{-2}$ & $1.56\times10^{-2}$ & $2.45\times10^{-1}$ & $8.21\times10^{-2}$ \\
    Rel.$L_2$ $T$ & - & - & - & - & $3.24\times10^{-2}$ \\
    \midrule
    MSE $u$ & $2.03\times10^{-5}$ & $2.32\times10^{-5}$ & $3.45\times10^{-5}$ & $3.29\times10^{-3}$ & $1.15\times10^{-4}$ \\
    MSE $v$ & $3.95\times10^{-6}$ & $6.38\times10^{-6}$ & $1.05\times10^{-5}$ & $4.00\times10^{-3}$ & $2.87\times10^{-4}$ \\
    MSE $V$ & $2.00\times10^{-5}$ & $2.42\times10^{-5}$ & $3.60\times10^{-5}$ & $3.52\times10^{-3}$ & $3.02\times10^{-4}$ \\
    MSE $p$ & $6.79\times10^{-5}$ & $7.29\times10^{-5}$ & $1.62\times10^{-4}$ & $5.55\times10^{-3}$ & $8.12\times10^{-4}$ \\
    MSE $T$ & - & - & - & - & $4.58\times10^{-4}$ \\
    \midrule
    Time (s) & 219 & 335 & 451 & 7269 & 6419 \\
    Sample points & $160 \times 72$ & $1026 \times 514$ & $881 \times 165$ & $401 \times 441 \times 83$ & $301 \times 128 \times 256$ \\
    Temporal behavior & Steady & Steady & Steady & Unsteady & Unsteady \\
    Flow domain & Channel & Open domain & Channel & Channel & Enclosed cavity \\
    \bottomrule
  \end{tabular}
  }
\end{table}

\subsection{Flow past a NACA0012 airfoil} 
\begin{figure}[!ht]
    \centering
    % 左图
    \begin{subfigure}[b]{0.49\linewidth}
        \centering
        \includegraphics[width=\linewidth]{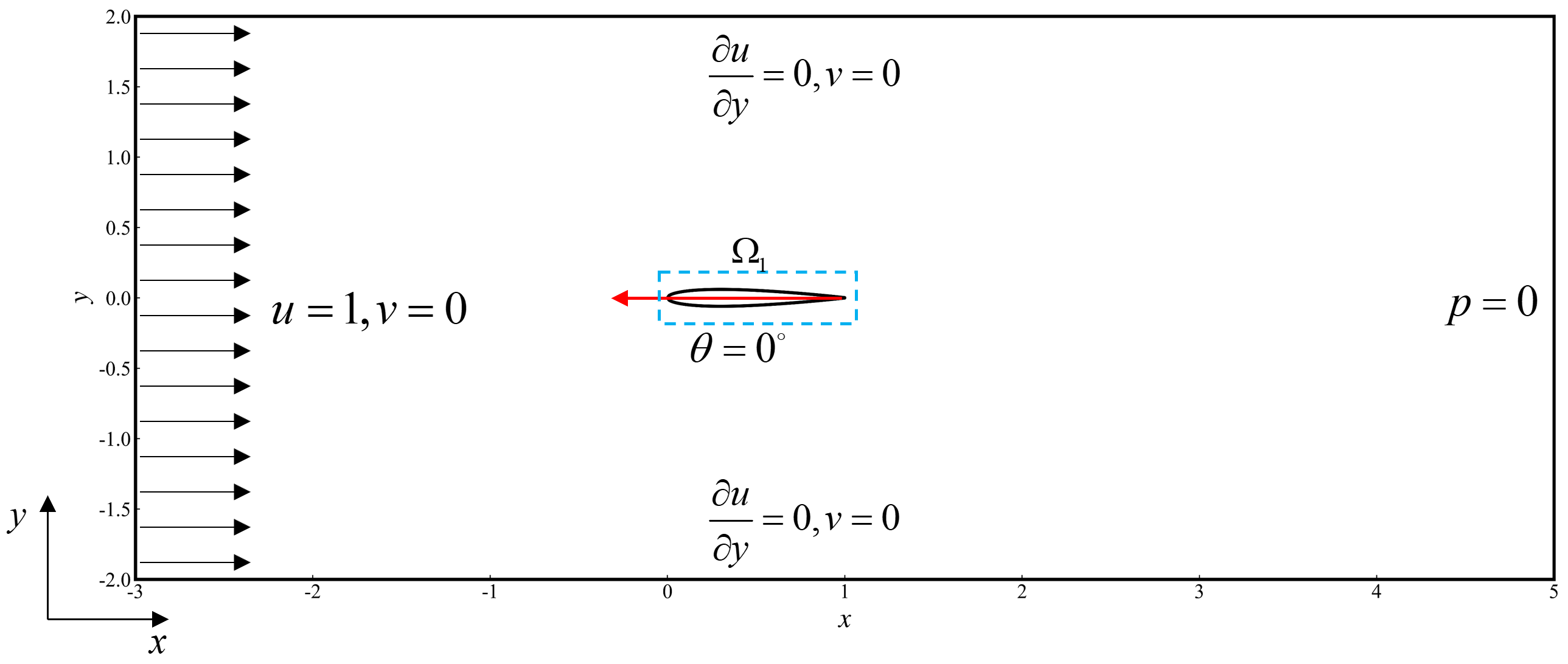}
        \caption{$Re=500$, $AOA=0^\circ$.}
        \label{fig:airfoil_Re500_geo}
    \end{subfigure}
    \hfill
    % 右图
    \begin{subfigure}[b]{0.49\linewidth}
        \centering
        \includegraphics[width=\linewidth]{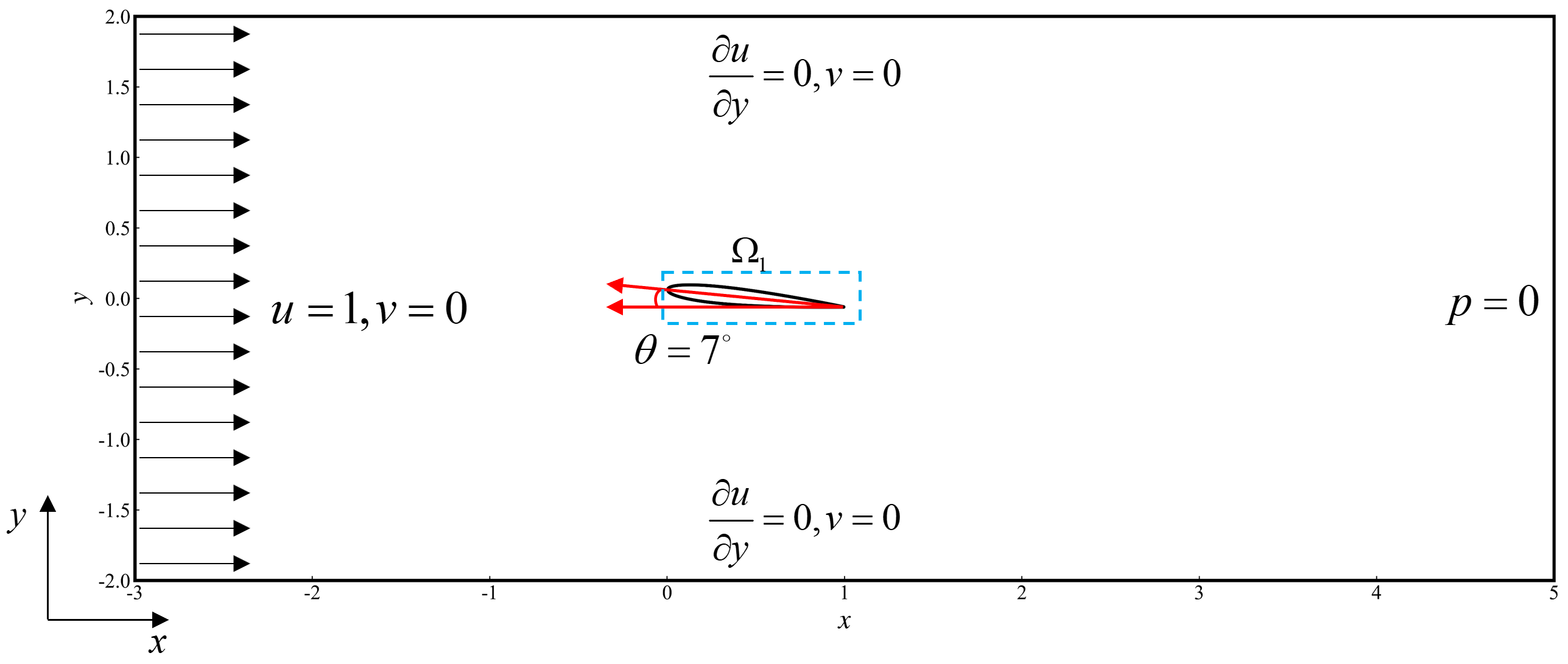}
        \caption{$Re=1000$, $AOA=7^\circ$.}
        \label{fig:airfoil_geo}
    \end{subfigure}
    
    \caption{Computational domains and boundary conditions for the NACA0012 airfoil at different $Re$ numbers and angles of attack.}
    \label{fig:airfoil_all}
\end{figure}

Flow past an airfoil is a fundamental topic in fluid mechanics, as it describes the interaction between fluids and solid bodies with complex shapes. The NACA0012 airfoil, a classical symmetric configuration, has been widely employed as a benchmark in aerodynamic studies. For PINNs, accurately capturing the pressure distribution around such airfoils remains a significant challenge.  To further assess the capability of SIMPLE-PINN,  this section considers the flow past a NACA0012 airfoil under two representative scenarios: (1) $Re=500$ with an angle of attack (AOA) of $0^\circ$, and (2) $Re=1000$ with an AOA of $7^\circ$. The airfoil is placed in an open domain and subjected to a uniform inflow from the right, as depicted in Fig.~\ref{fig:airfoil_all}.
% \begin{figure}[!ht]
%     \centering
%     \includegraphics[width=0.85\linewidth]{pictures/NACA airfoil 0.png}
%     \caption{Computational domain and boundary conditions for the NACA0012 airfoil at $Re=500$, $AOA=0^\circ$.}
%     \label{fig:airfoil_Re500_geo}
% \end{figure}

% \begin{figure}[!ht]
%     \centering
%     \includegraphics[width=0.9\linewidth]{pictures/NACA airfoil 7.png}
%     \caption{Computational domain and boundary conditions for the NACA0012 airfoil at $Re=1000$, $AOA=7^\circ$. A uniform inlet velocity of $u=1, v=0$ is applied on the left boundary, a zero-pressure condition ($p=0$) is set at the right boundary, and free-slip conditions ($\frac{\partial u}{\partial y}=0, v=0$) are applied to the top and bottom boundaries.}
%     \label{fig:airfoil_geo}
% \end{figure}

\begin{figure}[!ht]
    \centering
    % 上图
    \begin{subfigure}[b]{1.0\linewidth}
        \centering
        \includegraphics[width=\linewidth]{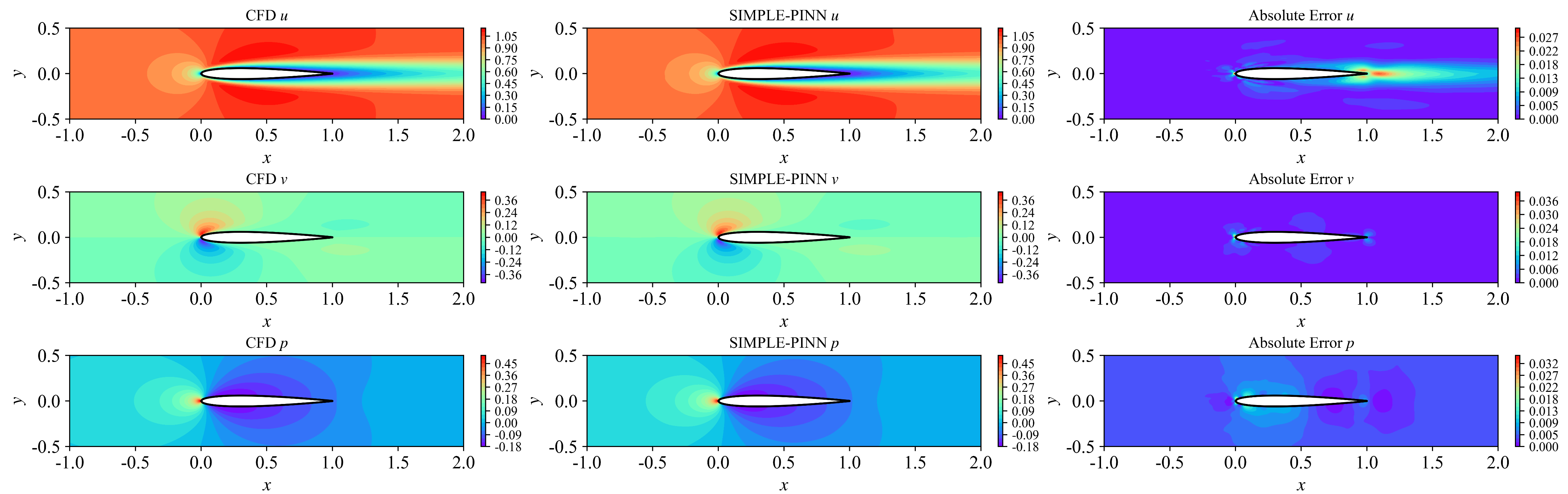}
        \caption{$Re=500$, $AOA=0^\circ$.}
        \label{fig:airfoil_Re500_uvp}
    \end{subfigure}
        \vspace{0.0em} 
    \begin{subfigure}[b]{1.0\linewidth}
        \centering
        \includegraphics[width=\linewidth]{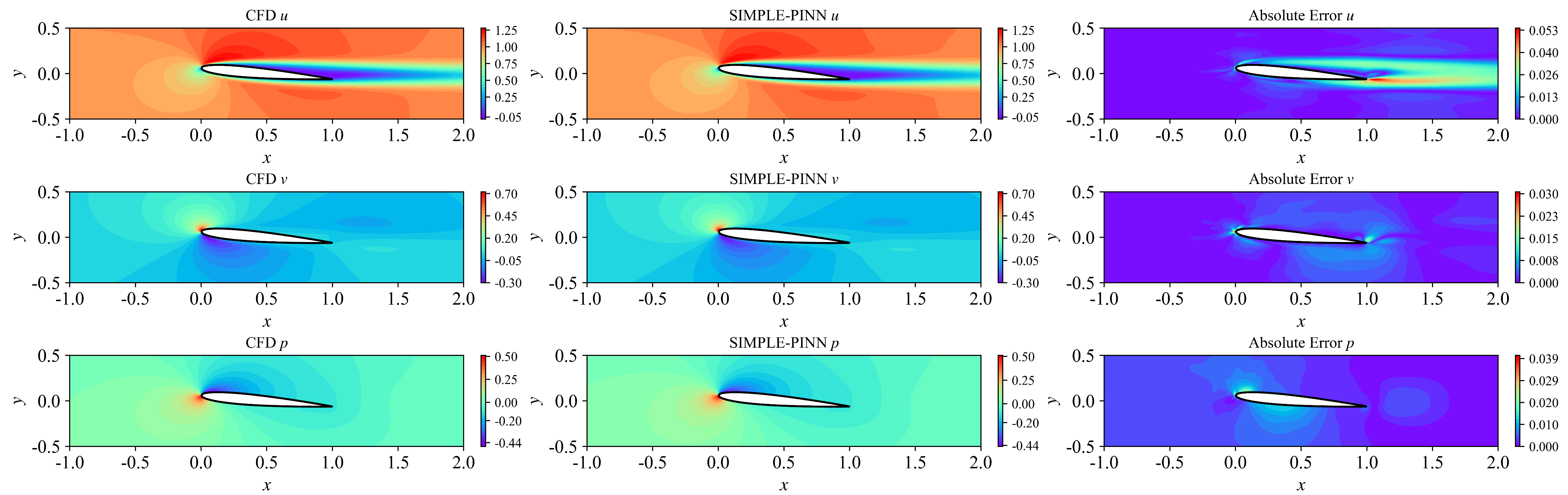}
        \caption{$Re=1000$, $AOA=7^\circ$.}
        \label{fig:airfoil_contour} 
    \end{subfigure}
    
    \caption{Comparison of CFD and SIMPLE-PINN predictions for the NACA0012 airfoil at different $Re$ numbers and angles of attack.}
    \label{fig:airfoil_uvp_all}
\end{figure}

To provide a detailed analysis of the local flow feature and pressure distributions around the airfoil, computational results are presented for the subdomain $\Omega_1$ depicted in Fig.~\ref{fig:airfoil_all}. As illustrated in Fig.\ref{fig:airfoil_uvp_all}, SIMPLE-PINN produces predictions that closely match the CFD solution \citep{chiu2023cdfib}, accurately capturing the velocity distribution near the leading edge as well as the corresponding pressure field. The absolute error remains on the order of $10^{-3}$, corresponding to a two-order-of-magnitude improvement over IBM-PINN \citep{xiao2025immersed}. The significant accuracy enhancement is particularly evident in regions with steep velocity and pressure gradients, highlighting SIMPLE-PINN's capability for resolving detailed aerodynamic features around complex geometries. In addition, SIMPLE-PINN consistently maintains its high accuracy even as the flow complexity increases with higher \textit{Re} and AOA, confirming its reliable performance across a range of flow conditions.  
% \begin{figure}[!ht]
%     \centering
%     \includegraphics[width=1.0\linewidth]{pictures/NACA Re500 uvp contour.png}
%     \caption{Comparison of CFD and SIMPLE-PINN results for the NACA0012 airfoil at $Re=500$, $AOA=0^\circ$.}
%     \label{fig:airfoil_Re500_uvp}
% \end{figure}

% \begin{figure}[!ht]
%     \centering
%     \includegraphics[width=1.0\linewidth]{pictures/NACA Re1000 uvp contour.png}
%     \caption{Comparison of CFD and SIMPLE-PINN results for the NACA0012 airfoil at $Re=1000$, $AOA=7^\circ$, showing the velocity components ($u$, $v$) and pressure field ($p$). From left to right: CFD solution, SIMPLE-PINN prediction, and the absolute error.}
%     \label{fig:airfoil_contour} 
% \end{figure}

\begin{figure}[!ht]
    \centering
    % 左图
    \begin{subfigure}[b]{0.49\linewidth}
        \centering
        \includegraphics[width=\linewidth]{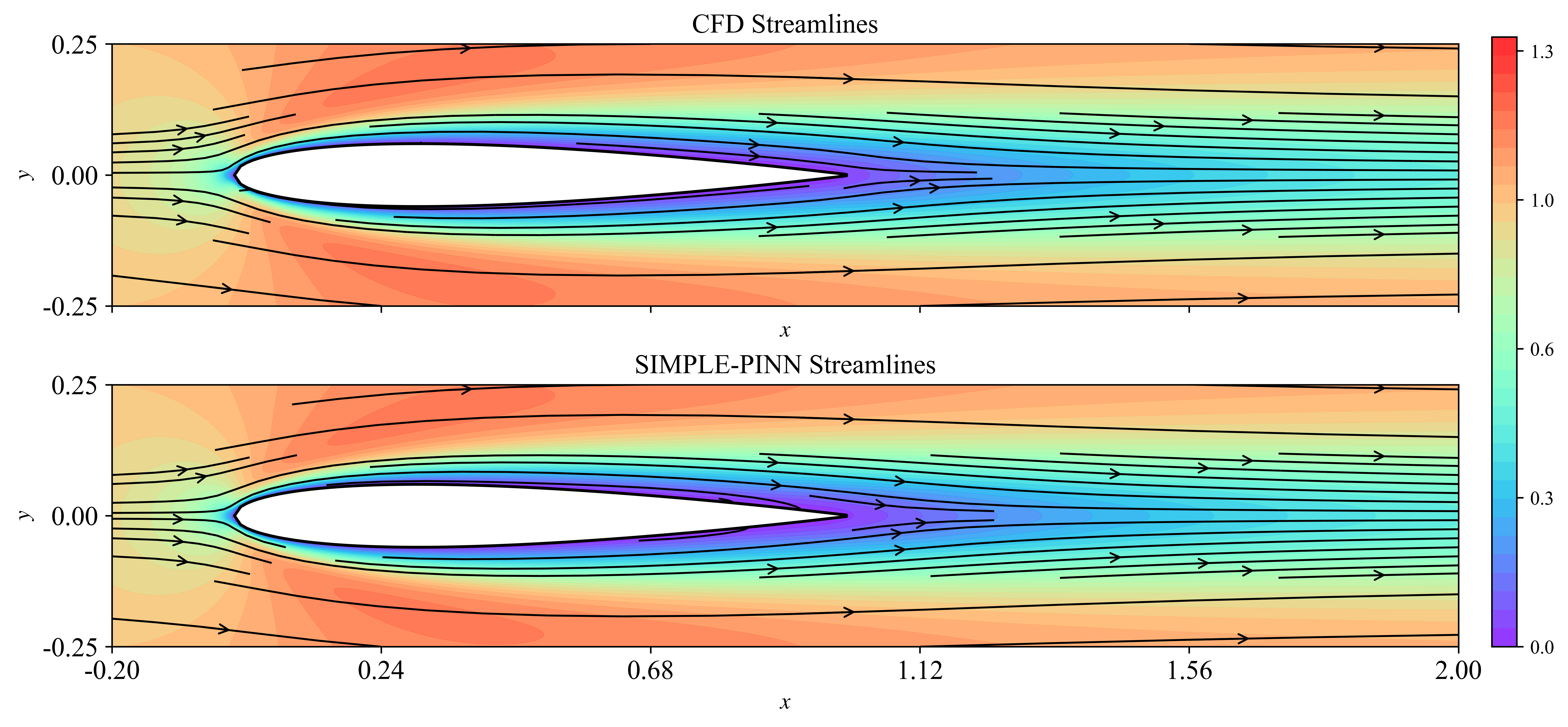}
        \caption{$Re=500$, $AOA=0^\circ$. }
        \label{fig:airfoil_stlines_Re500}
    \end{subfigure}
    \hfill
    % 右图
    \begin{subfigure}[b]{0.49\linewidth}
        \centering
        \includegraphics[width=\linewidth]{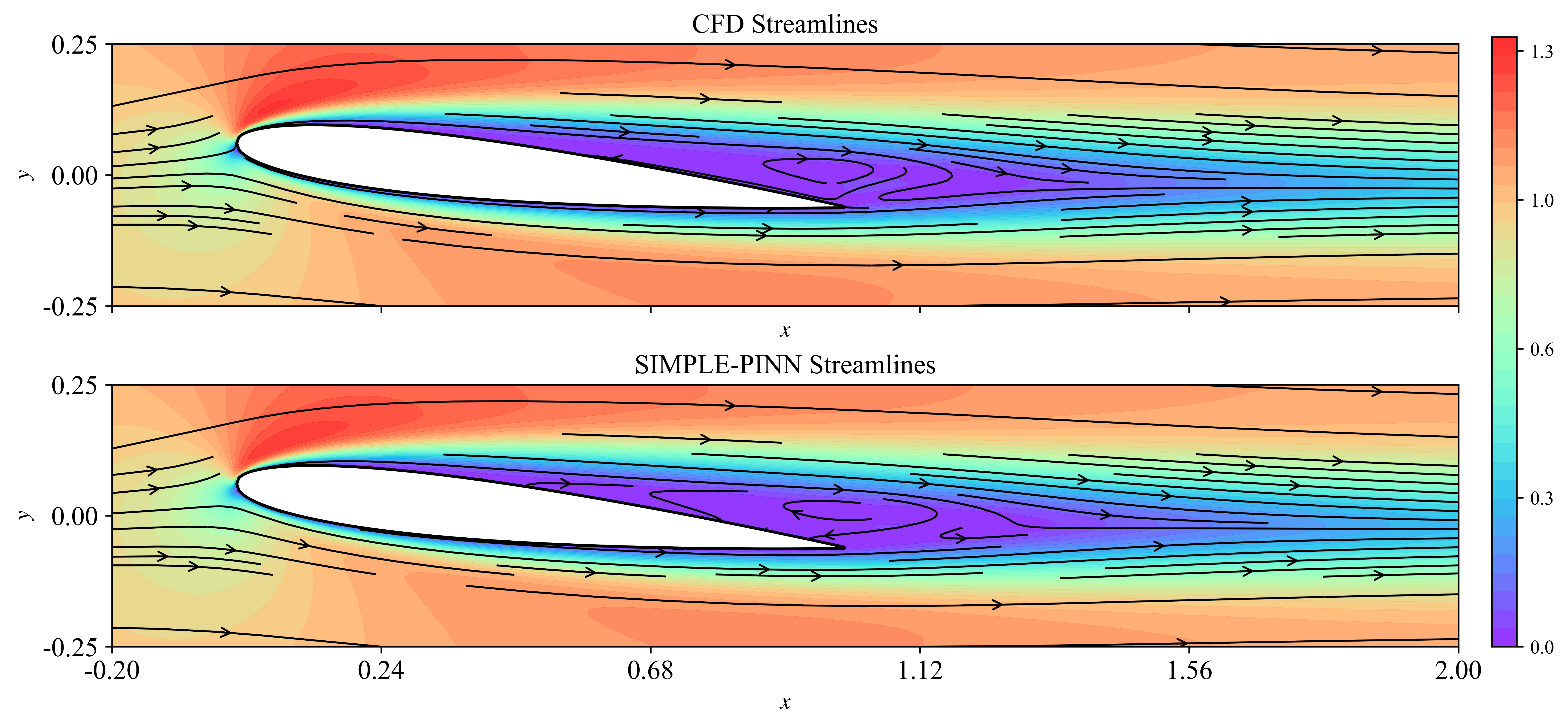}
        \caption{$Re=1000$, $AOA=7^\circ$. }
        \label{fig:airfoil_stlines_Re1000}
    \end{subfigure}
    
    \caption{Streamlines and velocity magnitude ($V$) for the NACA0012 airfoil at different $Re$ numbers and angles of attack.}
    \label{fig:airfoil_stlines_all}
\end{figure}

As shown in Fig.~\ref{fig:airfoil_stlines_all},  the streamlines predicted by SIMPLE-PINN closely match the CFD results \citep{chiu2023cdfib}, with velocity magnitude displayed in the background. The streamline patterns clearly depict the transition from smooth, laminar, attached flow ($Re = 500$, $\alpha = 0^\circ$) to complex, separated flow with distinct vortex formation ($Re = 1000$, $\alpha = 7^\circ$). SIMPLE-PINN successfully resolves both flow regimes, capturing the complex dynamics of flow separation and vortex formation, which are often challenging for conventional PINN models.
% \begin{figure}[!ht]
%     \centering
%     \includegraphics[width=0.9\linewidth]{pictures/NACA Re500 streamlines.png}
%         \caption{Streamlines and velocity magnitude ($V$) for the NACA0012 airfoil at $Re=1000$ and $AOA=7^\circ$. Top: CFD results; Bottom: SIMPLE-PINN predictions.}
%     \label{fig:airfoil_stlines}
% \end{figure}detailed flow features.

% \begin{figure}[!ht]
%     \centering
%     \includegraphics[width=0.9\linewidth]{pictures/NACA Re1000 streamlines.png}
%         \caption{Streamlines and velocity magnitude ($V$) for the NACA0012 airfoil at $Re=1000$ and $AOA=7^\circ$. Top: CFD results; Bottom: SIMPLE-PINN predictions.}
%     \label{fig:airfoil_stlines}
% \end{figure}

\begin{figure}[!ht]
    \centering
    % 左图
    \begin{subfigure}[b]{0.49\linewidth}
        \centering
        \includegraphics[width=\linewidth]{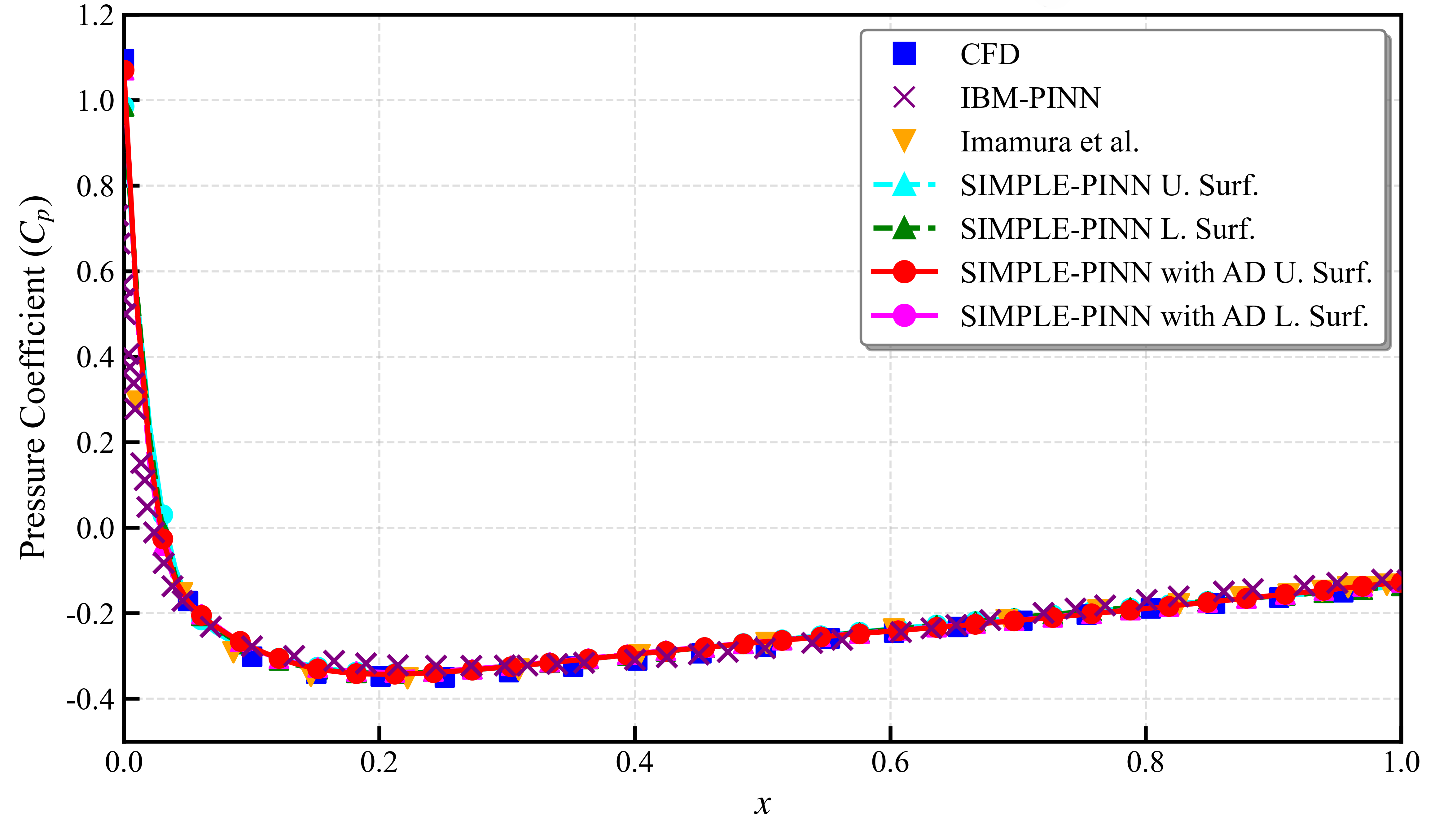}
        \caption{$Re=500$, $AOA=0^\circ$.}
        \label{fig:airfoil_Re500_cp}
    \end{subfigure}
    \hfill
    % 右图
    \begin{subfigure}[b]{0.49\linewidth}
        \centering
        \includegraphics[width=\linewidth]{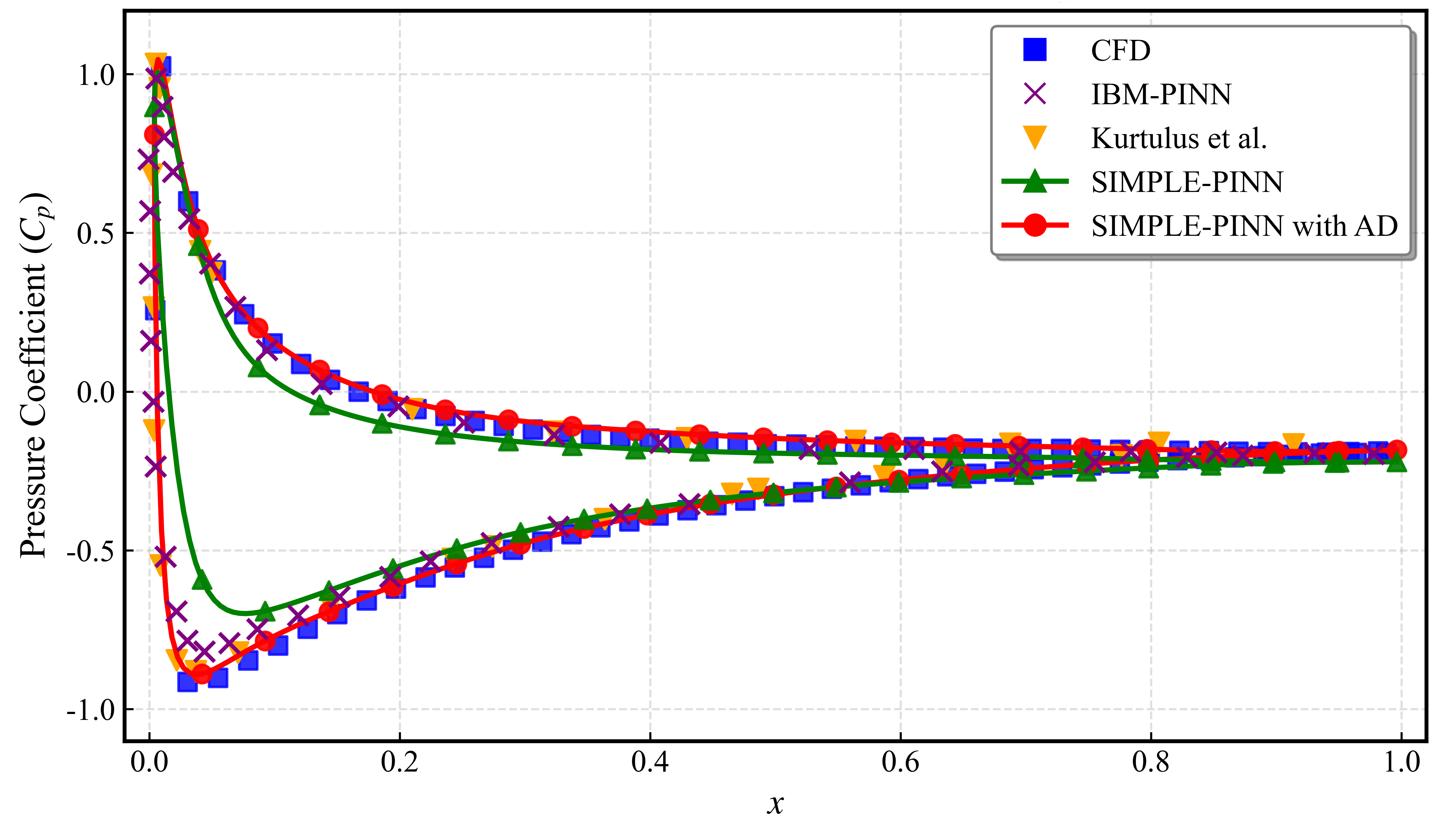}
        \caption{$Re=1000$, $AOA=7^\circ$.}
        \label{fig:airfoil_cp}
    \end{subfigure}
    
    \caption{Pressure coefficient ($C_p$) distributions on the NACA0012 airfoil surface at different $Re$ numbers and angles of attack.  }
    \label{fig:airfoil_cp_all}
\end{figure}

The effect of applying AD at near-wall points within the SIMPLE-PINN framework is assessed by analyzing the pressure coefficient ($C_p$) distribution along the airfoil surface, as shown in Fig.~\ref{fig:airfoil_cp_all}. The $C_p$ is defined as
\begin{equation}
C_p = \frac{p - p_\infty}{\tfrac{1}{2}\rho U_\infty^2},
\end{equation}
where $p$ is the pressure, $p_\infty$ is the free-stream pressure (set to $p_\infty = 1$), $\rho$ is the density (set to $\rho = 1$), and $U_\infty$ is the free-stream velocity (set to $U_\infty = 1$).

At $Re = 500$ and $AOA = 0^\circ$ (Fig. \ref{fig:airfoil_Re500_cp}), the $C_p$ distribution on the airfoil surface predicted by SIMPLE-PINN shows excellent agreement with both the CFD results and the reference data from Imamura et al. \citep{imamura2004flow}. This high fidelity is maintained irrespective of whether AD is applied to near-wall points.  Moreover, SIMPLE-PINN is able to capture the inherent symmetry of the $C_p$ distribution on the airfoil’s upper and lower surfaces, confirming the model’s ability to preserve essential physical properties.

At $Re = 1000$ and $AOA = 7^\circ$ (Fig. \ref{fig:airfoil_cp}), the predictions of SIMPLE-PINN using AD remain in strong agreement with CFD and reference data from Kurtulus et al. \citep{kurtulus2015unsteady} across the entire airfoil surface. This includes accurately reproducing the leading-edge suction peak and the subsequent pressure recovery toward the trailing edge. In contrast, without the application of AD, SIMPLE-PINN shows noticeable errors near the leading edge. In this region, the sharp pressure gradient is not well resolved, leading to an overprediction of the $C_p$. This comparison highlights the critical role of employing AD for near-wall points in SIMPLE-PINN, which enables the accurate resolution of steep pressure gradients and thereby promotes high-accuracy reconstruction of surface pressure. By comparison, the IBM-PINN method \citep{xiao2025immersed} requires additional labeled data to achieve similar accuracy; even then, it tends to overestimate the minimum pressure values.

Table~\ref{tab:airfoil_errors} provides a comprehensive summary of the errors and computational costs of the SIMPLE-PINN with and without AD in the subdomain $\Omega_1$. From a quantitative perspective, the integration of AD significantly enhances the model's predictive accuracy, particularly for the more complex airfoil flow case at $Re=1000$. For this case, the incorporation of AD results in a remarkable reduction in the relative $L_2$ error for the velocity component  $u$, decreasing by more than an order of magnitude compared to the non-AD configuration. Similarly, the MSE for the pressure field is reduced by nearly an order of magnitude. These results underscore the significance of AD in enhancing the capability of SIMPLE-PINN for complex flow regimes. Although the improvements are less pronounced at $Re=500$, AD still provides an enhancement in both velocity and pressure predictions.
\begin{table}[!ht]
\small
  \centering
  \caption{Evaluation of error and computational cost for SIMPLE-PINN with and without AD in airfoil test cases.}
  \label{tab:airfoil_errors}
  \renewcommand{\arraystretch}{1.2}
  \footnotesize
  % \resizebox{\linewidth}{!}{
  \begin{tabular}{lcccc}
    \toprule
     Model & \multicolumn{2}{c}{SIMPLE-PINN} & \multicolumn{2}{c}{SIMPLE-PINN with AD}\\
    \midrule
    $Re$ &  500  & 1000 &  500  & 1000  \\
Rel.$L_2$ $u$ & $9.78\times10^{-3}$ & $1.53\times10^{-1}$ & $7.37\times10^{-3}$ & $1.52\times10^{-2}$ \\
Rel.$L_2$ $v$ & $3.29\times10^{-2}$ & $2.17\times10^{-1}$ & $2.51\times10^{-2}$ & $3.41\times10^{-2}$ \\
Rel.$L_2$ $V$ & $9.83\times10^{-3}$ & $1.51\times10^{-1}$ & $7.33\times10^{-3}$ & $1.51\times10^{-2}$ \\
Rel.$L_2$ $P$ & $4.78\times10^{-2}$ & $1.95\times10^{-1}$ & $4.61\times10^{-2}$ & $4.20\times10^{-2}$ \\
\midrule
MSE $u$ & $6.24\times10^{-5}$ & $1.59\times10^{-2}$ & $3.54\times10^{-5}$ & $1.57\times10^{-4}$ \\
MSE $v$ & $1.21\times10^{-5}$ & $6.13\times10^{-4}$ & $7.06\times10^{-6}$ & $1.51\times10^{-5}$ \\
MSE $V$ & $6.42\times10^{-5}$ & $1.58\times10^{-2}$ & $3.57\times10^{-5}$ & $1.59\times10^{-4}$ \\
MSE $P$ & $3.09\times10^{-5}$ & $7.46\times10^{-4}$ & $2.86\times10^{-5}$ & $3.46\times10^{-5}$ \\
   \midrule
    AOA & 0$^\circ$ & 7$^\circ$ & 0$^\circ$ & 7$^\circ$ \\   
    Time (s)  & 383      & 397      & 416      & 427 \\ 
    Sample points & \multicolumn{4}{c}{801$\times$401} \\
    \bottomrule
  \end{tabular}
  % }
\end{table}

In terms of computational efficiency, the incorporation of AD incurs only a modest overhead, increasing the training time by approximately 5\% relative to the non-AD SIMPLE-PINN implementation. Specifically, SIMPLE-PINN completes training in just a few hundred seconds for both the $Re=500$ and $Re=1000$ cases, achieving a computational speedup of more than an order of magnitude compared to IBM-PINN \citep{xiao2025immersed}, which requires substantially longer training durations, approximately 8500 s for $Re=500$ and 7800 s for $Re=1000$. These findings underscore the favorable balance between accuracy and computational efficiency offered by SIMPLE-PINN, establishing it as a highly promising approach for simulating challenging flow scenarios.

\subsection{Flow past three square cylinders} 
\begin{figure}[!ht]
    \centering
    % 左图
    \begin{subfigure}[b]{0.33\linewidth}
        \centering
        \includegraphics[width=\linewidth]{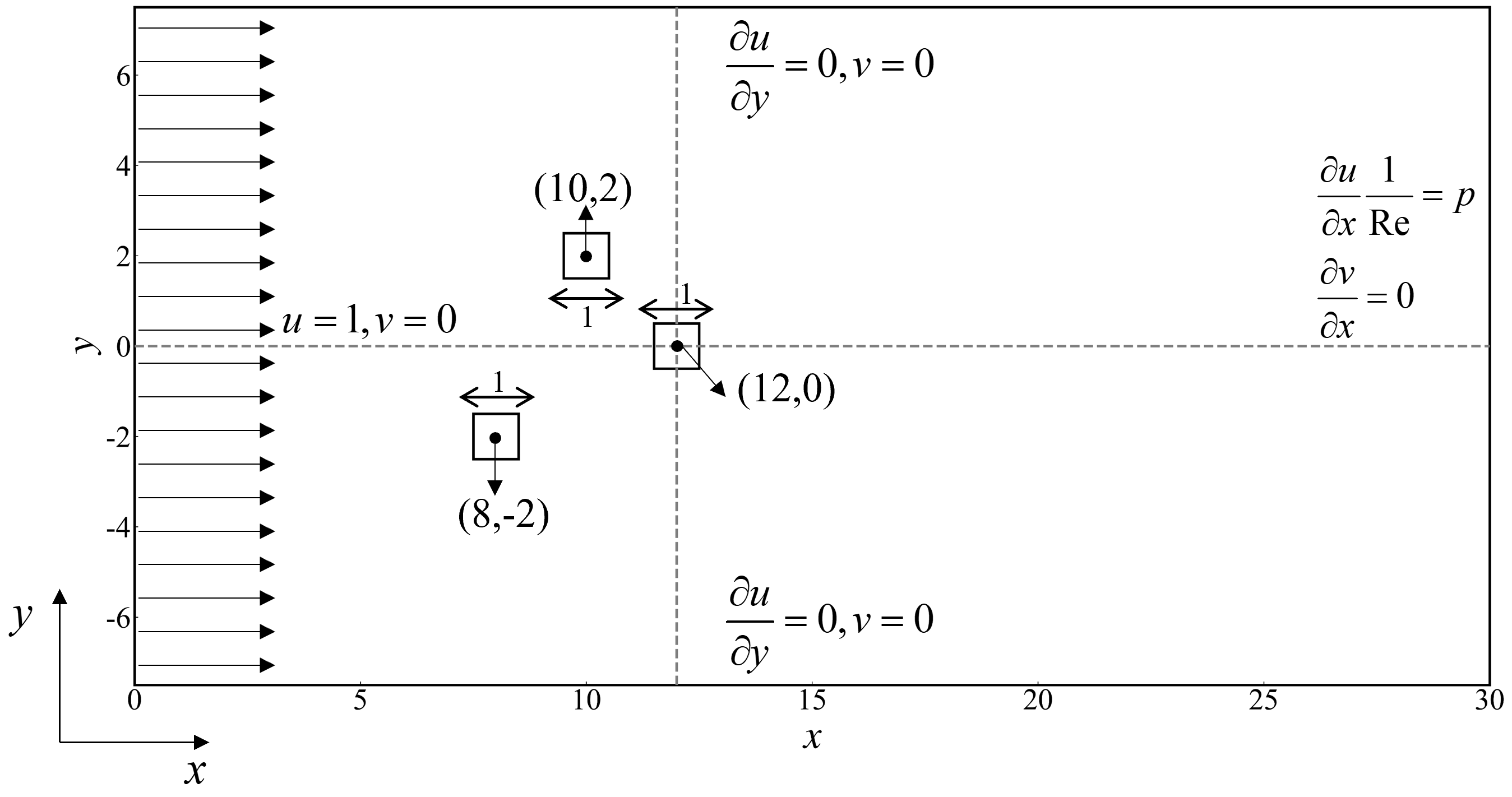}
        \caption{Open domain }
        \label{fig:3square_geo}
    \end{subfigure}
    \hfill
    % 右图
    \begin{subfigure}[b]{0.66\linewidth}
        \centering
        \includegraphics[width=\linewidth]{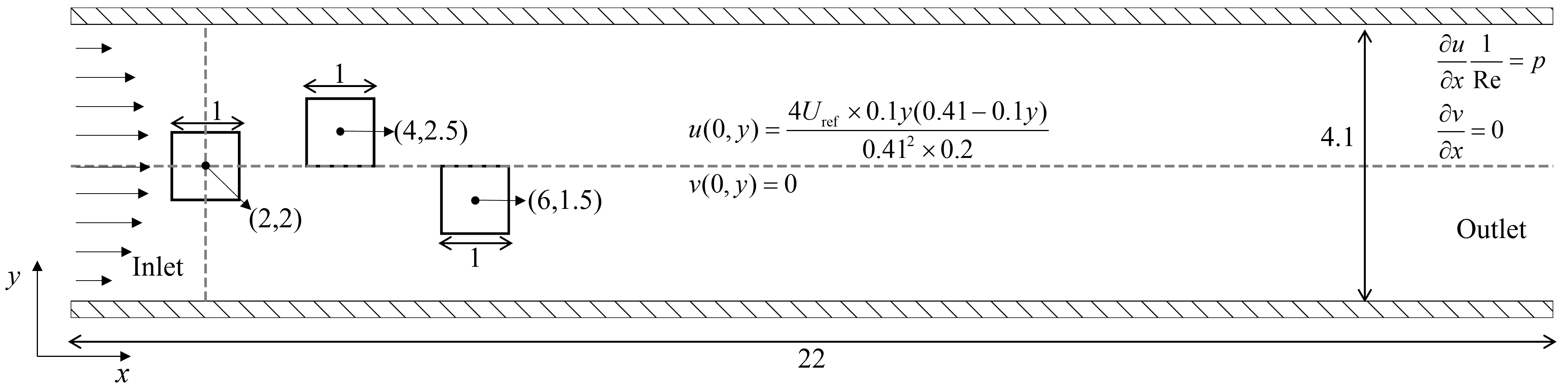}
        \caption{Channel flow}
        \label{fig:3square_Re40_geo}
    \end{subfigure}
    
    \caption{Computational domains and boundary conditions for flow past three square cylinders in two different configurations.}
    \label{fig:3square_all}
\end{figure}

This subsection examines the flow dynamics past three square cylinders obtained using the SIMPLE-PINN framework. Compared with flow past an airfoil, the presence of multiple cylinders gives rise to intricate vortex interactions and wake interference, markedly increasing the complexity of the flow field. Furthermore, the sharp corners of the square cylinders generate steep pressure and velocity gradients, presenting a formidable challenge for accurate flow resolution. The computational domain and corresponding boundary conditions are illustrated in Fig.~\ref{fig:3square_all}.
% \begin{figure}[!ht]
%     \centering
%     \includegraphics[width=0.85\linewidth]{pictures/3square in open domain.png}
%     \caption{ Computational domain and boundary conditions for the flow past three square cylinders. Three square cylinders with side length 1 are located at $(10,2)$, $(12,0)$, and $(8,-2)$. The top and bottom boundaries are free-slip ($v=0$, $\partial u / \partial y = 0$), the right boundary is an outflow ($\frac{\partial u}{\partial x}    \frac{1}{Re} =  p$, $\partial v / \partial x = 0$), and the left boundary serves as a uniform horizontal inflow ($u=1$, $v=0$).}
%     \label{fig:3square_geo}
% \end{figure}
% \begin{figure}[!ht]
%     \centering
%     \includegraphics[width=1.0\linewidth]{pictures/3square in channel.png}
%     \caption{Computational domain for the flow around three square cylinders in a channel. The flow enters from the left and exits on the right. Each square has a width of 1 and is positioned at $(2,2)$, $(4,2.5)$, and $(6,1.5)$, respectively, while the channel height is 4.1. The reference velocity is $U_\mathrm{ref} = 0.3$.}
%     \label{fig:3square_Re40_geo}
% \end{figure}

\begin{figure}[!ht]
    \centering
    % 上图
    \begin{subfigure}[b]{1.0\linewidth}
        \centering
        \includegraphics[width=\linewidth]{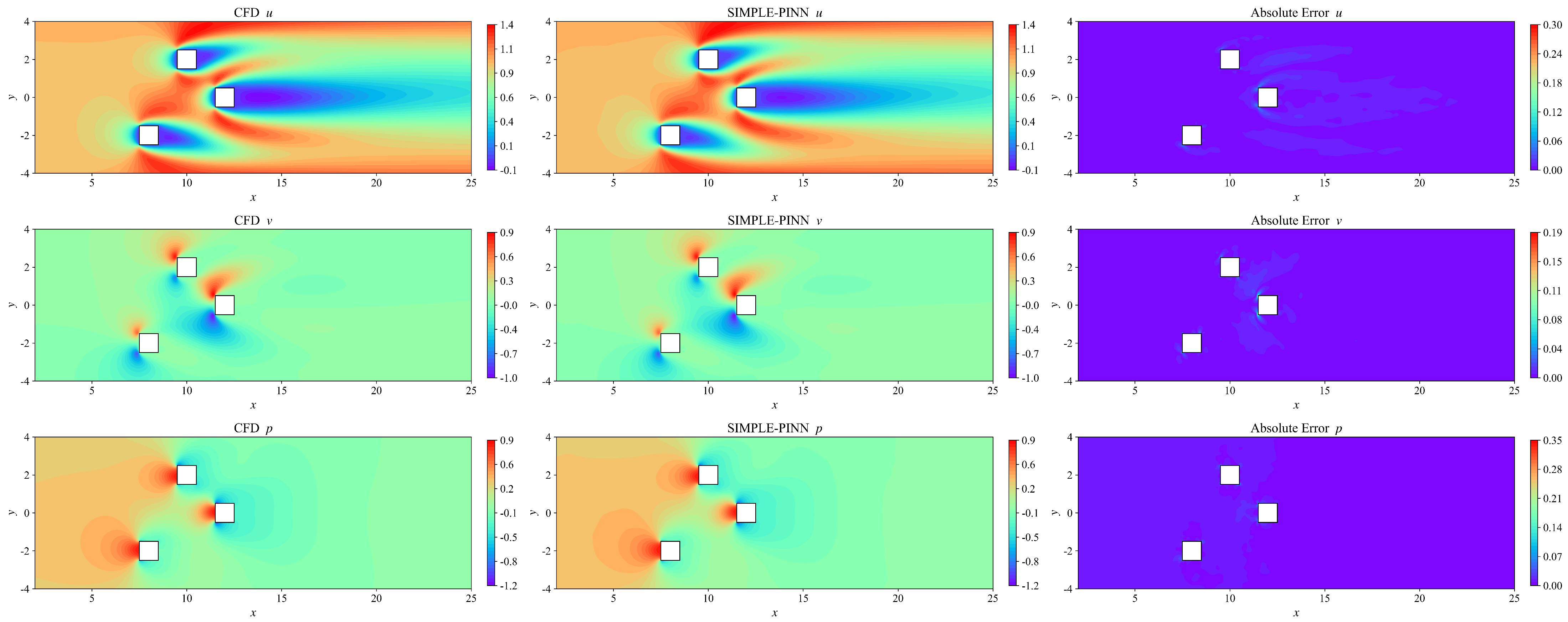}
        \caption{Flow past three square cylinders in an open domain at $Re=25$.}
        \label{fig:3square_uvp}
    \end{subfigure}
    
    \vspace{0.5em} % 上下间距
    
    % 下图
    \begin{subfigure}[b]{1.0\linewidth}
        \centering
        \includegraphics[width=\linewidth]{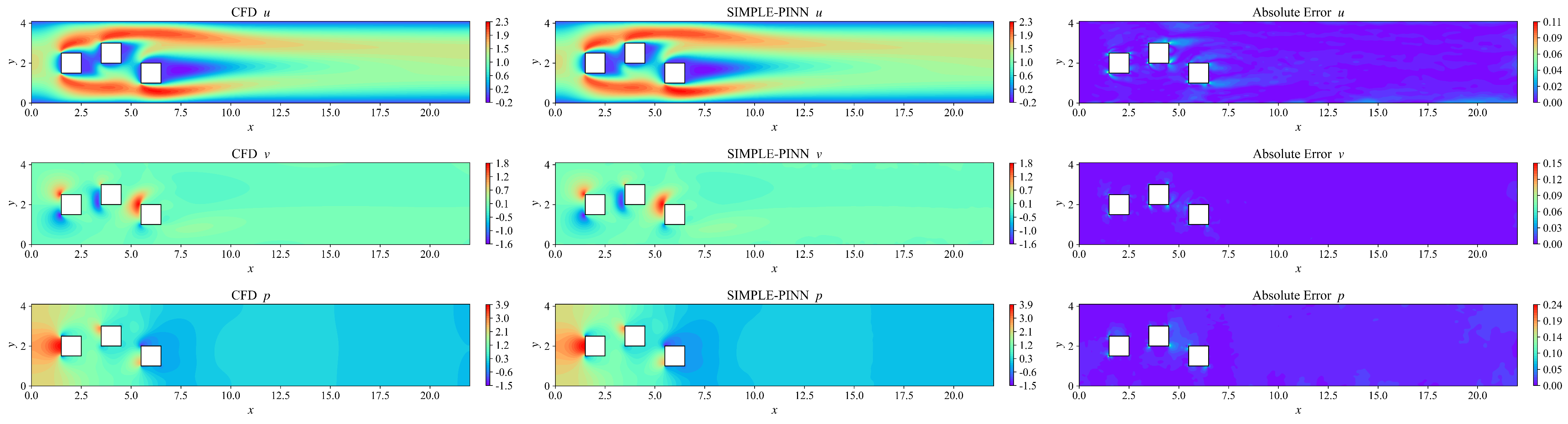}
        \caption{Flow past three square cylinders in a channel at $Re=40$.}
        \label{fig:3square_Re40_uvp}
    \end{subfigure}
    
    \caption{Comparison of CFD and SIMPLE-PINN predictions for velocity components and pressure fields in two configurations.}
    \label{fig:3square_uvp_all}
\end{figure}

As shown in Fig.~\ref{fig:3square_uvp_all}, predictions from the SIMPLE-PINN method exhibit strong agreement with CFD results \citep{chiu2023cdfib} in both the open domain ($Re=25$) and channel ($Re=40$) configurations. The corresponding absolute error, predominantly represented by dark blue regions, indicates minimal deviation from the CFD reference, highlighting the high predictive accuracy of SIMPLE-PINN. Notable discrepancies are observed primarily in the vicinity of the sharp corners of the square cylinders, where steep gradients pose significant challenges. 
% \begin{figure}[!ht]
%     \centering
%     \includegraphics[width=1.0\linewidth]{pictures/3square Re25 uvp contour.png}
%     \caption{Comparison of velocity and pressure contours for flow past three square cylinders at $Re=25$. The top, middle, and bottom rows show the streamwise velocity ($u$), vertical velocity ($v$), and pressure ($p$), respectively. The columns represent CFD, SIMPLE-PINN predictions, and the absolute error. }
%     \label{fig:3square_uvp}
% \end{figure}
% \begin{figure}[!ht]
%     \centering
%     \includegraphics[width=1.0\linewidth]{pictures/3square Re40 uvp contour.png}
%     \caption{ Comparison of velocity and pressure contours for the flow past three square cylinders in a channel at $Re=40$.}
%     \label{fig:3square_Re40_uvp}
% \end{figure}

\begin{figure}[!ht]
    \centering
    % 左图
    \begin{subfigure}[b]{0.42\linewidth}
        \centering
        \includegraphics[width=\linewidth]{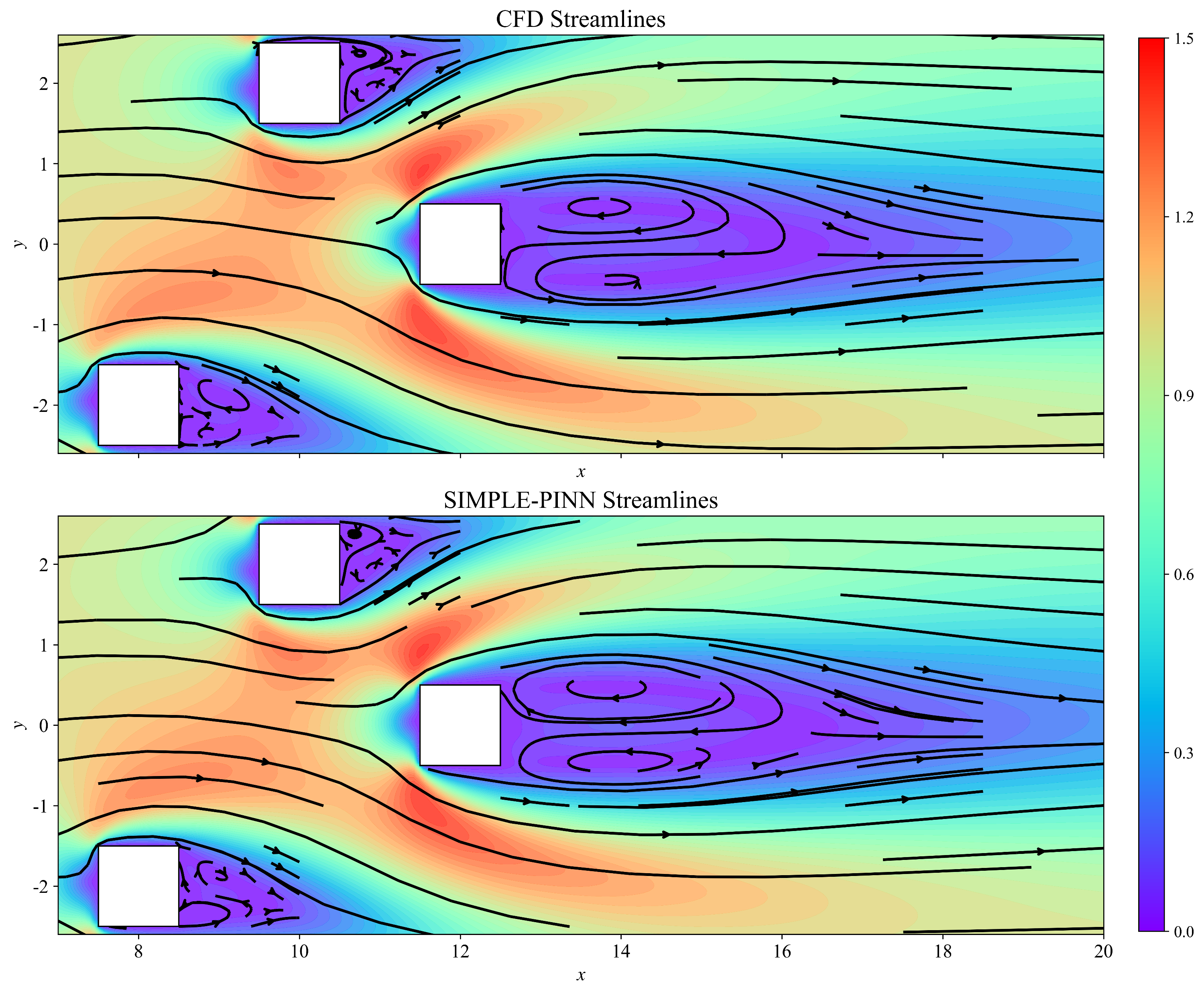}
        \caption{Streamlines in an open domain.}
        \label{fig:3square_stlines}
    \end{subfigure}
    \hfill
    % 右图
    \begin{subfigure}[b]{0.57\linewidth}
        \centering
        \includegraphics[width=\linewidth]{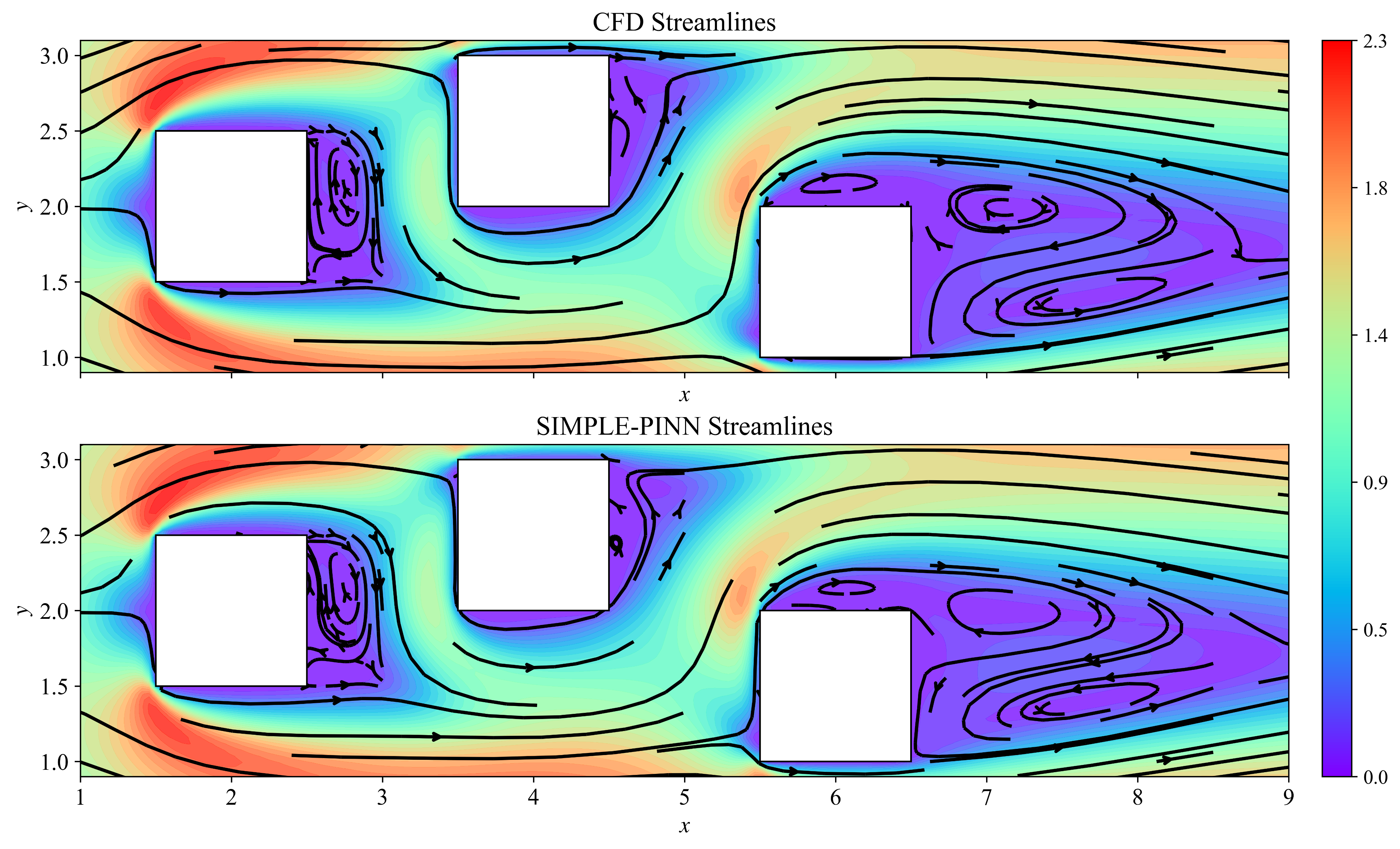}
        \caption{Streamlines in a channel.}
        \label{fig:3square_Re40_stlines}
    \end{subfigure}
    
    \caption{Comparison of streamlines and velocity magnitude ($V$) for the flow past three square cylinders at different configurations. Top: CFD results; Bottom: SIMPLE-PINN predictions.}
    \label{fig:3square_streamlines}
\end{figure}

Fig.~\ref{fig:3square_streamlines} presents the streamlines around the three square cylinders, with the velocity magnitude shown as the background color. The SIMPLE-PINN predictions exhibit strong qualitative agreement with the CFD results, successfully capturing detailed flow features. For instance, in the open-domain case, the model accurately reproduces the two large vortices that form downstream of the third cylinder, whereas in the channel case, it correctly predicts the emergence of three distinct vortices around the third cylinder. These results demonstrate SIMPLE-PINN’s predictive capability and its effectiveness in modeling underlying physics, including complex vortex formation, across different configurations, even for multiple obstacles with sharp corners.
% This demonstrates its efficacy for multi-obstacle flow problems and its potential for use in practical fluid dynamics applications.
% \begin{figure}[!ht]
%     \centering
%     \includegraphics[width=0.75\linewidth]{pictures/3square Re25 streamlines.png}
%     \caption{Streamlines and velocity magnitude ($V$) for the flow past three square cylinders at $Re=25$. Top: CFD results; Bottom: SIMPLE-PINN predictions.}
%     \label{fig:3square_stlines}
% \end{figure}
% \begin{figure}[!ht]
%     \centering
%     \includegraphics[width=0.7\linewidth]{pictures/3square Re40 streamlines.png}
%     \caption{Streamlines and velocity magnitude ($V$) for the flow past three square cylinders at $Re=40$. Top: CFD results; Bottom: SIMPLE-PINN predictions.}
%     \label{fig:3square_Re40_stlines}
% \end{figure}

\begin{figure}[!ht]
    \centering
    % 上图
    \begin{subfigure}[b]{1.0\linewidth}
        \centering
        \includegraphics[width=\linewidth]{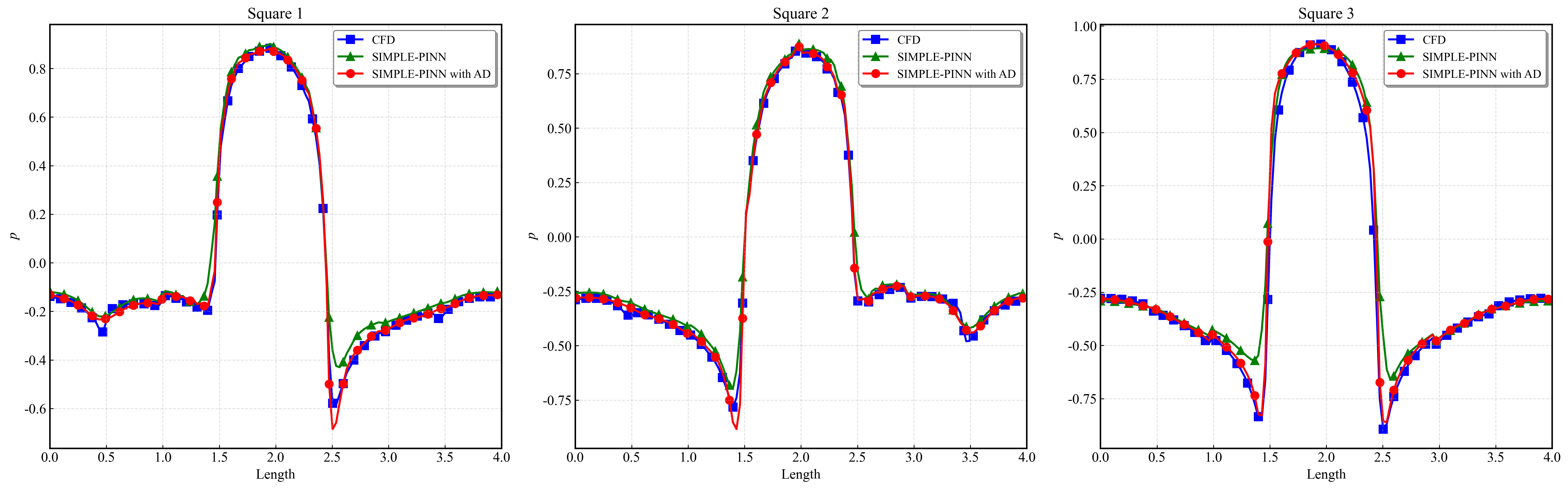}
        \caption{Pressure distributions along the surfaces of the three square cylinders in an open domain at $Re=25$.}
        \label{fig:3square_cp}
    \end{subfigure}
    
    % \vspace{0.5em} % 上下间距
    % 下图
    \begin{subfigure}[b]{1.0\linewidth}
        \centering
        \includegraphics[width=\linewidth]{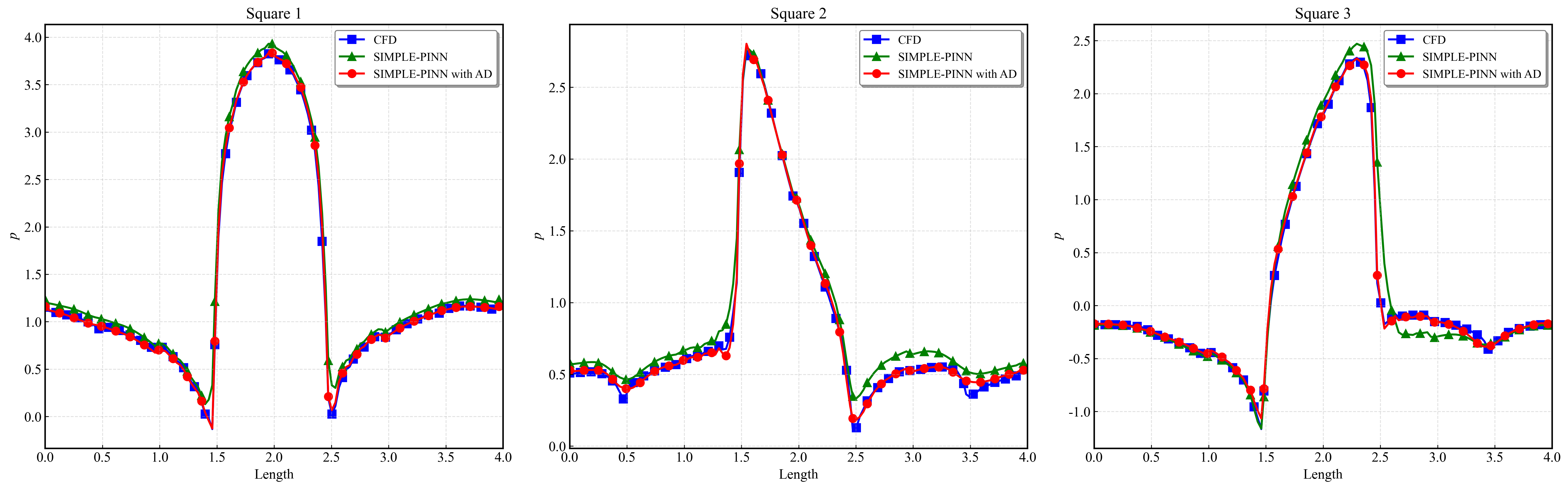}
        \caption{Pressure distributions along the surfaces of the three square cylinders in a channel at $Re=40$.}
        \label{fig:3square_Re40_cp}
    \end{subfigure}
    
    \caption{Comparison of pressure distributions on the surfaces of three square cylinders in two flow configurations.}
    \label{fig:3square_cp_all}
\end{figure}

Fig.~\ref{fig:3square_cp_all} presents the pressure distributions along the surfaces of the three square cylinders, comparing SIMPLE-PINN predictions with the corresponding CFD results. The SIMPLE-PINN with AD shows close agreement with CFD, effectively capturing the pressure peaks and troughs induced by flow interactions with the obstacles. In contrast, SIMPLE-PINN without AD exhibits deviations, particularly near the pressure minima at the sharp corners of the obstacles. These discrepancies indicate a reduced capability to capture pressure variations in regions with sharp geometric transitions. This comparison further underscores the pivotal role of AD, as the SIMPLE-PINN employing AD consistently demonstrates high accuracy in resolving detailed surface pressure distributions.

As shown in Table~\ref{tab:performance_SIMPLE-PINN}, SIMPLE-PINN demonstrates strong performance in simulating flow past three square cylinders, achieving both high accuracy and computational efficiency in the open domain and channel configurations. Its precision is confirmed by the consistently low relative $L_2$ error and MSE, which indicate reliable resolution of the complex flow fields generated by multiple obstacles. In terms of efficiency, the solution is obtained in just 335~s for the open-domain case and 451~s for the channel case. This speed positions SIMPLE-PINN as a promising tool for rapid fluid dynamics simulations. 
    
\subsection{Unsteady flow past a cylinder} 
\begin{figure}[!ht]
    \centering
    \includegraphics[width=1.0\linewidth]{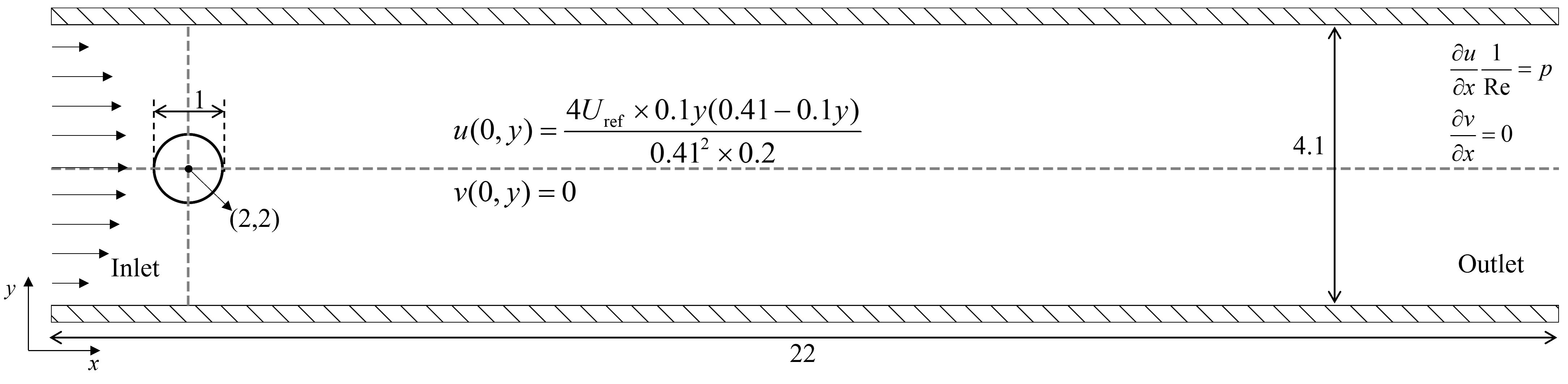}
    \caption{Computational domain for the unsteady flow past a cylinder. The cylinder is centered at $(2,2)$ with a diameter of 1, and the channel height is 4.1. A parabolic velocity profile with $U_\mathrm{ref}=0.3$ is applied at the inlet; the top and bottom boundaries are no-slip ($u=v=0$); the outlet enforces outflow conditions ($\frac{\partial u}{\partial x}\frac{1}{Re}=p$, $\frac{\partial v}{\partial x}=0$).}
    \label{fig:cylinder_geo}
\end{figure}

To further evaluate the capability of SIMPLE-PINN in predicting the dynamic evolution of flow fields, we consider the unsteady flow past a cylinder at $Re=100$. The computational domain and boundary conditions are shown in Fig.~\ref{fig:cylinder_geo}. The CFD solution \citep{chiu2023cdfib} at $t=90$ serves as the initial condition for SIMPLE-PINN, which then predicts the evolution of the velocity and pressure fields over the subsequent 10 s. Notably, unlike JAX-PI \citep{wang2023expert}, which requires dividing the temporal domain into 10 separate time windows, SIMPLE-PINN does not employ a time-marching strategy. Instead, it directly predicts the solution over the entire time interval. 

\begin{figure}[!ht]
    \centering
    \includegraphics[width=1.0\linewidth]{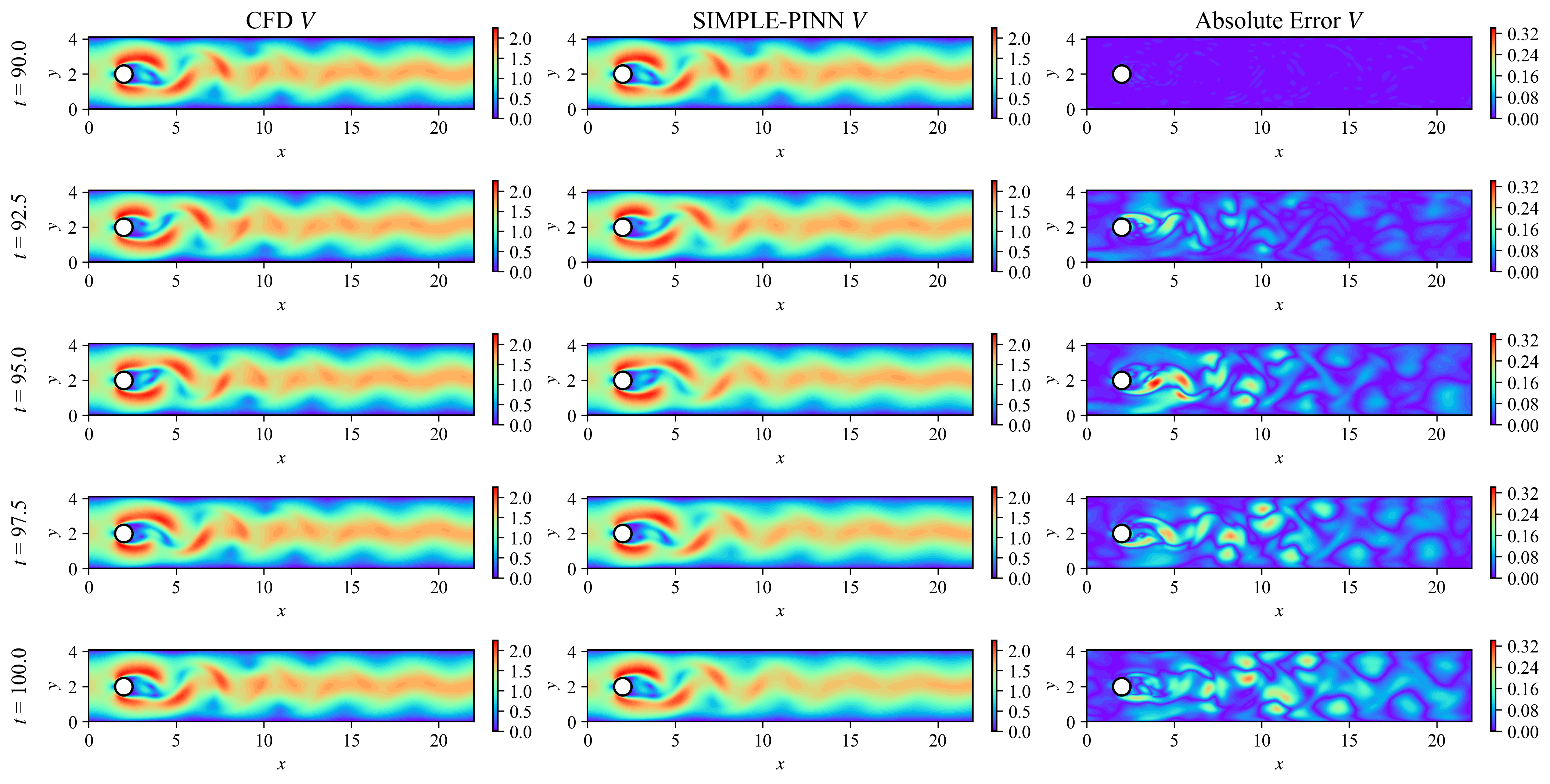}
    \caption{ Instantaneous velocity magnitude ($V$) contours for flow past a cylinder at $Re=100$ from CFD and SIMPLE-PINN, with absolute errors, shown at $t=90$, $92.5$, $95$, and $100$.}
    \label{fig:cylinder_Re100_Vmag}
\end{figure}

Fig.\ref{fig:cylinder_Re100_Vmag} compares the velocity magnitude at different time slices from SIMPLE-PINN and CFD over the interval $t = 90$ to $100$. The corresponding velocity components and pressure fields are presented in Fig.\ref{fig:cylinder_Re100_all} in \ref{sec:appendix_cylinders_100}. The velocity magnitude predicted by SIMPLE-PINN is in close agreement with the CFD results, successfully reproducing the unsteady Kármán vortex street. The absolute error distribution indicates that, although localized discrepancies exist, the overall error remains low, thereby demonstrating the high accuracy of SIMPLE-PINN in capturing the long-term evolution of complex flow phenomena.

\begin{figure}[!ht]
    \centering
    \includegraphics[width=1.0\linewidth]{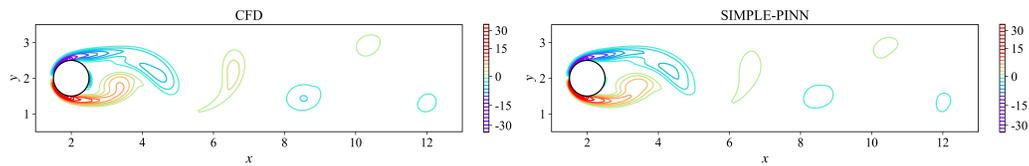}
    \caption{Comparison of vorticity for the flow past a cylinder at $Re=100$ at $t = 91$ between CFD and SIMPLE-PINN.}
    \label{fig:cylinder_Re100_vor}
\end{figure}

Fig.\ref{fig:cylinder_Re100_vor} presents a comparison of vorticity ($\omega$) at $t= 91$, which serves as key indicators of fluid motion and vividly illustrates the vortex structures. The SIMPLE-PINN predictions show good agreement with the CFD reference, accurately reproducing the alternating vortices shed from both sides of the cylinder and effectively capturing the detailed flow features. In particular, both the contours and internal structures of the primary vortices, as well as the weaker downstream secondary vortices, closely match the CFD results in terms of shape and position. 

Table \ref{tab:performance_SIMPLE-PINN} presents the high quantitative accuracy of SIMPLE-PINN for unsteady flow past a cylinder at $Re=100$. The model consistently achieves a low MSE, which remains on the order of $10^{-3}$ for all variables. This performance, coupled with a training time of 7269 s and the use of $401 \times 441 \times 83$ sample points, demonstrates the model's significant computational efficiency in predicting unsteady flow fields.

We further test SIMPLE-PINN by simulating the flow past a cylinder over the temporal domain from $t=0$ to $100$, with initial conditions set as $u=1$ and $v=0$. In this test case, the SIMPLE-PINN employs a time-marching strategy, a common approach for solving unsteady problems. The entire temporal domain was partitioned into ten consecutive segments, each spanning 10 non-dimensional time units. The prediction from one segment was then used as the initial condition for the subsequent segment.

\begin{figure}[!ht]
    \centering
    \includegraphics[width=1.0\linewidth]{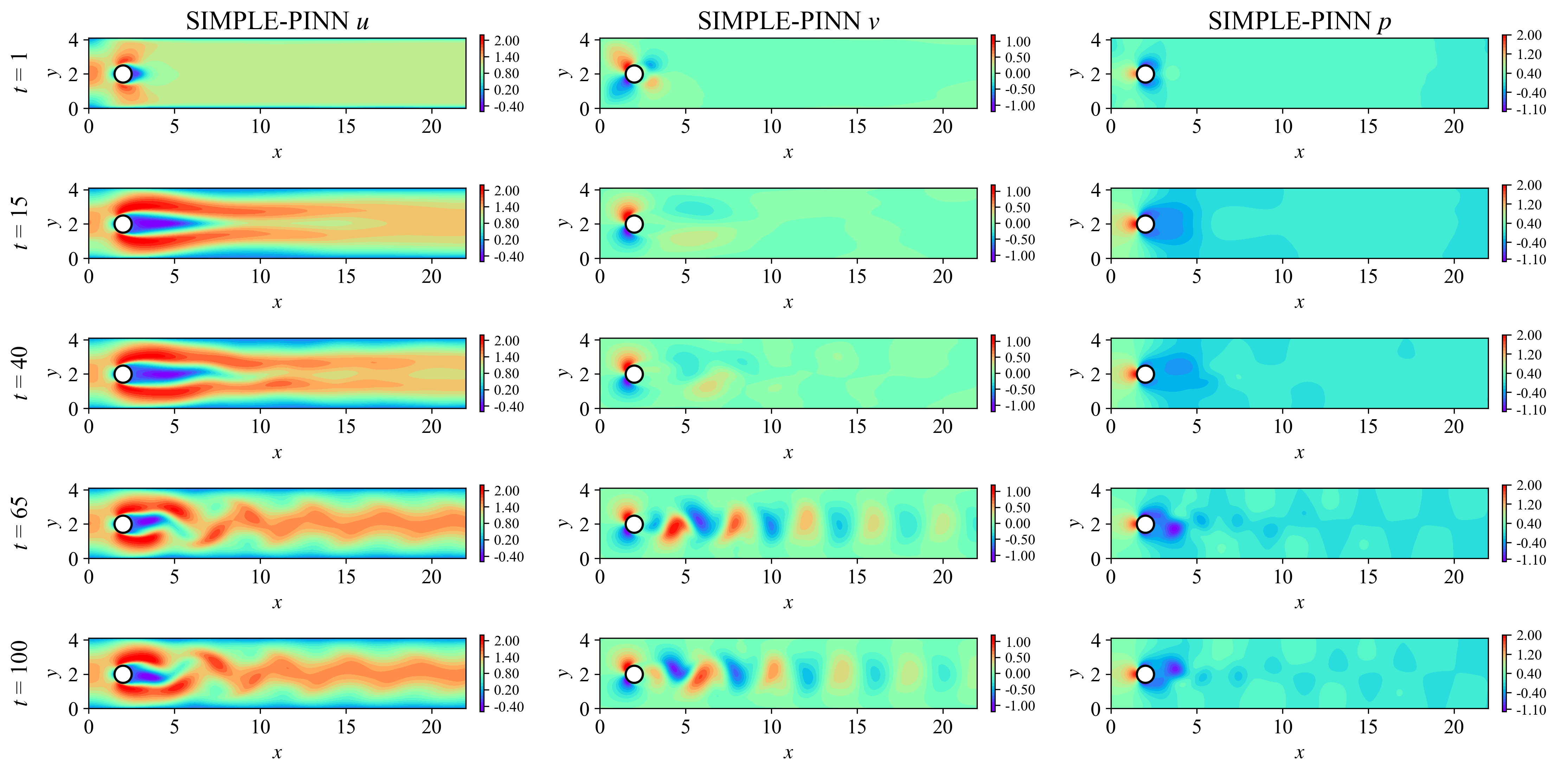}
    \caption{Temporal evolution of the flow variables ($u$, $v$, $p$) in the cylinder wake at $Re=100$, predicted by SIMPLE-PINN at $t = 1$, $15$, $40$, $65$, and $100$.}
    \label{fig:cylinder_Re100_uvp}
\end{figure}

Fig.~\ref{fig:cylinder_Re100_uvp} illustrates the temporal evolution of the flow field past a cylinder as predicted by SIMPLE-PINN. At the initial stages, the flow exhibits near-symmetry and stability, consistent with a steady state. For reference, the model’s performance in steady cylinder flows at $Re=20$ and $Re=40$ is provided in \ref{sec:appendix_cylinders_low_Re}.  As time advances, instabilities emerge within the wake region, ultimately leading to the formation of the von Kármán vortex street, characterized by alternating vortices shed from the upper and lower sides of the cylinder. The velocity and pressure distributions effectively capture these unsteady flow features, highlighting SIMPLE-PINN's proficiency in resolving complex, time-dependent flow structures. To the best of our knowledge, this study represents the first successful application of a PINN-based approach to capture the dynamic evolution of vortex shedding in flow past a cylinder, starting from uniform initial conditions ($u=1$, $v=0$) and simulating over a 100-second time interval.

\begin{figure}[!ht]
    \centering
    \includegraphics[width=1.0\linewidth]{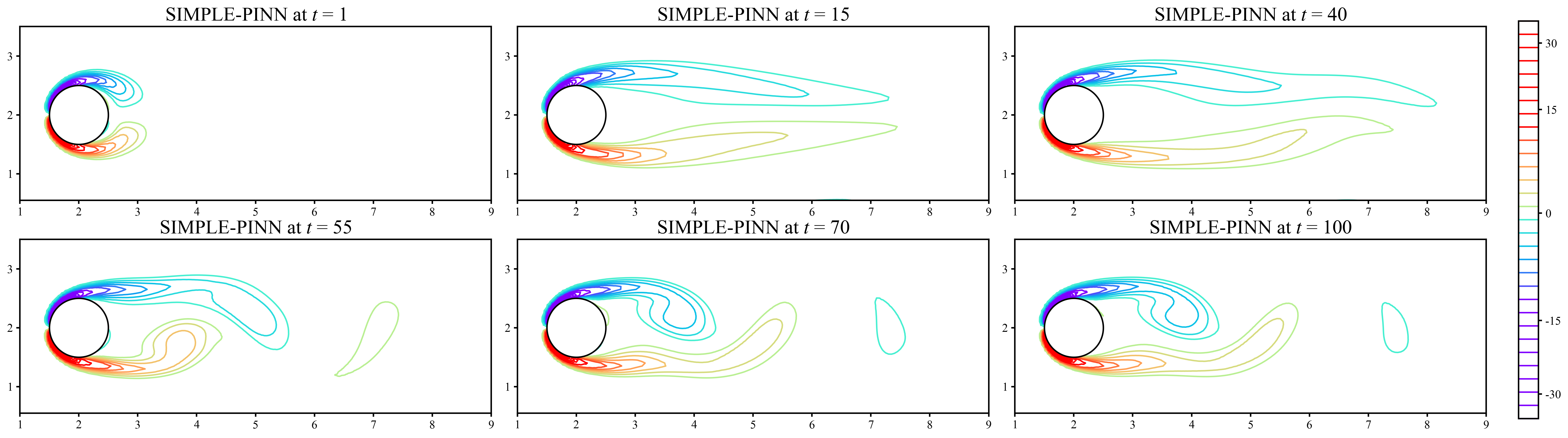}
    \caption{Temporal evolution of vorticity ($\omega$) in the cylinder wake at $Re=100$, predicted by SIMPLE-PINN at $t = 1$, $15$, $40$, $55$, $70$, and $100$.}
    \label{fig:cylinder_Re100_vor_0-100}
\end{figure}

Fig.~\ref{fig:cylinder_Re100_vor_0-100} illustrates the temporal evolution of the vorticity as simulated by SIMPLE-PINN. The model accurately reproduces the transition from an initially symmetric vortex structure to the onset of vortex instabilities, culminating in the shedding of alternating vortices from the cylinder’s top and bottom surfaces. These results demonstrate that SIMPLE-PINN not only captures the vortex generation process but also simulates its spatial propagation and evolution, highlighting the model’s capability to resolve the dynamic development and detachment of vortices over time.

\subsection{Rayleigh-Taylor instability}
\begin{figure}[!ht]
    \centering
    \includegraphics[width=0.35\linewidth]{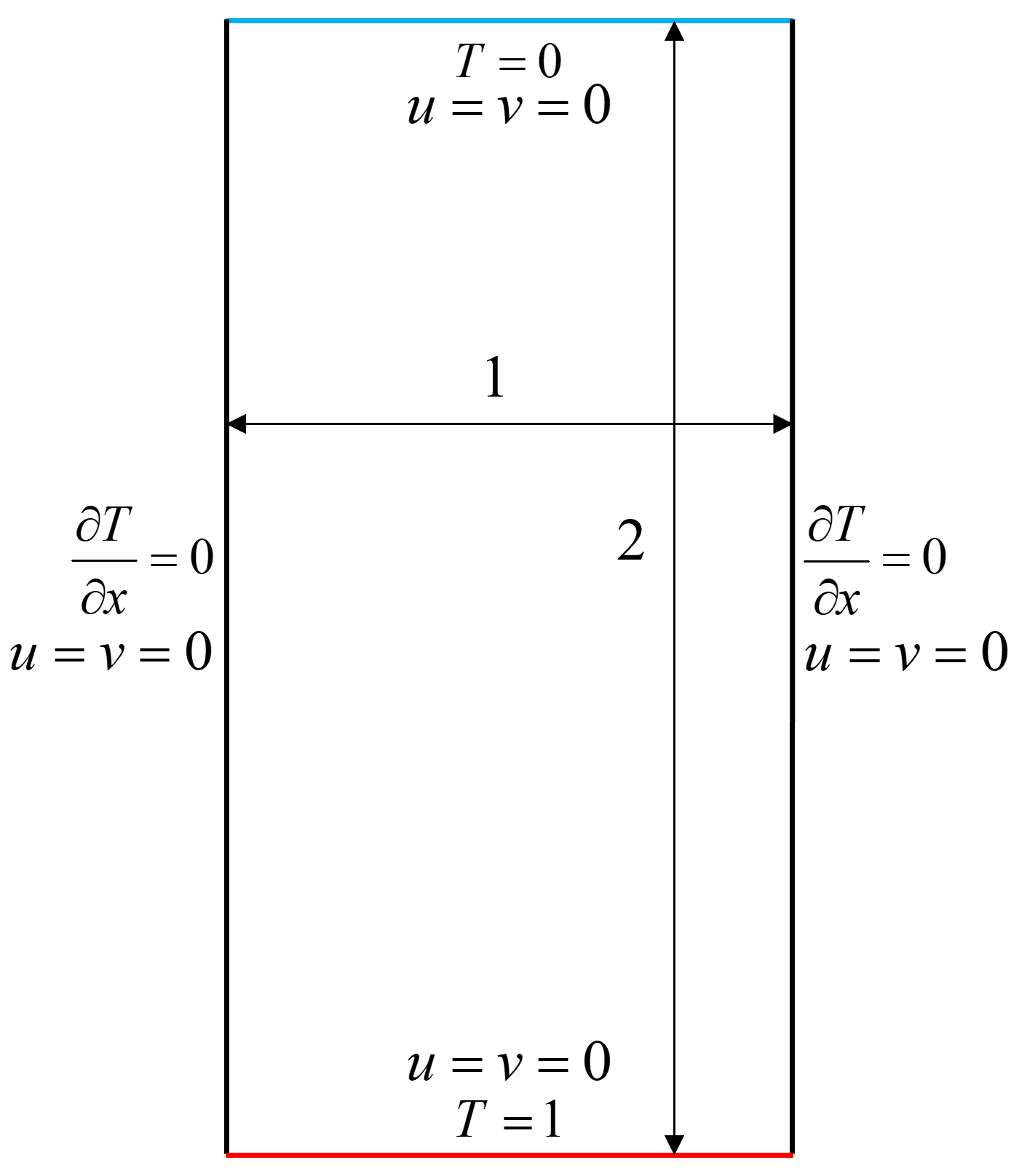}
\caption{The computational domain for the Rayleigh-Taylor instability problem is a vertical rectangle with height $2$ and width $1$. The top and bottom boundaries are maintained at $T=0$ and $T=1$, respectively, while the side walls are thermally insulated ($\partial T / \partial x = 0$). No-slip velocity conditions ($u=v=0$) are applied on all four boundaries.}
    \label{fig:RT_geo}
\end{figure}

In this section, we investigate a flow instability driven by a temperature gradient in a two-dimensional rectangular domain, as illustrated in Fig.~\ref{fig:RT_geo}. The system exhibits dynamics analogous to the Rayleigh-Taylor (RT) instability. However, unlike the classical RT problem, which involves two immiscible fluids, here a single incompressible fluid is considered, with the temperature field serving as a proxy for density under the Boussinesq approximation. This configuration represents a canonical multiphysics system, governed by the incompressible N-S equations coupled with the energy equation: 
\begin{subequations} \label{eq:2dRT}
\small
\begin{align}
    &\frac{\partial u}{\partial x} + \frac{\partial v}{\partial y} = 0, \label{eq:2dRT-div} \\ 
    &\frac{\partial u}{\partial t}+u \frac{\partial u}{\partial x} + v \frac{\partial u}{\partial y} 
    = \sqrt{\frac{Pr}{Ra}} \left( \frac{\partial^2 u}{\partial x^2} + \frac{\partial^2 u}{\partial y^2} \right) - \frac{\partial p}{\partial x}, \label{eq:2dRT-m1} \\ 
    &\frac{\partial v}{\partial t} + u \frac{\partial v}{\partial x} + v \frac{\partial v}{\partial y} 
    = \sqrt{\frac{Pr}{Ra}} \left( \frac{\partial^2 v}{\partial x^2} + \frac{\partial^2 v}{\partial y^2} \right) - \frac{\partial p}{\partial y} + T, \label{eq:2dRT-m2} \\ 
    &\frac{\partial T}{\partial t} + u \frac{\partial T}{\partial x} + v \frac{\partial T}{\partial y} 
    = \frac{1}{\sqrt{Pr\,Ra}} \left( \frac{\partial^2 T}{\partial x^2} + \frac{\partial^2 T}{\partial y^2} \right). \label{eq:2dRT-T}
\end{align}
\end{subequations}
\normalsize
where $Ra$ denotes the Rayleigh number, characterizing the relative strength of buoyancy against viscous and thermal diffusion, and $Pr$ represents the Prandtl number, which measures the ratio of momentum diffusivity to thermal diffusivity. In this study, the parameters are fixed at $Pr = 0.71$ and $Ra = 10^6$. Additionally, for the energy equation, we introduce a residual-based correction loss term \citep{wei2025ffv} for the temperature field.

\begin{figure}[!ht]
    \centering
    \includegraphics[width=1.0\linewidth]{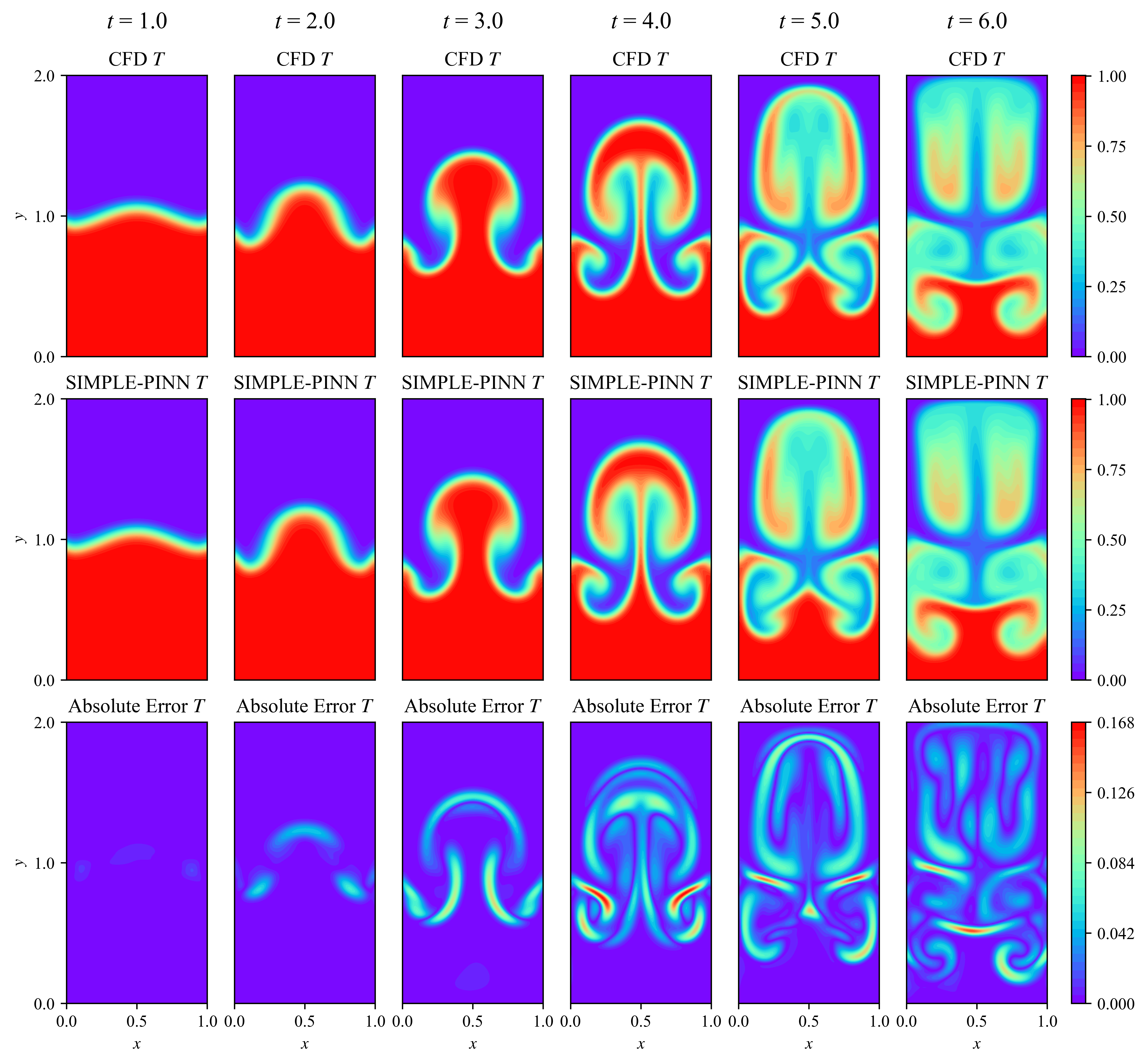}
  \caption{Comparison of the temporal evolution of the temperature field $T$ between CFD reference solutions and SIMPLE-PINN predictions, together with their absolute error. The first row reports the CFD results, the second row the SIMPLE-PINN predictions, and the third row the corresponding absolute error field.}
    \label{fig:RT_T}
\end{figure}

Fig.~\ref{fig:RT_T} compares the temporal evolution of the temperature field in the RT instability, as predicted by SIMPLE-PINN, with reference CFD results \citep{chiu2018improved} at several representative time instances. The corresponding velocity and pressure distribution are presented in Fig.~\ref{fig:RT_all} of ~\ref{sec:appendix_RT}. The SIMPLE-PINN predictions exhibit a high level of consistency with the reference CFD results. At $t=1.0$, the model accurately reproduces the sharp temperature gradient interface between the hot and cold fluids. As time progresses, the hot fluid undergoes upward convection and evolves into the characteristic “mushroom-shaped” vortical structures of the RT instability. SIMPLE-PINN successfully captures this evolution, including the formation, growth, and detachment of vortices, particularly during the highly nonlinear interactions occurring between $t=3.0$ and $6.0$.

The absolute error in Fig.\ref{fig:RT_T} further provides a quantitative measure of the model’s predictive performance, demonstrating its ability to resolve temperature distributions with high accuracy. In the early stages ($t=1.0$ and $2.0$), the temperature field is relatively simple, and the absolute error is nearly zero.  During $t=3.0$ and $4.0$, the absolute error increases but remains largely confined to regions with steep gradients, such as vortex edges. In the later stages ($t=5.0$ and $6.0$), the interactions between hot and cold fluids intensify and the flow evolves into intricate vortical structures. Although the extent of the error expands, its maximum value remains approximately 0.168, which is well within an acceptable range.

As summarized in Table~\ref{tab:performance_SIMPLE-PINN}, the discrepancies between SIMPLE-PINN and CFD results are minimal. The relative $L_2$ errors remain at the order of $10^{-2}$, and the MSE is consistently at the order of $10^{-4}$, demonstrating the reliability and accuracy of the proposed framework.

To assess the computational efficiency of SIMPLE-PINN, we compare it with a state-of-the-art SOAP-based PINN \citep{wang2025gradient}. The SOAP-based method solves the RT problem over the time interval $t=[0,8]$ by partitioning it into four temporal windows, requiring a total training time of 21.73~h. In contrast, SIMPLE-PINN computes the solution directly over the continuous time domain $t=[0,6]$ without temporal decomposition, completing the training in only 6419~s (1.78~h). A difference between the two lies in their approach to time. The SOAP-based PINN, like many traditional PINN methods, relies on temporal decomposition, solving the problem sequentially across discrete time windows. Conversely, SIMPLE-PINN allows for the direct learning of the solution over a continuous time interval. Hence, SIMPLE-PINN streamlines the training process and prevents the accumulation of errors at window boundaries, a common issue in decomposition-based methods. When normalized by the solved time span (training time per unit simulation time), SIMPLE-PINN achieves approximately 9$\times$ improvement in computational efficiency compared to the SOAP-based approach.

\section{Discussion and Conclusion}\label{sec: conclusion}
This paper presents a novel PINN framework for solving the N-S equations. The framework incorporates a velocity and pressure correction loss, a new component inspired by the SIMPLE algorithm, to significantly improve the network's convergence and accuracy. While SIMPLE-PINN can be categorized as a hybrid framework that combines PINNs with numerical methods, its contribution extends far beyond simply replacing AD with numerical discretization. The key innovation of SIMPLE-PINN lies in transforming classical numerical solution procedures into neural network-compatible optimization strategies. Within SIMPLE-PINN, traditional operations, such as velocity-pressure coupling, are reformulated as specialized loss functions, creating a hybrid solver that combines the flexibility of PINNs with the robustness of classical numerical algorithms. Furthermore, by applying AD at near-wall points in complex geometries, the framework enhances its capability to accurately resolve flow behavior in the vicinity of irregular boundaries.

To assess the performance of the proposed SIMPLE-PINN framework, we conduct a series of numerical experiments representing diverse flow scenarios. These cases are particularly challenging for most existing PINN models. The benchmark problems include strong nonlinear flows (lid-driven cavity flow at high \textit{Re} numbers), irregular channel flows (wavy channel flow), open domain flows (flow past a NACA0012 airfoil), flows with multiple obstacles (flow past three square cylinders), unsteady flows (flow past a cylinder), and multiphysics problem (Rayleigh-Taylor instability). SIMPLE-PINN achieves two significant breakthroughs in applying PINN-based solvers to fluid mechanics. First, for the first time, it successfully solves the lid-driven cavity flow at a high \textit{Re} number ($Re = 20000$) using a PINN approach, requiring only 448~s of training. This achievement demonstrates an improvement over PINNs in handling high-$Re$ flows. Second, for the first time, SIMPLE-PINN predicts the complete flow evolution past a cylinder from initial conditions ($\mathbf{u} = (u, v) = (1, 0)$) over the time interval $t=[0,100]$, accurately capturing the unsteady Kármán vortex street. This demonstrates its remarkable capability in long-term prediction of unsteady flows. In addition, for the RT instability, SIMPLE-PINN is able to capture the dynamic evolution of the temperature field without resorting to temporal decomposition. These results demonstrate that SIMPLE-PINN exhibits excellent stability, rapid convergence compared to other PINN models, and high accuracy across a wide range of challenging flow scenarios. Furthermore, SIMPLE-PINN does not currently incorporate advanced training strategies such as adaptive loss balancing, curriculum training, causal training, and other related techniques. These strategies are compatible with SIMPLE-PINN and could further enhance its performance.

Notably, although the proposed velocity and pressure correction loss was originally developed in the context of the N-S equations, its underlying principles suggest potential applications to a broader class of problems. The N-S system is merely one instance of constrained physical systems that arise pervasively across science and engineering. Such systems are often governed by dynamic, transport, or force-balance equations that are coupled with divergence-free or conservation-based constraints. For example, in solid mechanics, incompressible elastic materials are described by momentum balance equations subject to a volume-conservation condition, directly analogous to the divergence-free velocity condition. Likewise, in electromagnetism, Maxwell’s equations impose the divergence-free property of the magnetic flux density \citep{sheu2010development}, again coupling field dynamics with strict conservation laws. Building on this shared mathematical structure, SIMPLE-PINN provides a systematic framework for enforcing constraints in neural network-based solvers by embedding the correction mechanism directly into the loss function. This universality underscores the broader potential of the SIMPLE-PINN approach, with extensions to diverse scientific and engineering domains representing a promising avenue for future research.

Although SIMPLE-PINN has produced promising outcomes on several benchmark problems, certain challenges remain. First, in multi-scale flow scenarios, the framework shows limited capability in accurately resolving fine-scale flow structures. Second, for unsteady flow simulations, while the training time is reduced compared to existing PINN models, it remains relatively longer than that required by conventional CFD approaches.  Third, in deriving the velocity and pressure correction loss terms, the contributions from neighboring nodes were neglected, which may affect the convergence rate. Future efforts will be directed toward resolving these limitations by examining techniques such as adaptive mesh refinement to enhance multi-scale flow, adaptive time-stepping algorithms to improve computational efficiency in unsteady simulations, and the adoption of advanced strategies such as SIMPLER (Semi-Implicit Method for Pressure-Linked Equations Revised) or PISO (Pressure-Implicit with Splitting of Operators) to further accelerate convergence and stability.

% \section*{Acknowledgment}
% Chang Wei and Yuchen Fan would like to acknowledge support from the China Scholarship Council for the scholarship to conduct research at Agency for Science, Technology and Research (A*STAR). This research was in part supported by the National Research Foundation, Singapore through the AI Singapore Programme, under the project ``AI-based urban cooling technology development" (Award No. AISG3-TC-2024-014-SGKR), in part supported by the National Research Foundation, Singapore, and Ministry of National Development, Singapore under its Weather Science Research Programme (Award No.: WSRP-2025-1R-02-02). 

% under xxx: [Award No. xxx].

\clearpage
\appendix

\section{Momentum equation coefficients in simplified FVM}\label{sec: appendix coefficient fvm}
When applying the simplified FVM to the momentum equations, the coefficients in Eqs.~\eqref{eq:momentum-u_fvm_implicit} and \eqref{eq:momentum-v_fvm_implicit} are given as follows:
\begin{subequations} \label{eq_coff}
\begin{align}
\small
&a_E = a_W = a_N = a_S = -\frac{1}{Re}, \\[1mm]
&a_e = u_e \, \Delta y, \\ 
&a_w = -u_w \, \Delta y, \\ 
&a_n = v_n \, \Delta x, \\ 
&a_s = -v_s \, \Delta x, \\ 
&a_P = \frac{4}{Re}.
\end{align}
\end{subequations}
\normalsize

\section{Evaluation of $b$ terms at control faces}\label{sec: b_terms}
The terms $b$ at the control faces, evaluated at both the current iteration ($n$) and the next iteration  ($n+1$), are expressed as follows:
\begin{subequations}\label{eq:b-face}
\small
\begin{align}
b_{e,u}^n &= (p_E^n - p_P^n) \, \Delta y - \frac{u_e^{\,n,t-\delta t}}{\delta t} \, \Delta x \, \Delta y, \\
b_{w,u}^n &= (p_P^n - p_W^n) \, \Delta y - \frac{u_w^{\,n,t-\delta t}}{\delta t} \, \Delta x \, \Delta y, \\
b_{n,v}^n &= (p_N^n - p_P^n) \, \Delta x - \frac{v_n^{\,n,t-\delta t}}{\delta t} \, \Delta x \, \Delta y, \\
b_{s,v}^n &= (p_P^n - p_S^n) \, \Delta x - \frac{v_s^{\,n,t-\delta t}}{\delta t} \, \Delta x \, \Delta y. \\
b_{e,u}^{\,n+1} &= (p_E^{\,n+1} - p_P^{\,n+1}) \, \Delta y - \frac{u_e^{\,n+1,t-\delta t}}{\delta t} \, \Delta x \, \Delta y, \\
b_{w,u}^{\,n+1} &= (p_P^{\,n+1} - p_W^{\,n+1}) \, \Delta y - \frac{u_w^{\,n+1,t-\delta t}}{\delta t} \, \Delta x \, \Delta y, \\
b_{n,v}^{\,n+1} &= (p_N^{\,n+1} - p_P^{\,n+1}) \, \Delta x - \frac{v_n^{\,n+1,t-\delta t}}{\delta t} \, \Delta x \, \Delta y, \\
b_{s,v}^{\,n+1} &= (p_P^{\,n+1} - p_S^{\,n+1}) \, \Delta x - \frac{v_s^{\,n+1,t-\delta t}}{\delta t} \, \Delta x \, \Delta y.
\end{align}
\end{subequations}
\normalsize

\clearpage
\section{Hyperparameter configurations for benchmark PDEs}\label{app_hyper}
\begin{table}[!ht]
\centering
\begin{threeparttable}
\caption{Training parameters and loss weights for different cases.}
\label{tab:training_parameters}
\begin{tabular}{@{}lcccccc@{}}
\toprule
Cases & LDC & Wavy & Airfoil & Square & Cylinder & RT \\
& $Re=20000$ &$ Re=100 $& $Re=1000$ & $Re=40$ & $Re=100$ & $Ra=10^6$ \\
\midrule
Init. LR & $1\times10^{-3}$ & $5\times10^{-3}$ & $5\times10^{-3}$ & $5\times10^{-3}$ & $5\times10^{-3}$ & $3\times10^{-3}$ \\
Max iter. & $10^5$ & $5\times10^4$ & $5\times10^4$ & $5\times10^4$ & $10^5$ & $1.5\times10^5$ \\
Relax. factor $\alpha$ & 0.65 & 0.95 & 0.95 & 0.85 & 0.90 & 0.95 \\
\midrule
\multicolumn{7}{@{\hspace{0.0em}}l}{\textbf{Loss weights} ($\lambda_i$)}\\
$\lambda_{\text{FVM,c}}$ & 3 & 1 & 10 & 10 & 1 & 1 \\
$\lambda_{\text{FVM,m}}$ & 3 & 1 & 1 & 1 & 1 & 1 \\
$\lambda_{\text{FVM,e}}$ & -- & -- & -- & -- & -- & 1 \\
$\lambda_{\text{AD,c}}$ & -- & 1 & 0.001 & 0.01 & 0.001 & -- \\
$\lambda_{\text{AD,m}}$ & -- & 1 & 0.001 & 0.001 & 0.001 & -- \\
$\lambda_{\text{RC}}$ & 3 & 1 & 10 & 15 & 50 & 5 \\
$\lambda_{\text{BC}}$ & 1 & 1 & 1 & 1 & 1 & 1 \\
$\lambda_{\text{IC}}$ & -- & -- & -- & -- & 50 & 5 \\
\bottomrule
\end{tabular}
\begin{tablenotes}[flushleft]
\footnotesize
\item 1. The LDC is the abbreviation for lid-driven cavity.
\item 2. The subscripts “FVM,c”, “FVM,m”, and “FVM,e” denote the residuals of the continuity, momentum, and energy equations, respectively, computed using the simplified FVM.
\item 3. The subscripts “AD,c” and “AD,m” denote the continuity- and momentum-equation residuals computed using AD.
\item 4. The subscript “RC” represents the residual-correction term. 
\end{tablenotes}
\end{threeparttable}
\end{table}

\section{Supplementary velocity and pressure fields for unsteady flow past a cylinder }\label{sec:appendix_cylinders_100}
\begin{figure}[!ht]
    \centering
    \captionsetup[subfigure]{labelformat=empty,skip=0pt,aboveskip=0pt,belowskip=-10pt}

    \begin{subfigure}[b]{0.9\linewidth}
        \centering
        \includegraphics[width=\linewidth]{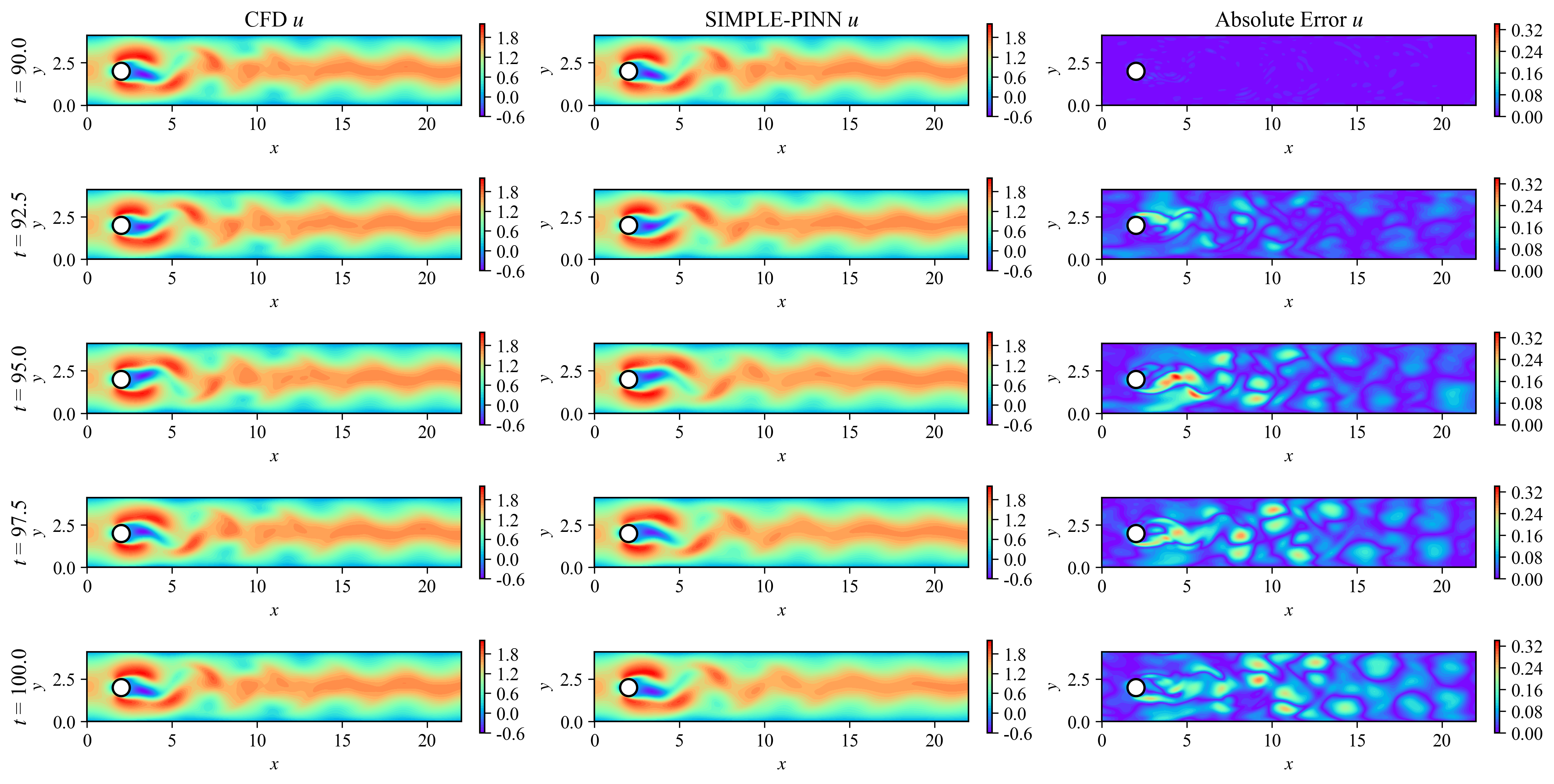}
        \caption*{} % 恢复子图标题（可自定义文字）
        \label{fig:cylinder_Re100_u}
    \end{subfigure}

    \begin{subfigure}[b]{0.9\linewidth}
        \centering
        \includegraphics[width=\linewidth]{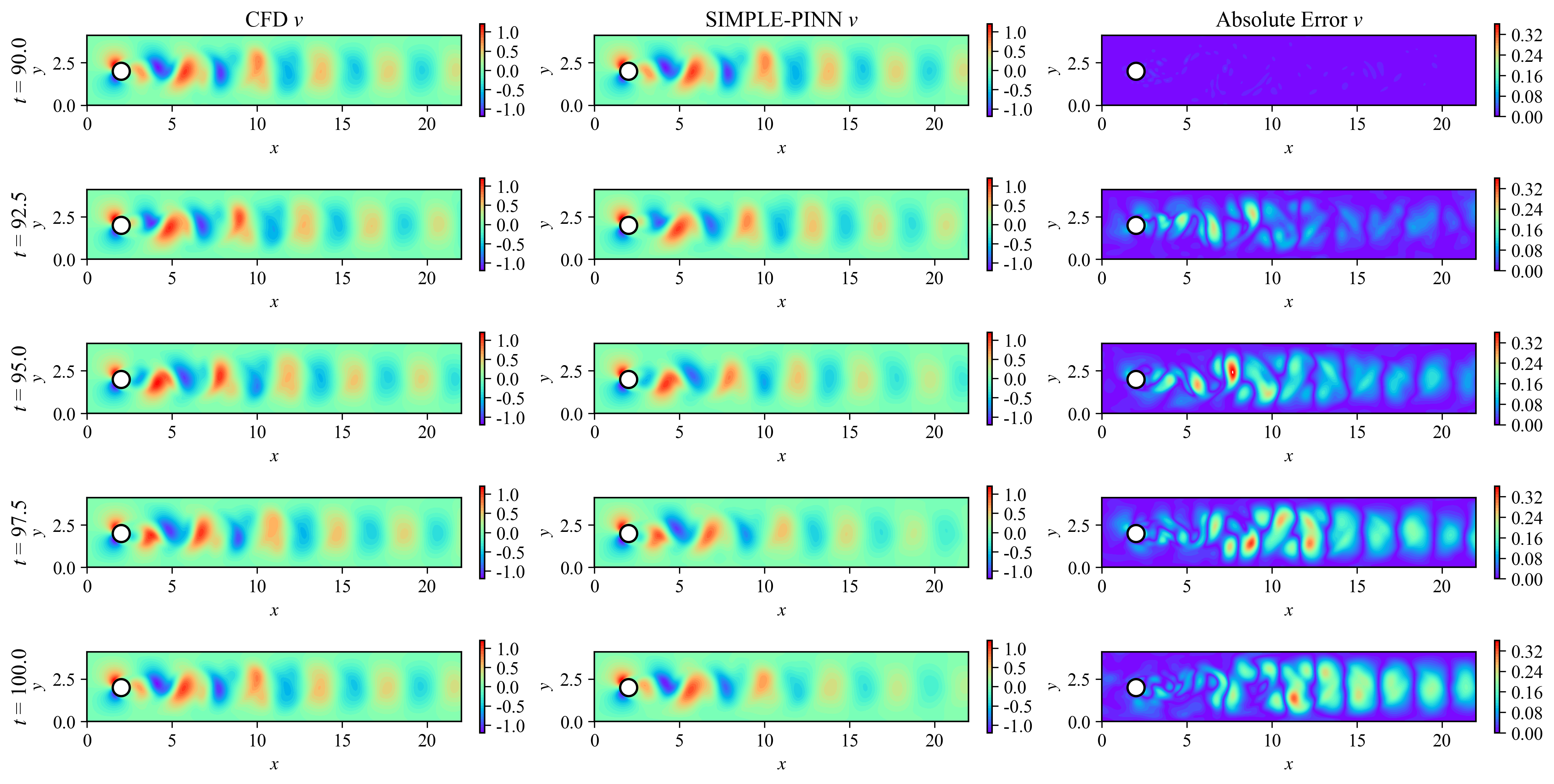}
        \caption*{}
        \label{fig:cylinder_Re100_v}
    \end{subfigure}

    \begin{subfigure}[b]{0.9\linewidth}
        \centering
        \includegraphics[width=\linewidth]{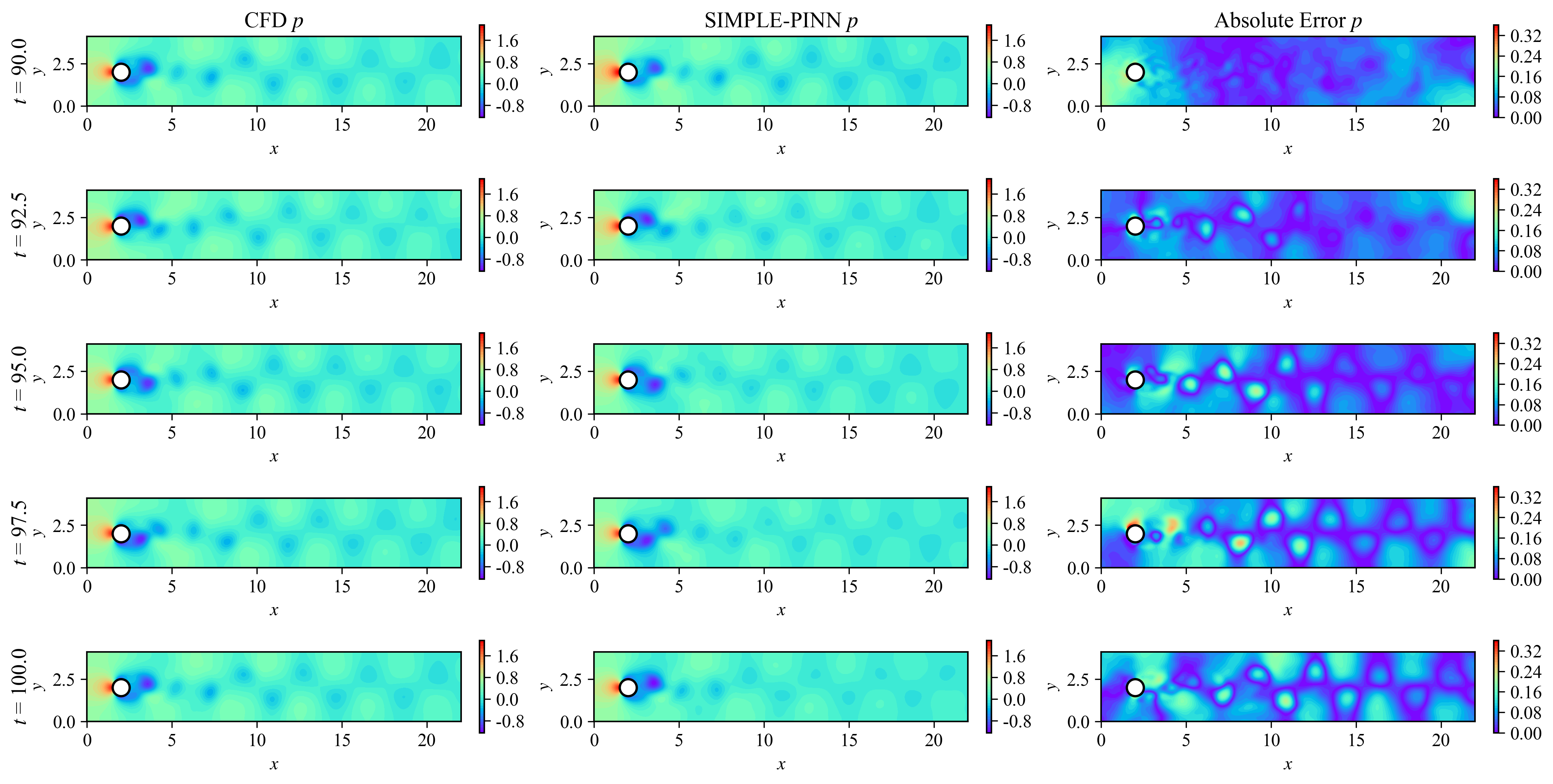}
        \caption*{}
        \label{fig:cylinder_Re100_p}
    \end{subfigure}

    \caption{Instantaneous velocity and pressure fields for flow past a cylinder at $Re=100$ as obtained from CFD and predicted by SIMPLE-PINN.}
    \label{fig:cylinder_Re100_all}
\end{figure}

\clearpage
\section{Steady flow past a cylinder}\label{sec:appendix_cylinders_low_Re}
\begin{figure}[!ht]
    \centering
    % 上方图片
    \begin{subfigure}[b]{1.0\linewidth}
        \centering
        \includegraphics[width=\linewidth]{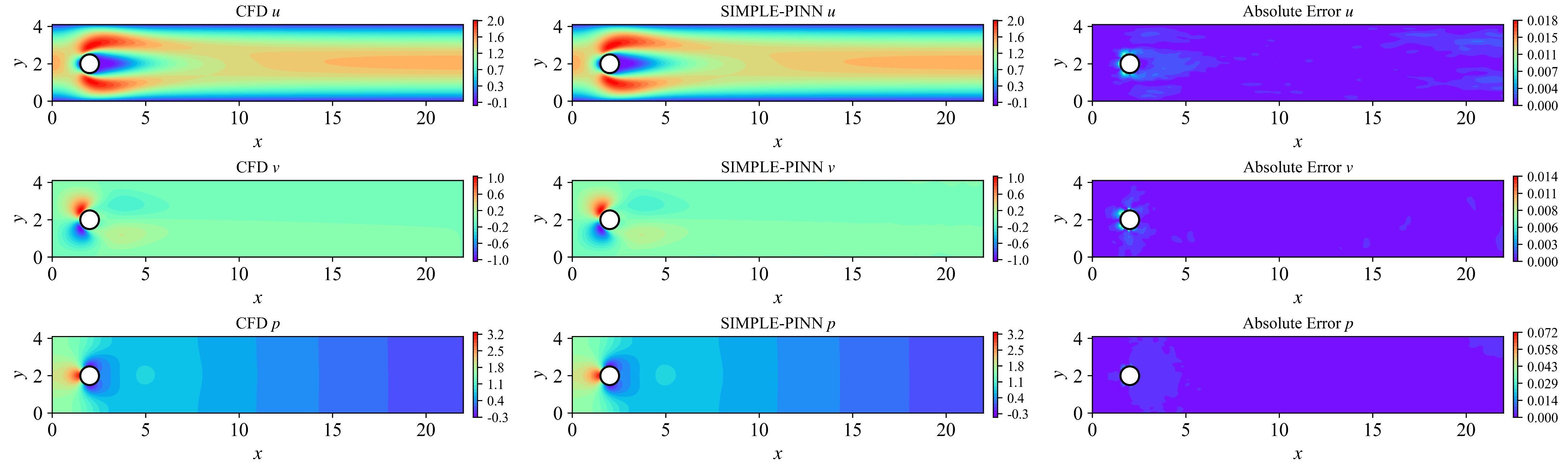}
        \caption{$Re=20$}
        \label{fig:cylinder_Re20_uvp}
    \end{subfigure}
    \vspace{0.5cm} 
    \begin{subfigure}[b]{1.0\linewidth}
        \centering
        \includegraphics[width=\linewidth]{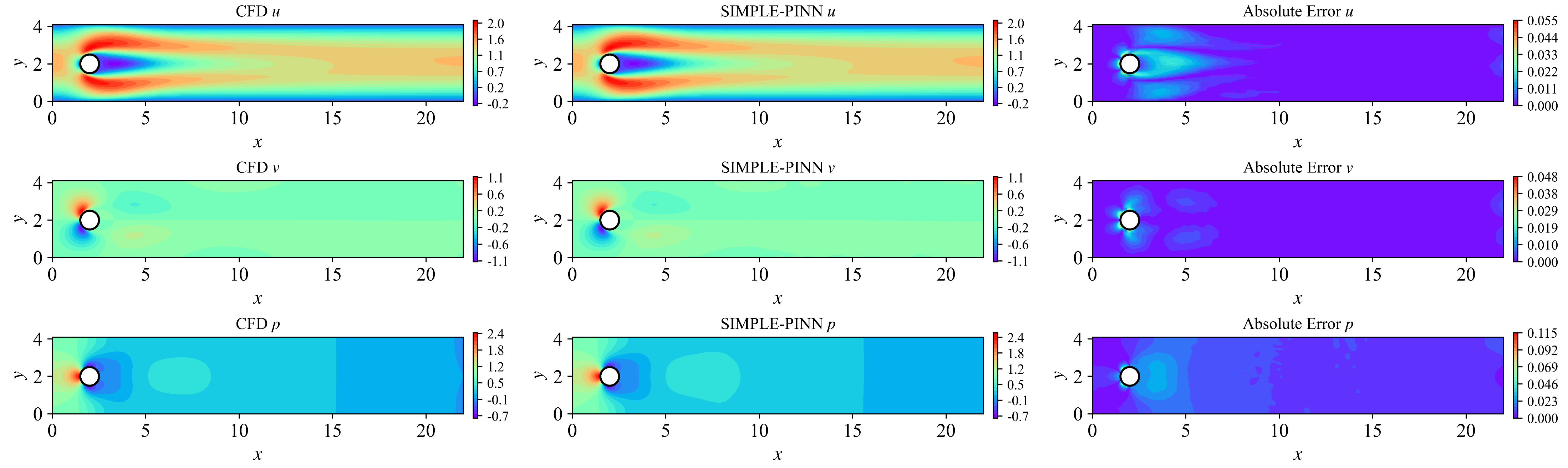}
        \caption{$Re=40$}
        \label{fig:cylinder_Re40_uvp}
    \end{subfigure}

    \caption{Velocity and pressure contours of flow past a cylinder at different \textit{Re} numbers.}
    \label{fig:cylinder_uvp_steady}
\end{figure}

\begin{figure}[!ht]
    \centering
    \begin{subfigure}[b]{0.48\linewidth}
        \centering
        \includegraphics[width=\linewidth]{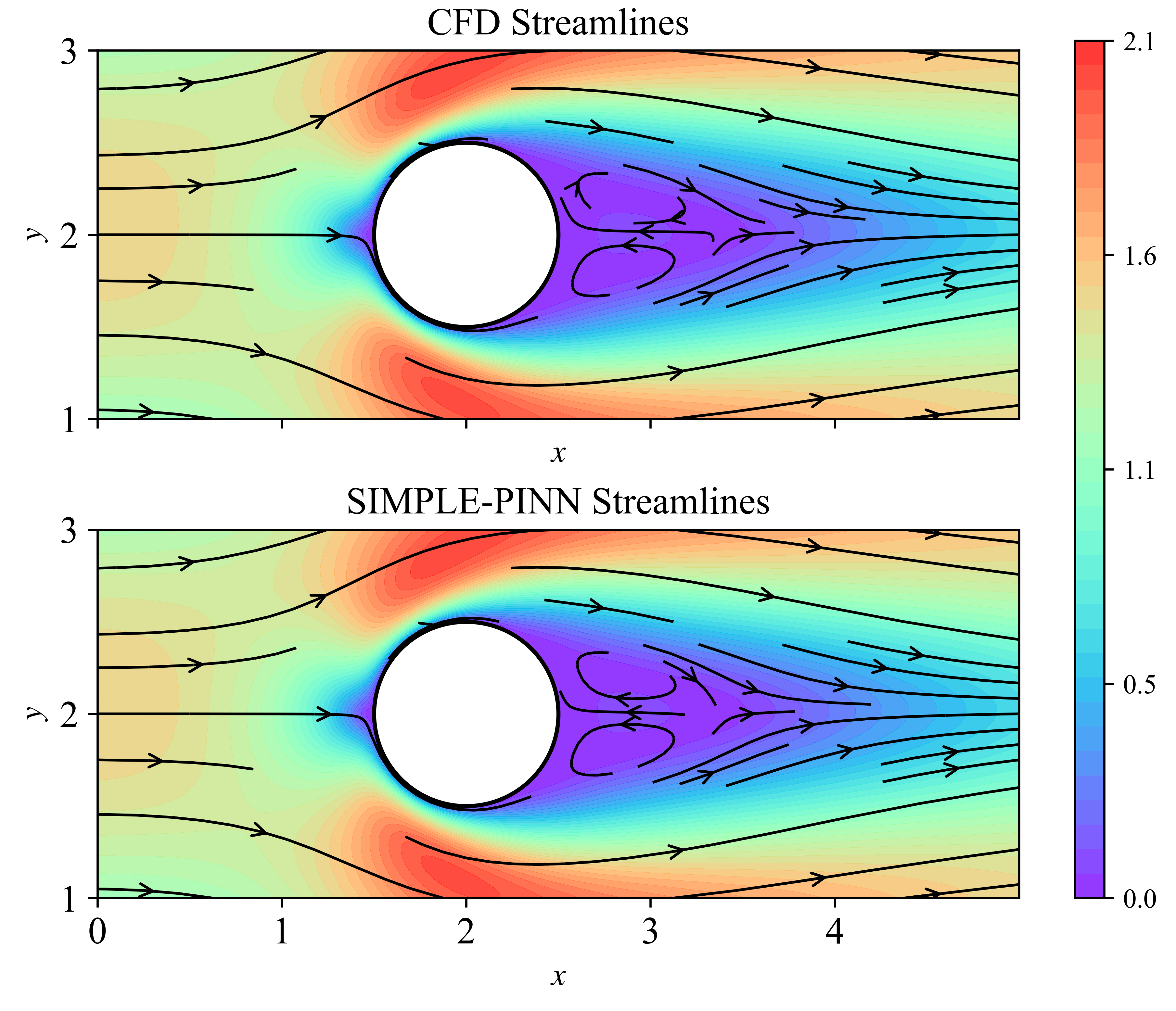}
        \caption{$Re=20$}
        \label{fig:cylinder_Re20_stlines}
    \end{subfigure}
    \hfill
    \begin{subfigure}[b]{0.48\linewidth}
        \centering
        \includegraphics[width=\linewidth]{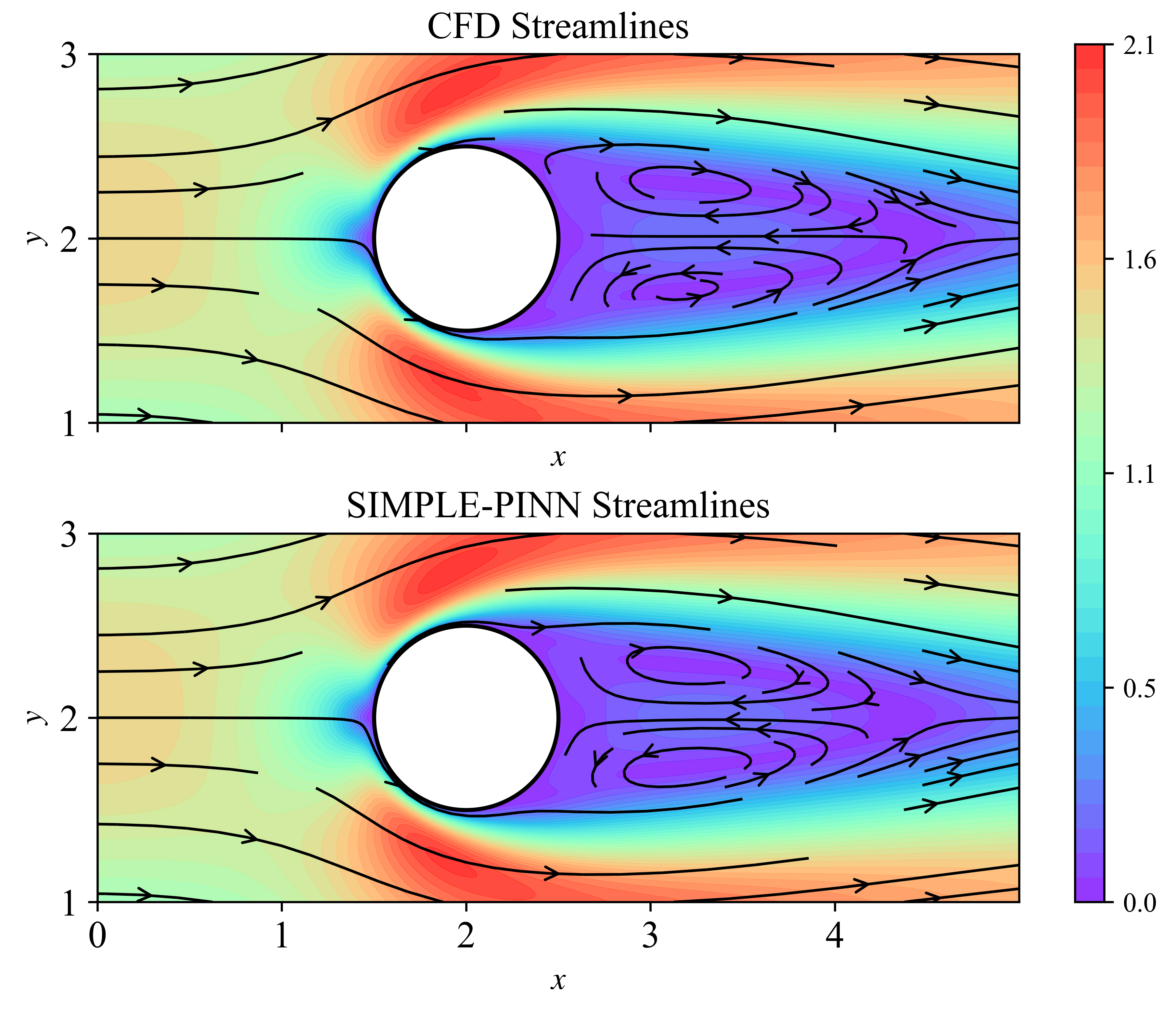}
        \caption{$Re=40$}
        \label{fig:cylinder_Re40_stlines}
    \end{subfigure}

    \caption{Streamlines of flow past a cylinder at different \textit{Re} numbers.}
    \label{fig:cylinder_stlines_steady}
\end{figure}

\begin{figure}[!ht]
    \centering
    \begin{subfigure}[b]{0.9\linewidth}
        \centering
        \includegraphics[width=\linewidth]{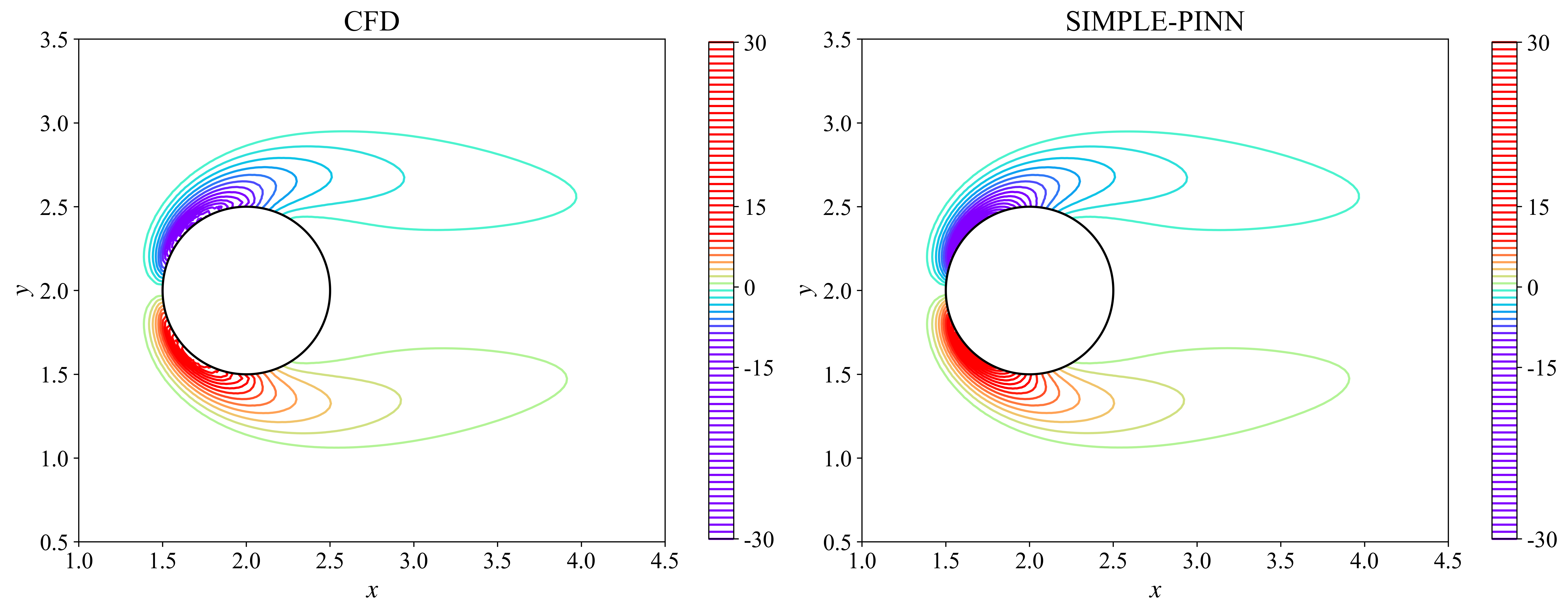}
        \caption{$Re=20$}
        \label{fig:cylinder_Re20_vorticity}
    \end{subfigure}
    
    \vspace{0.5cm} 
    
    \begin{subfigure}[b]{0.9\linewidth}
        \centering
        \includegraphics[width=\linewidth]{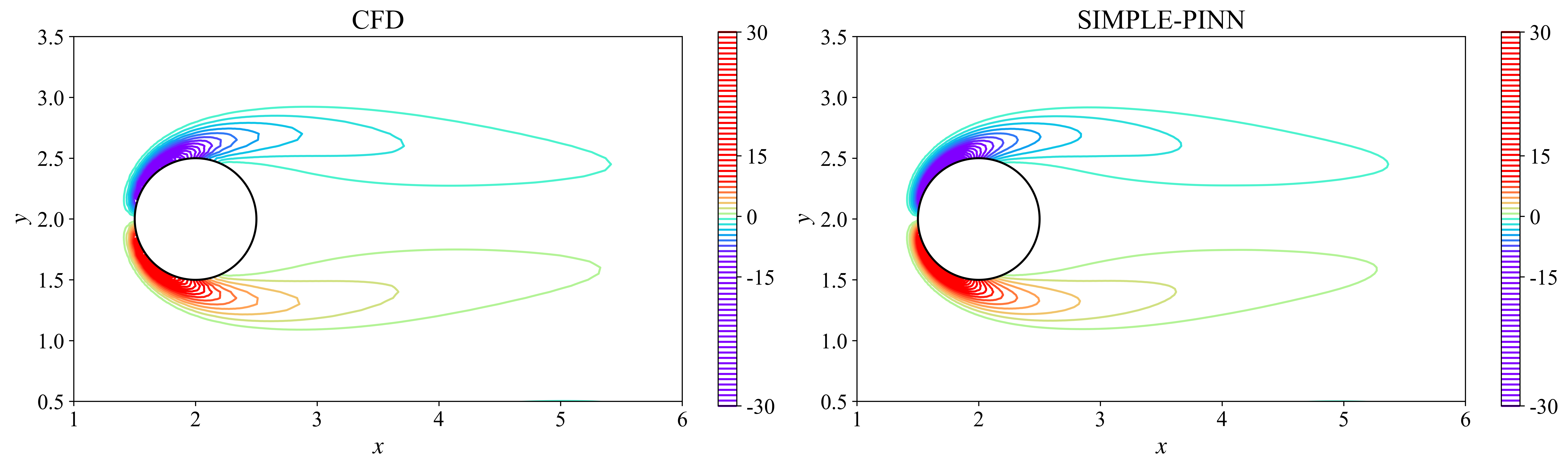}
        \caption{$Re=40$}
        \label{fig:cylinder_Re40_vorticity}
    \end{subfigure}

    \caption{Vorticity of flow past a cylinder at different \textit{Re} numbers.}
    \label{fig:cylinder_vorticity_steady}
\end{figure}

In this appendix, we present supplementary results for the steady flow past a cylinder at $Re=20$ and $Re=40$. The accompanying figures and tables provide detailed visualizations and quantitative assessments, including velocity and pressure contours (Fig.\ref{fig:cylinder_uvp_steady}), streamline patterns (Fig.\ref{fig:cylinder_stlines_steady}), vorticity fields (Fig.\ref{fig:cylinder_vorticity_steady}), and error metrics (Table\ref{tab:steady_cylinder_errors}) from the SIMPLE-PINN predictions. These results complement the main text by further demonstrating the accuracy and computational efficiency of the SIMPLE-PINN framework for steady flows past a cylinder under varying $Re$ numbers.

% \begin{table}[H]
%   \centering
%   \small
%   \caption{Quantitative evaluation of the SIMPLE-PINN for steady flow past a cylinder cases }
%   \label{tab:steady_cylinder_errors}
%   \renewcommand{\arraystretch}{1.2}
%   \resizebox{\textwidth}{!}{
%   \begin{tabular}{lccccccccc}
%     \toprule
%     $Re$ & Rel. $L_2$ $u$ & Rel. $L_2$ $v$ & Rel. $L_2$ $V$ & Rel. $L_2$ $p$ & MSE $u$ & MSE $v$ & MSE $V$ & MSE $p$ & Time (s) \\
%     \midrule
%     $20$ & $8.10\times10^{-4}$ & $4.07\times10^{-3}$ & $8.10\times10^{-4}$ & $3.85\times10^{-3}$ & $7.89\times10^{-7}$ & $1.85\times10^{-7}$ & $7.97\times10^{-7}$ & $6.34\times10^{-6}$ & 378  \\
%     $40$ & $4.26\times10^{-3}$ & $2.08\times10^{-2}$ & $4.14\times10^{-3}$ & $3.31\times10^{-2}$ & $2.19\times10^{-5}$ & $4.69\times10^{-6}$ & $2.09\times10^{-5}$ & $1.37\times10^{-4}$ & 374  \\
%     \bottomrule
%   \end{tabular}
%   }
% \end{table}

\begin{table}[!ht]
  \centering
  % \small
  \caption{Quantitative evaluation of the SIMPLE-PINN for steady flow past a cylinder}
  \label{tab:steady_cylinder_errors}
  \renewcommand{\arraystretch}{1.2}
  % \resizebox{\textwidth}{!}{
  \begin{tabular}{lcc}
    \toprule
    Metric & $Re=20$ & $Re=40$ \\
    \midrule
    Rel.$L_2$ $u$ & $8.10\times10^{-4}$ & $4.26\times10^{-3}$ \\
    Rel.$L_2$ $v$ & $4.07\times10^{-3}$ & $2.08\times10^{-2}$ \\
    Rel.$L_2$ $V$ & $8.10\times10^{-4}$ & $4.14\times10^{-3}$ \\
    Rel.$L_2$ $p$ & $3.85\times10^{-3}$ & $3.31\times10^{-2}$ \\
    \midrule
    MSE $u$        & $7.89\times10^{-7}$ & $2.19\times10^{-5}$ \\
    MSE $v$        & $1.85\times10^{-7}$ & $4.69\times10^{-6}$ \\
    MSE $V$        & $7.97\times10^{-7}$ & $2.09\times10^{-5}$ \\
    MSE $p$        & $6.34\times10^{-6}$ & $1.37\times10^{-4}$ \\
    \midrule
    Time (s)       & 378 & 374 \\
    Sample points  & $881\times165$  & $881\times165$  \\
    \bottomrule
  \end{tabular}
  % }
\end{table}

\clearpage
\section{Supplementary visualizations of the Rayleigh-Taylor instability}\label{sec:appendix_RT}

\begin{figure}[!ht]
    \centering
    % 第一行
    \begin{subfigure}[b]{0.49\linewidth}
        \centering
        \includegraphics[width=\linewidth]{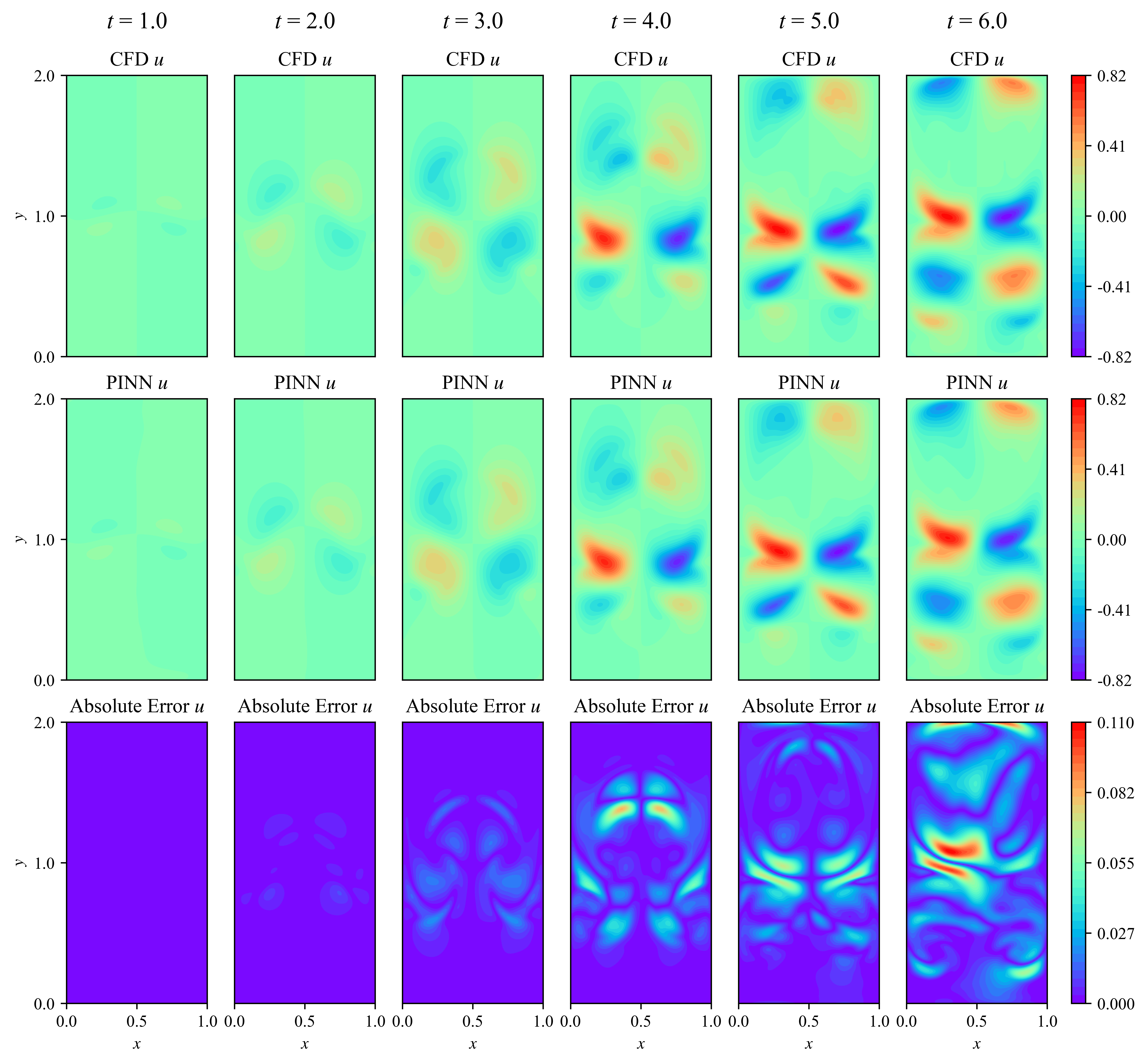}
        \caption{}
        \label{fig:RT_u}
    \end{subfigure}
    \hfill
    \begin{subfigure}[b]{0.49\linewidth}
        \centering
        \includegraphics[width=\linewidth]{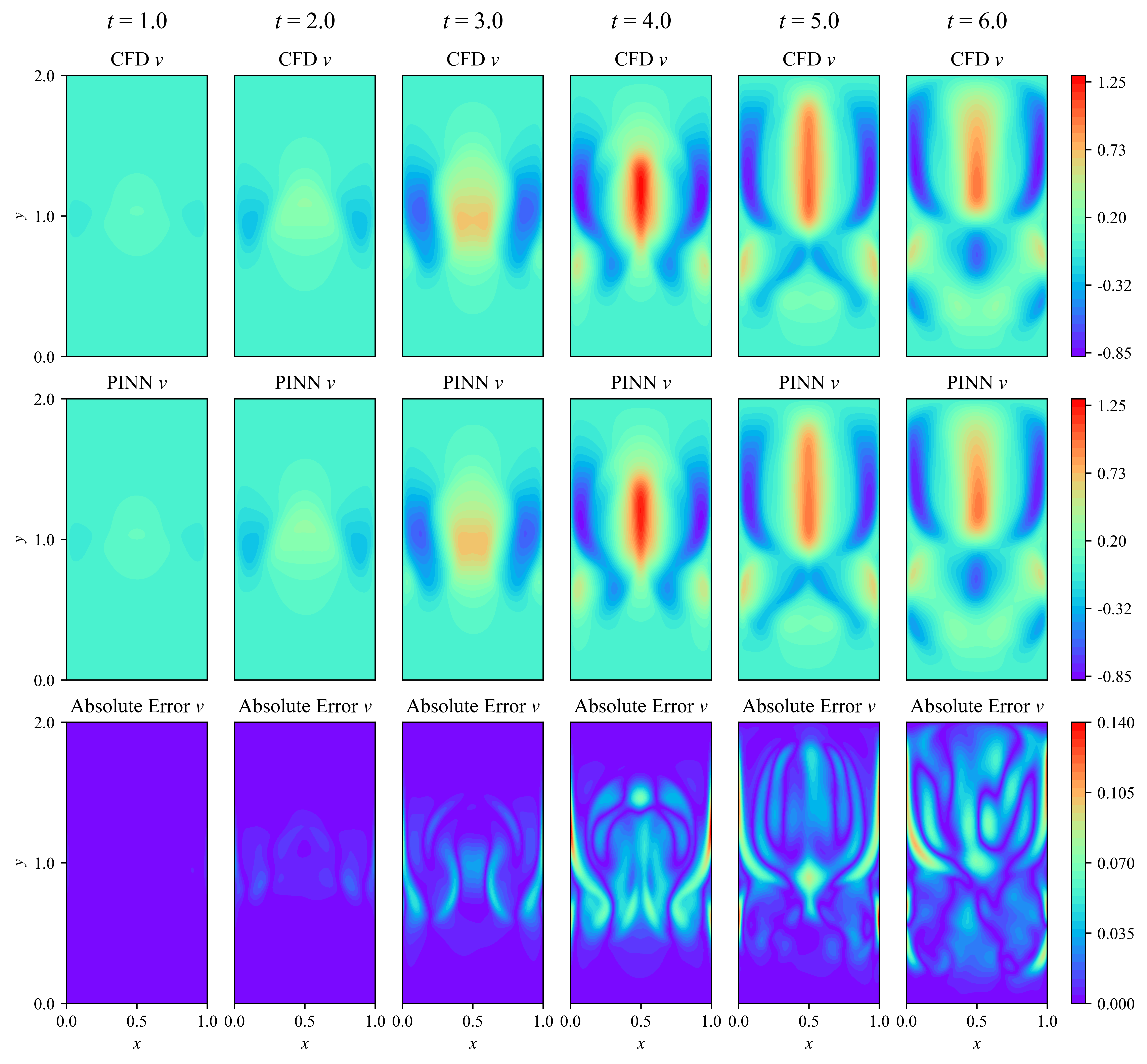}
        \caption{}
        \label{fig:RT_v}
    \end{subfigure}
    
    % 第二行
    \vspace{0.5em}
    \begin{subfigure}[b]{0.49\linewidth}
        \centering
        \includegraphics[width=\linewidth]{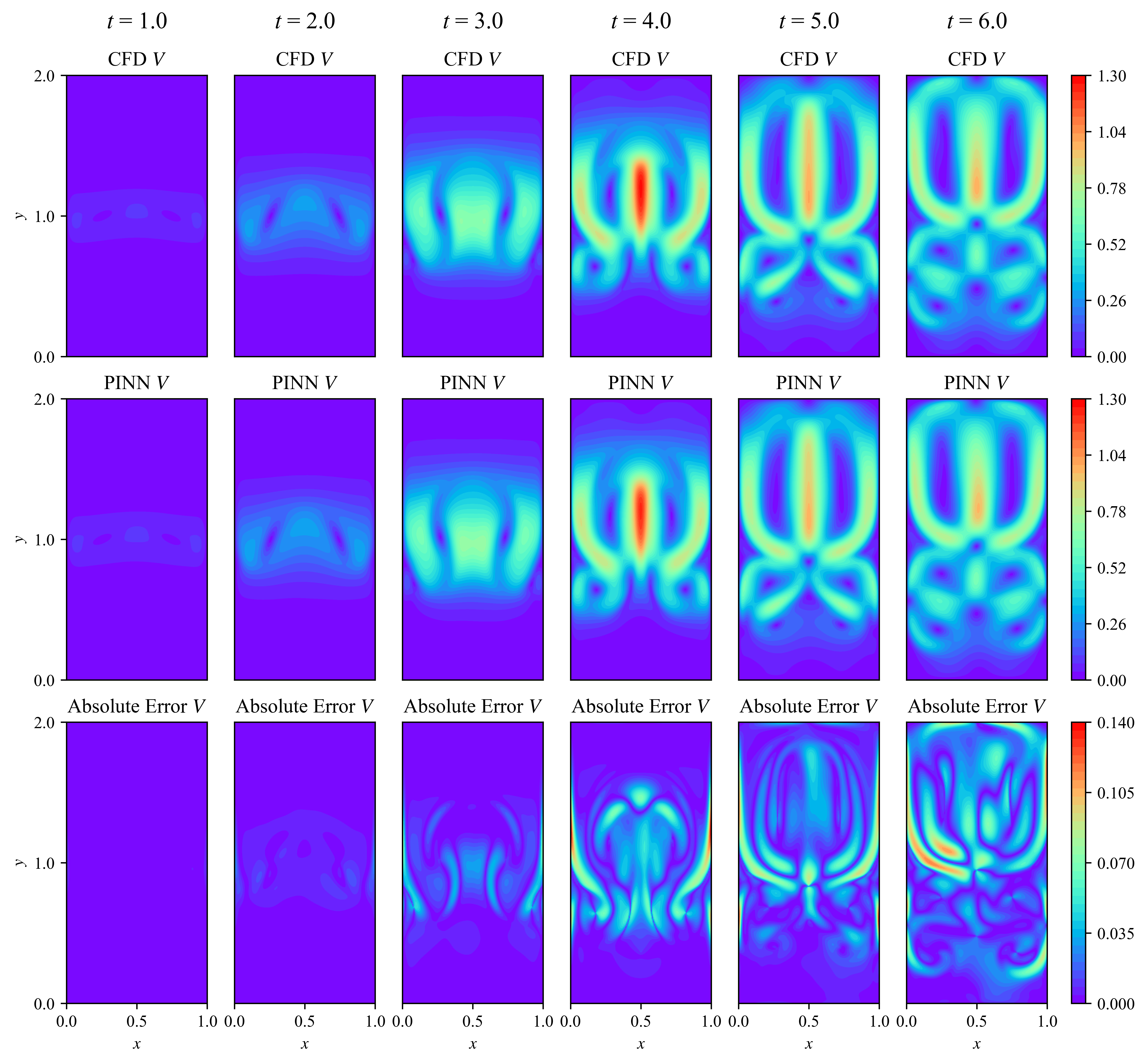}
        \caption{}
        \label{fig:RT_vmag}
    \end{subfigure}
    \hfill
    \begin{subfigure}[b]{0.49\linewidth}
        \centering
        \includegraphics[width=\linewidth]{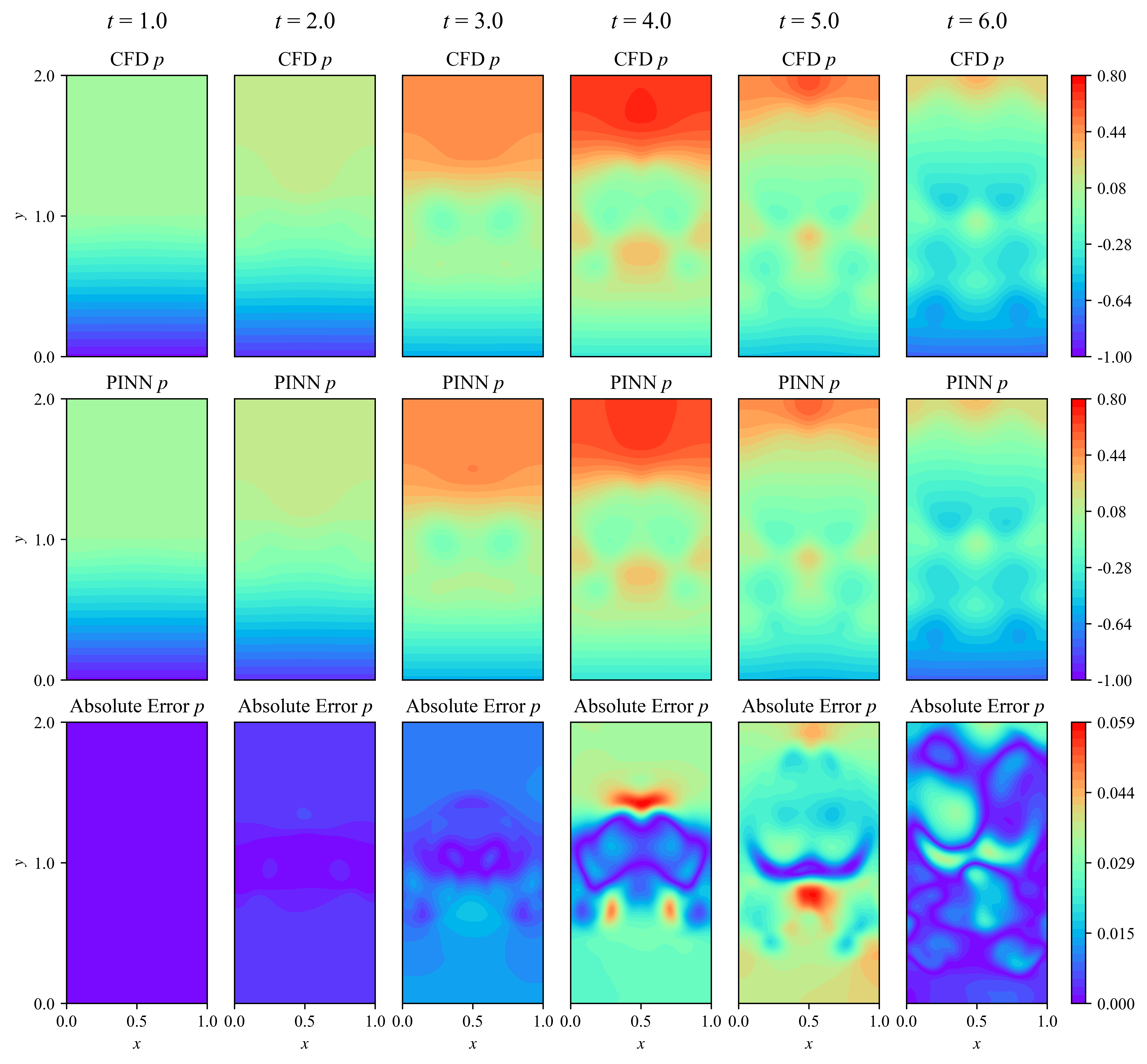}
        \caption{}
        \label{fig:RT_p}
    \end{subfigure}
    
    % 总 caption
    \caption{Comparison of flow fields predicted by SIMPLE-PINN and CFD over the time interval \(t = 1.0\)-\(6.0\). Each subfigure presents the temporal evolution of a specific flow variable: (a) horizontal velocity \(u\), (b) vertical velocity \(v\), (c) velocity magnitude \(V\), and (d) pressure \(p\). For each variable, the top, middle, and bottom rows correspond to CFD reference solutions, SIMPLE-PINN predictions, and the absolute error, respectively.}
    \label{fig:RT_all}
\end{figure}

\clearpage
\bibliographystyle{elsarticle-num} 

\bibliography{cas-refs}





% \end{thebibliography}
\end{document}